\documentclass[a4paper,10pt]{article}

\pdfoutput=1
\usepackage{jheppub}
\usepackage{array}
\usepackage{multirow}
\usepackage{amsmath}
\usepackage{makecell}
\usepackage{tensor}
\usepackage{bigints}
\usepackage{physics}

\definecolor{green}{rgb}{0.1,0.8,0.2}

\usepackage{cancel}
\usepackage{amsmath,amssymb} 
\usepackage{graphicx} 
\usepackage[dvipsnames,x11names]{xcolor} 

\usepackage{hyperref}
\usepackage{cleveref}
\usepackage{float}
\usepackage{mathtools}
\usepackage{subcaption}

\makeatletter
\newcommand{\footnoteref}[1]{\protected@xdef\@thefnmark{\ref{#1}}\@footnotemark}
\makeatother

\begin{document}

	\preprint{}
	\title{Entropy-current for dynamical black holes in Chern-Simons theories of gravity}
	\author[a]{Ishan Deo}
    \affiliation[a]{Department of Physics, Indian Institute of Technology Kanpur, Kalyanpur, Kanpur 208016, India}

    \author[a]{, Prateksh Dhivakar}
    \author[a]{, and Nilay Kundu}
    
	\emailAdd{(ishandeo, prateksh, nilayhep)@iitk.ac.in}

	\abstract{We construct an entropy current and establish a local version of the classical second law of thermodynamics for dynamical black holes in Chern-Simons (CS) theories of gravity. We work in a chosen set of Gaussian null coordinates and assume the dynamics to be small perturbations around the Killing horizon. In explicit examples of both purely gravitational and mixed gauge gravity CS theories in $(2+1)$ and $(4+1)$-dimensions, the entropy current is obtained by studying the off-shell structure of the equations of motion evaluated on the horizon. For the CS theory in $(2+1)$ dimensions, we argue that the second law holds to quadratic order in perturbations by considering it as a low energy effective field theory with the leading piece given by Einstein gravity. In all such examples, we show that the construction of entropy current is invariant under the reparameterization of the null horizon coordinates. Finally, extending an existing formalism for diffeomorphism invariant theories, we construct an abstract proof for the linearised second law in arbitrary Chern-Simons theories in any given odd dimensions by studying the off-shell equations of motion. As a check of consistency, we verify that the outcome of this algorithmic proof matches precisely with the results obtained in explicit examples.}


\maketitle


\section{Introduction} \label{intro}

Classical black hole solutions in Einstein's theory of general relativity (GR) obey a set of geometric relations that are analogous to the laws of thermodynamics \cite{Bardeen:1973gs, Bekenstein:1973ur}. Hawking's independent calculation of a black hole's temperature and radiation \cite{Hawking:1974sw} asserts that these laws are not just analogies.  In such a theory, the area of the black hole horizon defines its entropy. However, GR is not the complete framework to explain all the gravitational phenomena in our universe. Also, any UV complete theory of quantum gravity, upon taking the consistent low energy effective theory limit, will generate corrections to the leading Einstein-Hilbert term in the Lagrangian \cite{Donoghue:1994dn, PhysRevLett.55.2656, Zwiebach:1985uq}\footnote{GR is a two-derivative theory of gravity since the Einstein-Hilbert Lagrangian contains two derivatives of the space-time metric variable. These correction terms involve more than two derivatives of the metric theories of gravity.}. These theories beyond Einstein's GR are generally called higher derivative theories of gravity.  Interestingly, to prove the laws of black hole mechanics, we need to know the theory of gravitation from which the black holes are obtained as classical solutions\footnote{It is expected that the black hole solutions in higher derivative gravity theories should continue behaving like thermodynamic objects. However, the geometric definitions of various thermodynamic quantities, like entropy, should be modified from their definitions in GR.}. The works of Iyer and Wald \cite{wald1993, Iyer:1994ys} derived a definition of entropy for stationary black holes, called Wald entropy, consistent with the first law of thermodynamics for any theory of gravity with a diffeomorphism invariant Lagrangian density. However, the Wald entropy suffers from ambiguities for dynamical black holes \cite{Jacobson:1993vj}, and we still lack one complete proof of the second law for such dynamical black holes in a general diffeomorphism invariant theory of gravity. 

Recently, significant progress was made in \cite{Wall:2015raa} on the question of whether we can define a black hole entropy in arbitrary diffeomorphism invariant theories of gravity that obey the classical second law. In this work, the dynamics of the black hole were approximated to be small fluctuations around a stationary black hole, and the analysis was performed perturbatively to the linear order in the amplitude of the fluctuations. Previously, this question of the second law was pursued for model-specific theories \cite{Jacobson:1993xs, Chatterjee:2013daa, Sarkar:2013swa, Bhattacharjee:2015yaa, Bhattacharjee:2015qaa, Sarkar:2019xfd, Wang:2020svl}. Building on \cite{Wall:2015raa}, in \cite{Bhattacharya:2019qal, Bhattacharyya:2021jhr}, an entropy current was constructed on the null horizon of the dynamical black holes up to linear order in fluctuations. The null component of this entropy current gives us the local entropy density, and its spatial component signifies a flux of in and out flow of entropy on the spatial sections of the horizon. Working to the linear order in the amplitude expansion, it was shown that this entropy current is divergenceless and thereby establishing an ultra-local (i.e., local both in the temporal and spatial extent of the horizon) version of the linearized second law \footnote{To derive these results, it was assumed that the matter sector was minimally coupled and satisfied the null energy condition (NEC). In a follow-up paper \cite{Biswas:2022grc}, this construction of entropy current at linear order was extended to theories involving non-minimally coupled matter as well.}. 

The main idea behind the construction of an entropy current in \cite{Bhattacharya:2019qal, Bhattacharyya:2021jhr} was to study the off-shell structure of the equations of motion (EoM) on the null horizon to the linear order in the fluctuations. After choosing a gauge for the metric and using a boost symmetry of the near horizon region, a specific off-shell form of the null projected EoM was established, from which the components of the entropy current can be read off \footnote{We would like to highlight that obtaining this specific off-shell form of the EoM components is an independent result with possible applications beyond the linearized second law.}. 

Furthermore, in \cite{Hollands:2022fkn, Davies:2022xdq}, the second law was established to the quadratic order in the fluctuations by treating the higher derivative terms in the Lagrangian in an effective field theory (EFT) perspective. In such an EFT approach, the Lagrangian of the gravitational theory appears as a summation of various terms (each of which is a scalar under diffeomorphism), all arranged according to the increasing number of derivatives present in each of these terms, starting with the leading two-derivative Einstein-Hilbert term for GR. 

It is important to note that the derivation of black hole entropy by Iyer and Wald  \cite{LeeWald, wald1993, Iyer:1994ys}, consistent with the first law of thermodynamics, and all the subsequent developments mentioned above in \cite{Wall:2015raa, Bhattacharya:2019qal, Bhattacharyya:2021jhr, Hollands:2022fkn}, are based on Noether formalism which heavily relies on diffeomorphism invariance. However, as low energy effective theories of gravity beyond GR, we have another class of very interesting theories which are not invariant under diffeomorphisms. These theories are known as the Chern-Simons (CS) theories \cite{Chern:1974ft}. For example, they appear naturally as low-energy effective theories resulting from compactifications of string theories \cite{ALVAREZGAUME1984269, GREEN1984117}. CS theories can be of the purely gravitational type or purely gauge type, or of mixed gauge gravity type. The Lagrangians of CS theories are written in odd space-time dimensions, and they explicitly involve the Christoffel symbols $\Gamma^\mu_{\nu\rho}$ or the gauge field $A_\mu$, thus justifying why they are not diffeomorphism invariant \footnote{As we will see later, strictly speaking, CS theories are diffeomorphism invariant up to total derivative terms involving non-covariant $\Gamma^\mu_{\rho\nu}$ and $A_{\mu}$. Additionally, although the Lagrangian is not a diffeomorphism covariant object, the EoMs turn out to be covariant.}. Previously, in the literature, people have already studied the particular case of three-dimensional gravitational CS terms from various different perspectives, most importantly in the context of topologically massive gravity theories \cite{Deser:1981wh} and from holographic perspectives \cite{Solodukhin:2005ah, Kraus:2006wn, Sahoo:2006vz}. 

We will, however, focus on CS theories in the context of black hole entropy. In \cite{wald1993, Iyer:1994ys}, the authors mainly used Noether principles to compute black hole entropy based on covariant phase space formalism applicable for manifestly diffeomorphism covariant theories \cite{LeeWald}. In an important work, Tachikawa \cite{Tachikawa:2006sz} suggested an extension of the Lee-Iyer-Wald prescription to CS theories by pointing out necessary modifications in defining a stationary black hole entropy consistent with the first law. Later in \cite{Bonora:2011gz}, and subsequently in \cite{Azeyanagi:2014sna}, Tachikawa's proposal was studied in detail, particularly focusing on refinements of the covariant phase space formalism, making it suitable for CS theories. They worked out a general formula for black hole entropy in theories with arbitrary CS terms in the Lagrangian (for both purely gravitational CS and mixed gauge gravity CS terms).

In this paper, our aim is to go beyond the first law and argue for the second law for dynamical black holes in CS theories, both in purely gravitational and mixed gauge gravity cases. It will be assumed that there are dynamical black hole solutions in CS theories \footnote{Such dynamical black hole solutions were constructed in \cite{Yunes:2009hc, Alexander:2009tp}. An important class of solutions was also constructed in \cite{Gauntlett:1998fz, Gauntlett:2003fk, PhysRevD.73.044006,Grumiller:2007rv}.}, such that the dynamics can be considered as small fluctuations around the stationary black holes, and they can be treated perturbatively with the amplitude of the fluctuations being the small parameter. This way, we can see that the formalism developed in \cite{Bhattacharya:2019qal, Bhattacharyya:2021jhr, Biswas:2022grc} can be directly employed. Following that set up, we will analyze the null-projected components of the EoM and will show that it has the desired off-shell structure when evaluated on the horizon (refer to eq.\eqref{eom1}). Consequently, we will obtain an entropy current on the horizon of dynamical black holes in CS theories. One consistency of our entropy current would be to verify that the entropy density reproduces Tachikawa's result for stationary black holes. Additionally, we will get the spatial component of the entropy current, suggesting possible in/out flux of entropy density on a spatial slice of the horizon.  Before summarizing the outline of our paper, we note that the concept of an entropy current for Chern-Simon theories has been considered before in \cite{Chapman:2012my}. However, there it was constructed in the context of fluid gravity correspondence \cite{Bhattacharyya:2008jc,Bhattacharyya:2008xc}, which is different from what we do in this paper. Also, see \cite{Eling:2012xa,A:2022qgc} for similar constructions of entropy current in fluid-gravity correspondence.

Let us now outline the organization of the rest of this paper. In \S\ref{msec:basicsetup}, we will briefly overview the background setup and the working principle. Following that in \S\ref{sec:bruteforce}, we will consider specific examples of CS theories in $(2+1)$ and $(4+1)$ dimensions, for which we know the EoMs explicitly, and we will work out the required components of EoM in a brute-force manner up to linearized order in the fluctuations. In working out the off-shell structure of the EoM, the boost symmetry of the near horizon stationary black hole and its perturbative breaking due to the dynamics will be used. This will readily give us the components of the entropy current on the horizon. Using the same logic as in \cite{Bhattacharya:2019qal, Bhattacharyya:2021jhr}, this entropy current will have zero entropy production by construction for on-shell configurations upto linearized order in the fluctuations. Consequently, the linearized second law will follow automatically, at least in those examples of CS theories we have studied. 

It is also important to note that working within the linearized approximation for the dynamics is very restricted in its scope to prove a second law since there is no actual entropy production to this order. One can, at the most, show that entropy is not destroyed. The effects of non-linear perturbations around stationarity must be incorporated to observe entropy production. For Gauss-Bonnet theory and other members of the Lovelock family, the second law was attempted in \cite{Bhattacharyya:2016xfs} to all non-linear orders of the fluctuation. Recently in \cite{Hollands:2022fkn}, the authors  looked at diffeomorphism invariant higher derivative theories in an EFT setup discussed above and proved a second law to the quadratic order in perturbations. In \S\ref{nlCS2+1}, we will consider CS theory in $(2+1)$ dimensions in the same EFT setup. There will be two terms in the Lagrangian, the leading one being the standard two derivative Einstein Hilbert term plus the sub-leading three derivative CS term. As we discussed before, terms with a higher number of derivatives will become less and less important. Thereafter, following the arguments presented in \cite{Hollands:2022fkn}, we will see that the second law can be proved in this theory within a double perturbation series, i.e., to the quadratic order in the fluctuations but ignoring terms that involve more than three derivatives on the metric.

The construction of entropy current relies heavily on a specific choice of coordinates and signifies entropy production at the horizon. It is quite natural then to expect that such a physical process should not depend on the choice of coordinate system. In other words, the covariance of entropy production under reparametrization of the horizon slicing should be a consistency check for the expressions of the entropy current. This has been verified in \cite{Bhattacharyya:2022njk} for a specific theory of higher derivative gravity, namely the Gauss-Bonnet gravity. Later, it has also been studied in \cite{Hollands:2022fkn} for generic diffeomorphism-invariant theories. Following these methods, in \S\ref{sec:reparam2+1d}, we will also check that for CS theories, both in $(2+1)$ and $(4+1)$ dimensions, the entropy current that we obtained previously indeed transforms covariantly under reparametrization of the horizon slicing. As we will argue later, this consistency check also justifies the need to consider the spatial components of the entropy current on the horizon. 

Going beyond working with specific examples, in \S\ref{sec:absprf}, we will abstractly prove that for generic CS theories in odd space-time dimensions, the null-projected components of the EoMs will always have the desired off-shell structure. In other words, going beyond the brute force calculation of EoMs, which is to be performed case by case in a theory with a given Lagrangian, an algorithm can be developed to construct an entropy current for CS theories with non-negative divergence up to linearized order in the dynamical fluctuations around a stationary black hole. For diffeomorphism invariant theories, such a proof has been worked out in \cite{Bhattacharyya:2021jhr}. The analysis in \cite{Bhattacharyya:2021jhr} essentially used elements from covariant phase space formalism adapted to a specific coordinate system and the chosen metric gauge but only applicable to diffeomorphism invariant theories. Thus, we will follow a similar set of principles outlined in \cite{Bhattacharyya:2021jhr} but use results from a modified covariant phase space formalism for CS theories, which was developed in \cite{Tachikawa:2006sz, Bonora:2011gz, Azeyanagi:2014sna}. This exercise will also lead us to an alternative definition of the components of the entropy current written in terms of objects from the covariant phase space formalism (e.g., pre-symplectic potential, current, Noether charge, etc.). In \S\ref{sec:proofverf}, we will explicitly check that the entropy current obtained from this abstract algorithm matches the results obtained from a brute force calculation of EoM for specific model examples of CS theories we have already studied. 

Finally, we will conclude with a summary of our results and some comments in \S\ref{sec:concl}. There are additional appendices containing technical details of our calculations that will be omitted in the main text. 
\section{Basic set up, and the working principle to construct an entropy current} \label{msec:basicsetup}
This section will review various essential elements of constructing an entropy current for dynamical black holes. In \S\ref{sec:basicsetup}, we will present the basic setup drawing from the formalism developed for diffeomorphism invariant theories, mainly focusing on \cite{Wall:2015raa, Bhattacharya:2019qal, Bhattacharyya:2021jhr, Biswas:2022grc, Bhattacharyya:2022njk, Hollands:2022fkn}. Next, in \S\ref{sec:cscovphase}, we review the elements of covariant phase space, highlighting the modifications one must consider for working with CS theories.

\subsection{Review of the setup for diffeomorphism invariant theories}
\label{sec:basicsetup}

\subsection*{A generic choice of horizon adapted coordinates:} 
Without any loss of generality, we make a choice of coordinates and also work in a chosen gauge for the metric, known as Gaussian normal coordinates, in $d$ dimensions as mentioned below 
\begin{equation} \label{nhmetric}
ds^2 = 2 \, dv \, dr -r^2 X(r, \, v, \, x^i) \, dv^2 + 2 \, r \, \omega_i(r, \, v, \, x^i) dv dx^i + h_{ij}(r, \, v, \, x^i) \, dx^i dx^j \, . 
\end{equation} 
This metric describes the space-time in the neighborhood of a co-dimension one null hypersurface, denoted by $\mathcal{H}$ and placed at $r=0$. The co-dimension two constant $v$-slices of the horizon, indicated by $\mathcal{H}_v$, is spanned by the $(d-2)$ spatial coordinates $x^i$, and has the induced metric given by $h_{ij}$. Within our choice, $v$ and $r$ are affine parameters, and thus $\partial_v$ denotes the generators for affinely parametrized null geodesics on $\mathcal{H}$ \footnote{See \S 2.1 of \cite{Bhattacharyya:2021jhr} for details.}. With the understanding that the event horizon of a black hole is a null hypersurface, we can consider the metric gauge in eq.\eqref{nhmetric} describing the near horizon region of that black hole close to a final equilibrium configuration. 
\subsection*{Stationary black holes and linearized perturbations around it at $\mathcal{O}(\epsilon)$:}
An equilibrium configuration of a black hole is denoted by a stationary metric. With the choice of metric gauge mentioned above, a stationary black hole (at least the near horizon region of it) can always be written in the form eq.\eqref{nhmetric}, with further restrictions on the functional dependence of the metric coefficients ($X$, $\omega_i$, and $h_{ij}$) on the coordinates as follows 
\begin{equation} \label{eqlbmMetric}
X^{eq} = X^{eq} (rv, \, x^i), \quad \omega^{eq}_i = \omega^{eq}_i (rv, \, x^i), \quad h^{eq}_{ij} = h^{eq}_{ij} (rv, \, x^i). 
\end{equation}
For stationary black holes, with eq.\eqref{eqlbmMetric}, we get a Killing vector 
\begin{equation} \label{KillVec}
\xi = v \, \partial_v -r \,  \partial_r \, ,
\end{equation}
such that the Lie derivative with respect to $\xi$ vanishes for the equilibrium metric, $\mathcal{L}_\xi g^{eq}_{\mu\nu}|_{r=0}=0$. Also, the norm of $\xi$ vanishes on $r=0$, and, hence, it becomes a Killing horizon; see \S2.2 of \cite{Bhattacharyya:2021jhr} for details. Notably, in generic higher derivative gravity theory, there is no general proof that the event horizon of a stationary black hole is a Killing horizon. The technical setup that will be followed applies to a Killing horizon, and we are assuming that the metric in eq.\eqref{eqlbmMetric} with eq.\eqref{nhmetric} describes a stationary black hole with its event horizon being a Killing horizon \footnote{See the discussion in \S 1.5 of \cite{Hollands:2022fkn}.}. Actually, we have a bifurcate Killing horizon at $r=0$ \footnote{Recently, in \cite{Bhattacharyya:2022nqa} the Zeroth law was proved for a diffeomorphism invariant theory considered in an EFT expansion. This implies that once the Zeroth law is imposed, that is established, the space-time can always be brought to eq.\eqref{nhmetric} with eq.\eqref{eqlbmMetric}. See Appendix A of \cite{Bhattacharya:2019qal} for details.}. 

To define a stationary configuration involving a $U(1)$ gauge field $A_\mu$, we need to specify if our definition of stationarity is consistent with the $U(1)$ gauge transformations, $A_\mu \rightarrow A_\mu + D_\mu \Lambda$, where $D_\mu$ is the covariant derivative with respect to the full dynamical metric eq.\eqref{nhmetric}. Following the convention mentioned in \cite{Biswas:2022grc}, we define the following definition of stationarity that is also $U(1)$ gauge invariant,
\begin{equation}
    \mathcal{L}_\xi A^{eq}_{\mu}+ D_\mu \Lambda = 0 \, ,
\end{equation}
where $\Lambda$ is the parameter of $U(1)$ gauge transformations. In our horizon adapted coordinates eq.\eqref{nhmetric} and eq.\eqref{eqlbmMetric}, this becomes the following 
\begin{equation} \label{eqlbmA}
     (A^{eq}_{\mu} \xi^\mu + \Lambda) |_{r=0}= (v \, A^{eq}_{v} + \Lambda) |_{r=0}=0 \, ,
\end{equation}
 on the horizon (see \S3 in \cite{Biswas:2022grc} for details).

Having discussed the equilibrium configuration for both the metric and the gauge field, we can now define the black hole's out-of-equilibrium (i.e., non-stationary) dynamics. In our work, we consider only linearized dynamical fluctuations around the stationary background configuration as follows
\begin{equation} \label{defnoneqlbm}
\begin{split}
X =& \, X^{eq} (rv, \, x^i) +\epsilon \, \delta X (r, \, v, \, x^i), \quad  \omega_i =\, \omega_i^{eq} (rv, \, x^i) +\epsilon \, \delta \omega_i (r, \, v, \, x^i),  \\ h_{ij} = &\,h_{ij}^{eq} (rv, \, x^i) +\epsilon \, \delta h_{ij} (r, \, v, \, x^i), \quad
A_{\mu} = A_\mu^{eq} (rv, \, x^i) +\epsilon \, \delta A_\mu(r, \, v, \, x^i) \, ,
\end{split}
\end{equation}
where $\epsilon$ is the amplitude of the fluctuations. Working to the linearized order in a perturbative expansion in the amplitude essentially means that we keep terms of $\mathcal{O}(\epsilon)$ but neglect all terms of $\mathcal{O}(\epsilon^n)$ with $n\ge 2$. It should be noted that the difference between the equilibrium quantities and the non-stationary fluctuations is in their functional dependence on the $v$ coordinate, e.g., $X^{eq}$ can depend only on the product $vr$ as opposed to $\delta X$ depending arbitrarily on $v$ and $r$.
\subsection*{Boost symmetry of near horizon stationary metric and classifying terms with boost weight:}
The near horizon metric of a stationary black hole, as in eq.\eqref{eqlbmMetric}, has the following symmetry, more specifically, an isometry preserving the gauge eq.\eqref{nhmetric}, 
\begin{equation} \label{boosttrns}
    r \rightarrow \tilde{r} = \lambda \, r , \, \, v \rightarrow \tilde{v} = v/\lambda \, , \quad (\text{with} \, \, \lambda \, \, \text{being a free constant parameter}) \, ,
\end{equation}
which we call the boost symmetry \footnote{Actually, there is a larger class of isometry preserving our choice of metric gauge where $\lambda$ can be $x^i$ dependent function, see \S2.1 of \cite{Bhattacharyya:2021jhr} for details. For our purpose of constructing the entropy current on the horizon with a fixed choice of coordinates, it is sufficient to work with constant $\lambda$. However, as we will see later, the $x^i$ dependence of $\lambda$ will be important to understand the reparametrization covariance of the entropy production with the change of coordinates on the horizon.}. An infinitesimal boost transformation is generated by the Killing vector $\xi$ defined in eq.\eqref{KillVec}. The product $rv$ is invariant under the boost transformation eq.\eqref{boosttrns}, and, therefore, it is justified why the $v$ dependence of all equilibrium quantities (e.g., $X^{eq}$), if any, must always be through $rv$. 

As explained in detail in \cite{Bhattacharyya:2021jhr} (see \S2.2 and \S2.3 therein), this boost symmetry is crucial for us to classify various terms depending on how they transform under eq.\eqref{boosttrns}. To felicitate this classification, we assign boost-weight to a generic term, say $\mathcal{P}$, in the following way
\begin{equation}  \label{defboostwt} 
\text{Boost weight of $\mathcal{P}$ is $w$ if,} \quad \mathcal{P} \rightarrow \widetilde{\mathcal{P}} = \lambda^w \, \mathcal{P}\, , \quad \text{under eq.\eqref{boosttrns}} \, .
\end{equation}
Now, it is straightforward to see that $X^{eq}, \, \omega_i^{eq}, \, h_{ij}^{eq}$ are all boost-invariant objects. Additionally, the derivative operator $\nabla_i$ (the covariant derivative compatible with the induced metric $h_{ij}$) is also boost-invariant, but $\partial_v$ and $\partial_r$ have boost weights $+1$ and $-1$, respectively. 

The basic working principle to analyze the off-shell equations of motion is to consider any covariant tensor to be built out of the following basic elements: the metric coefficient functions ($X, \, \omega_i, \, h_{ij}$) and various derivatives ($\partial_v, \, \partial_r, \, \nabla_i$) acting on them. Consequently, one can immediately know the boost-weight of $\mathcal{P}$, a generic covariant tensor, once we know its explicit construction in terms of the basic elements mentioned above. Equivalently, one can look at the Lie derivative of $\mathcal{P}$ along $\xi$. For $\mathcal{P}$ evaluated on equilibrium configurations, its Lie derivative should vanish. Once we demand this, we can interpret eq.\eqref{defboostwt} in terms of the following alternative definition of boost weight of $\mathcal{P}$: \textit{it is the difference between the number of lower $v$ indices and the number of lower $r$ indices in $\mathcal{P}$, with all its indices lowered}. 

Apart from knowing the boost-weight of arbitrary covariant terms, we also need to decide, when evaluated at the horizon, which terms are in linear order at amplitude expansion (i.e., at $\mathcal{O}(\epsilon)$) and which are at higher orders. We should determine the explicit appearance of $\partial_v$ and $\partial_r$ in any generic term. For example, let us assume that we schematically know\footnote{We are suppressing the possible tensor components of $\mathcal{P}$.} $\mathcal{P} \sim (\partial_r)^{n_1} (\partial_v)^{n_2} \mathcal{Q}$, with $n_2 > n_1$, and $\mathcal{Q}$ depending on $X, \, \omega_i, \, h_{ij}$ and only $\nabla_i$ acting on them. Since, generally, we also know $\mathcal{P} (r,\, v, \, x^i) = \mathcal{P}^{eq} (rv, \, x^i)+ \epsilon \, \delta\mathcal{P} (r,\, v, \, x^i)$, it is obvious that $\mathcal{P}^{eq}|_{r=0}$ vanishes when evaluated at the horizon, and $\mathcal{P}|_{r=0}$ is non-zero only for non-stationary configurations. Additionally, we must note that, in $\mathcal{P} \sim (\partial_r)^{n_1} (\partial_v)^{n_2} \mathcal{Q}$, all the extra\footnote{In $\mathcal{P} \sim (\partial_r)^{n_1} (\partial_v)^{n_2} \mathcal{Q}$, $n_1$ number of $\partial_v$ are paired with an equal number of $\partial_r$, but $(n_2-n_1)$ number of $\partial_v$'s are uncompensated, and hence, they are the extra ones that we are referring to.} $\partial_v$'s act on a single quantity $\mathcal{Q}$. Hence, we can argue that $\mathcal{P}|_{r=0} \sim \mathcal{O}(\epsilon)$. On the contrary, consider that we are dealing with a generic term, say $\mathcal{R}$, with a schematic structure given by the product of two such terms (each of which is like $\mathcal{P}$), i.e., $\mathcal{R} \sim (\partial_r)^{n_1} (\partial_v)^{n_2} \mathcal{Q}_1 \, \times (\partial_r)^{n_3} (\partial_v)^{n_4} \mathcal{Q}_2$, with $n_2 > n_1$ and $n_4 > n_3$. Then, following the same logic mentioned above, we must conclude that $\mathcal{R}|_{r=0} \sim \mathcal{O}(\epsilon^2)$. 

Also, with the input of boost weight specifications for a gauge field, eq.\eqref{eqlbmA}, which is a statement about equilibrium $A^{eq}_\mu$, can now be refined for a generic $A_\mu$ as
\begin{equation} \label{eqlbmA1}
     (A_{\mu} \xi^\mu + \Lambda) |_{r=0}= (v \, A_{v} + \Lambda) |_{r=0} = \mathcal{O}(\epsilon) \, ;
\end{equation}
see Appendix-B of \cite{Biswas:2022grc} for a derivation. This can alternatively be interpreted as a generalized Zeroth law involving the gauge fields. 
 
Thus, we have a consistent algorithm to decide whether a covariant tensor, built out of the metric coefficients, will contribute to $\mathcal{O}(\epsilon)$ or to $\mathcal{O}(\epsilon^2)$. As mentioned above, we need to keep track of explicit $\partial_v$ and $\partial_r$ present in any given term. It is also important to remember that the statements are being made on the horizon at $r=0$\footnote{We are being brief here, but the reader is requested to see \S2 of \cite{Bhattacharyya:2021jhr} for all the details.}. 
\subsection*{Off-shell structure of the equations of motion and the linearized second law:}
With the rules written down so far, it is straightforward now to write down the components of the entropy current by looking at the off-shell structure of the gravity equations of motion.
Consider the Lagrangian of a theory given to us; then, we readily have the equations of motion. It was shown in \cite{Bhattacharya:2019qal, Bhattacharyya:2021jhr, Biswas:2022grc} that the $vv$-component of the equations of motion for $U(1)$ gauge invariant and diffeomorphism invariant theories of gravity, when evaluated on the horizon, has the following structure
\begin{equation} \label{eom1}
     E_{vv} \, |_{r=0} = \partial_v \left[ \frac{1}{\sqrt{h}}  \partial_v \left( \sqrt{h} \, \mathcal{J}^v\right) + \nabla_i \mathcal{J}^i\right] + \mathcal{O}(\epsilon^2) \, .
\end{equation}
One can consider eq.\eqref{eom1} as the definitions of the quantities $\mathcal{J}^v$ and $\mathcal{J}^i$. 

As of now, eq.\eqref{eom1} is a statement about the off-shell structure of the EoM. Let us now briefly mention the chain of logic that connects establishing eq.\eqref{eom1} to the linearized second law. This also justifies the significance of $\mathcal{J}^v$ and $\mathcal{J}^i$ constituting the components of the entropy current. The reader is referred to \S2.4 of \cite{Bhattacharyya:2021jhr} for the details. 

The entropy of a dynamical non-stationary black hole can be written as 
\begin{equation} \label{stotdyn}
    S_{tot} = \int_{\mathcal{H}_v}d^{d-2}x \, \sqrt{h} \left(s_{wald} + s_{cor} \right) \, , 
\end{equation}
where $s_{wald}$ is the entropy density given by the Wald formula. Therefore, by construction, $s_{wald}$ corresponds to the entropy density of a stationary black hole consistent with the first law. We should remember that $s_{wald} = 1 + s^{HD}_{wald}$, where the factor of $1$ is the contribution due to two-derivative Einstein-Hilbert Lagrangian in GR and $s^{HD}_{wald}$ signifies the contribution solely coming from the higher derivative part of the Lagrangian. Finally, $s_{cor}$ in eq.\eqref{stotdyn} corresponds to a possible non-equilibrium contribution to black hole entropy density, which the Wald entropy piece does not capture. Therefore, it should vanish upon taking the equilibrium limit, $s_{cor} |_{eq} \rightarrow 0$.   

The Raychaudhuri equation for the affinely parameterized null generator of the horizon takes the following form 
\begin{equation} \label{RCeqn}
    \partial_v \vartheta = E^{HD}_{vv} + \partial_v \left[ \frac{1}{\sqrt{h}}  \partial_v \left( \sqrt{h} \,  (s^{HD}_{wald}+ s_{cor})\right) \right] -T_{vv} +  \mathcal{O}(\epsilon^2) \, ,
\end{equation}
where $\vartheta$ is the local expansion parameter for the null congruence, defined by 
\begin{equation} \label{deftheta}
\frac{\partial S_{tot}}{\partial v} = \int_{\mathcal{H}_v}d^{d-2}x \, \sqrt{h} \, \vartheta \, .
\end{equation} 
Also, $E^{HD}_{vv}$ is the EoM for the higher derivative part of the gravity Lagrangian (the Einstein-Hilbert part is being treated separately), and $T_{vv}$ is the stress-energy tensor for the matter part in the Lagrangian in eq.\eqref{RCeqn}. In deriving eq.\eqref{RCeqn} we have used the EoM $R_{vv}+E^{HD}_{vv}=T_{vv}$. At this stage, we assume that the matter fields, present if any, satisfy the null energy condition (NEC), $T_{vv} \ge 0$ \footnote{Within the classical setup that we are concerned with, for minimally coupled matter fields NEC is satisfied, due to $T_{vv} \sim \mathcal{O}(\epsilon^2)$. But for non-minimally coupled matter fields, NEC can be violated as $T_{vv} \sim \mathcal{O}(\epsilon)$. However, as shown in \cite{Biswas:2022grc}, an entropy current structure can be formulated even for such non-minimal coupling.}. Now, if the first two terms on the RHS of eq.\eqref{RCeqn} cancel each other upto $\mathcal{O}(\epsilon^2)$, we get $\partial_v \vartheta \le 0$, meaning $\vartheta$ is a monotonically decreasing function of $v$. Additionally, we are considering situations where at late times, the perturbations die off and the black hole settles down to a stationary one, i.e., $\vartheta \rightarrow 0$ as $v \rightarrow \infty$. This proves that $\vartheta \ge 0$, implying $S_{tot}$ is an increasing function of $v$ upto corrections at $\mathcal{O}(\epsilon^2)$. Thus, the linearized second law is proved \footnote{Truly speaking, there is no entropy production at $\mathcal{O}(\epsilon)$. So at the linearized order, we are showing $\vartheta = 0$, most importantly, confirming that $\vartheta$ cannot be negative.}. 

From the arguments presented above, we can clearly see that once eq.\eqref{eom1} establishes the off-shell structure of $E^{HD}_{vv}$, the first two terms on the RHS of eq.\eqref{RCeqn} indeed cancel each other upto $\mathcal{O}(\epsilon^2)$, provided we identify the equilibrium limit of $\mathcal{J}^v$ as the Wald entropy density, 
\begin{equation} \label{Jveq}
    \mathcal{J}^v \, |_{eq} = s_{wald} \, .
\end{equation}
On the other hand, for out-of-equilibrium configurations, we learn that $\mathcal{J}^i$ signifies the spatial flow of entropy density along a constant $v$-slice $\mathcal{H}_v$. For a compact slice, it does not contribute to $S_{tot}$. Also, $s_{cor}$ captures the terms known as the JKM ambiguities, which do not survive the equilibrium limit. 

In this paper, our primary goal is to establish the off-shell structure of $E_{vv}$ given by eq.\eqref{eom1} for Chern-Simons theories, both in the case of purely gravitational or mixed gauge gravity Chern-Simons theories, working only with $U(1)$ gauge fields. We will be constructing the entropy current ($\mathcal{J}^v$ and $\mathcal{J}^i$) for such theories. Following the arguments presented here, the linearized second law will also then be proved. 

\subsection*{Reparametrization covariance of entropy production at the horizon:} 
It is clear from the discussion so far that the construction of entropy current from the off-shell structure of $E_{vv}$ given in eq.\eqref{eom1} relies heavily on the specific choice of the coordinates in eq.\eqref{nhmetric}, in other words, it depends on how the constant $v$-slices are being chosen. This is obvious because the boost symmetry in eq.\eqref{boosttrns} played a crucial role in obtaining eq.\eqref{eom1}, and we could make use of the boost symmetry since the near horizon space-time has been written in the form eq.\eqref{nhmetric}, where both $v$ and $r$ have been taken to be affinely parameterized. However, as we have seen, the main content of eq.\eqref{eom1} is to validate the linearized second law. It is quite natural then to expect that such a physical process should not depend on the choice of coordinate system. This question has been verified in \cite{Bhattacharyya:2022njk} for a specific theory of higher derivative gravity, namely the Gauss-Bonnet gravity. Following that, skipping the details, here we will briefly list out the steps that should be followed to verify the covariance of entropy production under reparametrization of the horizon slicing. 

 The main idea is to verify the covariance of eq.\eqref{eom1} under possible coordinate transformation that preserves the gauge choice of eq.\eqref{nhmetric}. Such residual transformations that take $(v, \, r, \, x^a)$ to $(\tau, \, \rho, \, y^a)$ are the following 
 \begin{equation} \label{reslice}
     v = e^{\zeta(\vec{y})} \tau + \mathcal{O} (\rho) \, , \quad r = e^{-\zeta(\vec{y})} \rho + \mathcal{O} (\rho^2) \, , \quad x^a = y^a + \mathcal{O} (\rho) \, .
 \end{equation}
The coordinate transformations written in eq.\eqref{reslice} (including the sub-leading pieces at $\mathcal{O} (\rho)$ and higher, which we have not written explicitly) get constrained by the facts that $\tau$ and $\rho$ remain to be affinely parameterized. Also, the horizon is positioned at $\rho =0$. Under eq.\eqref{reslice}, the near-horizon metric would transform as given below 
\begin{equation} \label{nhmetslice}
ds^2 = 2 \, d\tau \, d\rho -\rho^2 \widetilde{X}(\rho, \, \tau, \, y^i) \, d\tau^2 + 2 \, \rho \, \widetilde{\omega}_i(\rho, \, \tau, \, y^i) d\tau dy^i + \widetilde{h}_{ij}(\rho, \, \tau, \, y^i) \, dy^i dy^j \, ,
\end{equation}
such that $\widetilde{X}, \, \widetilde{\omega}_i, \, \widetilde{h}_{ij}$ can be obtained in terms of the old quantities \footnote{See \cite{Bhattacharyya:2022njk} for derivations of these.}
\begin{equation} \label{relslice}
\begin{split}
\widetilde{X}(\rho, \, \tau, \, y^i) &= X +\omega _i \xi _j h^{ij}+\xi _i \xi _j h^{ij} -v \left(\xi _i h^{ij} \partial_v \omega _j+2 \omega _i \xi _j K^{ij}+2 \xi _i \xi _j K^{ij}\right) \\ & \quad \quad +v^2\left(\xi _i \xi _l h^{kl} h^{ij}\partial_v K_{jk}\right) + \mathcal{O}(\rho) \, ,\\
\widetilde{\omega}_i(\rho, \, \tau, \, y^i) &= \omega_i-2 \xi _i + 2 v  \xi _k h^{jk} K_{ij}+\mathcal{O}(\rho ) \, ,\\
\widetilde{h}_{ij}(\rho, \, \tau, \, y^i) &= h_{ij}(v,  r,  x^i) + \mathcal{O} (\rho) \, .    
\end{split}
\end{equation}
Here $\xi_i = \partial_i \zeta(x^a)$. This shouldn't be confused with the Killing vector $\xi$ given by eq.\eqref{KillVec}. 

Let us now focus on a particular theory with a given Lagrangian, e.g., Gauss-Bonnet in \cite{Bhattacharyya:2022njk}. Once we have performed the exercise of establishing eq.\eqref{eom1}, we know explicit expressions of $E_{vv}$, $\mathcal{J}^v$ and $\mathcal{J}^i$ for that given theory. Now, under eq.\eqref{reslice}, $E_{vv}$ should transform homogeneously in a straightforward way as the components of a rank two covariant tensor. However, $\mathcal{J}^v$ and $\mathcal{J}^i$ are expected to transform in a non-trivial way, which we can track by knowing how the metric coefficients ($X, \, \omega_i, \, h_{ij}$) and their derivatives would transform. When we input that on the RHS of eq.\eqref{eom1}, they should cancel among themselves such that eventually, both sides of that equation transform covariantly. We should note that this is a highly non-trivial check of the consistency of eq.\eqref{eom1} and also immediately proves the covariance of the entropy production on the horizon. In \S \ref{sec:reparam2+1d}, we will follow this principle to check this explicitly for the Chern-Simons theories of gravity. 
\subsection*{Non-linear second law within the EFT setup:} 
In \S\ref{intro}, we have already mentioned that we will work within the EFT approximation for CS theories in $(2+1)$ dimensions to prove the second law beyond linear orders in the perturbations. Following the description mentioned in \cite{Hollands:2022fkn}, we know that in an EFT setup, the Lagrangian of the gravitational theory is organized as a sum of scalar terms with an increasing number of derivatives acting on the metric. Each of them is built out of the Riemann tensor and its derivatives (for pure gravitational CS theories, it can also involve Christoffel symbol $\Gamma^\mu_{\nu \alpha}$, and, additionally, the gauge fields $A_\mu$ for mixed gauge gravity CS theory) appropriately contracted. The more the number of derivatives in such a term, the more suppressed it becomes. The dimensionful coefficients of the higher derivative terms in the Lagrangian can be called coupling constants, signifying the length scale (say $l$) where these terms become comparable to the leading two derivative GR term. Another inherent length scale (say $L$) present in the discussion is associated with the variation of the dynamical configuration. The validity of the EFT approximation lies in assuming that $l \ll L$, implying the dynamics are slow enough such that the higher derivative terms in the Lagrangian have smaller contributions compared to the Einstein-Hilbert term. The small parameter of the EFT expansion thus becomes $(l/L)$, which is the second perturbative expansion along with the amplitude of the dynamical expansion $\epsilon$ \footnote{Thus, one works with a double perturbation theory - one in the amplitude of the fluctuations and the other one being the EFT expansion.}. From dimensional analysis, we know that a term in the Lagrangian with $(k+2)$ derivatives will have a coefficient $l^k$. To set our working precision in the $(l/L)$ expansion, we will truncate the Lagrangian with a chosen, say, $N$ derivatives. It can then be argued that the EoM will have $E_{\mu\nu} \sim \mathcal{O} (l^N/L^{N+2})$ \footnote{Without any loss of generality we will choose units where $L =1$, such that we get  $E_{\mu\nu} \sim \mathcal{O} (l^N)$.}. 

With all the ingredients for an EFT setup explained in detail, let us now mention the important result that we will apply for studying CS theories in $(2+1)$ dimensions. For diffeomorphism invariant pure gravity theories without any matter fields, it has been shown in \cite{Hollands:2022fkn} that one can generalize eq.\eqref{eom1} with inputs from EFT approximation as follows 
\begin{equation} \label{eom2}
     E_{vv} \, |_{r=0} = \partial_v \left[ \frac{1}{\sqrt{h}}  \partial_v \left( \sqrt{h} \, \mathcal{J}^v\right) + \nabla_i \mathcal{J}^i\right] + (K_{ij} + X_{ij})(K^{ij} + X^{ij})+ \nabla_i \mathcal{Y}^i + \mathcal{O}( l^N) \, , 
\end{equation}
which for on-shell configurations (i.e., $E_{vv} = 0$)  leads to 
\begin{equation} \label{RCeqn1}
    \partial_v \left[ \frac{1}{\sqrt{h}}  \partial_v \left( \sqrt{h} \, \mathcal{J}^v\right) + \nabla_i \mathcal{J}^i\right] = - (K_{ij} + X_{ij})(K^{ij} + X^{ij}) - \nabla_i \mathcal{Y}^i + \mathcal{O}( l^N) \, . 
\end{equation}
Here $K_{ij}= (1/2)\partial_v h_{ij}$ is the extrinsic curvature of the horizon slice. It should be noted that 
 eq.\eqref{RCeqn1} is not generally exact up to $\mathcal{O}(\epsilon^2)$ (there maybe terms that are $\mathcal{O}(\epsilon^2 \, l^N)$ that we neglect), and also  both $X_{ij}$ and $\mathcal{Y}^{i}$ have boost weights $1$ and $2$ respectively. Also, $\mathcal{Y}^{i}$ contains terms that are the product of at least two terms, each being with boost weight $1$, implying $\mathcal{Y}^{i} \sim \mathcal{O}(\epsilon^2)$ (see \S 1.7 of \cite{Hollands:2022fkn}). One can now manipulate eq.\eqref{deftheta} such that we arrive at the following
\begin{equation} \label{deftheta1}
\begin{split} 
\frac{\partial S_{tot}}{\partial v} &= \int_{\mathcal{H}_v}d^{d-2}x \, \sqrt{h} \, \vartheta
 = - \int_{\mathcal{H}_v}d^{d-2}x \, \sqrt{h} \int_v^{\infty} dv' \, \partial_{v'} \vartheta\\
& =  \int_{\mathcal{H}_v}d^{d-2}x \, \sqrt{h} \int_v^{\infty} dv' \, \left[\partial_{v'}\left( \nabla_i \mathcal{J}^i \right) + (K_{ij} + X_{ij})(K^{ij} + X^{ij}) + \nabla_i \mathcal{Y}^i \right] \, ,
\end{split}
\end{equation}
where in the last step, we have used eq.\eqref{RCeqn1} and also\footnote{See \S5 of \cite{Bhattacharyya:2021jhr}, leading to the derivation of eq.(5.6) there, for justification.} $\vartheta =  (1/\sqrt{h}) \, \partial_v ( \sqrt{h} \, \mathcal{J}^v )$. 
From eq.\eqref{deftheta1} we can get 
\begin{equation} \label{deftheta2}
\begin{split} 
\frac{\partial S_{tot}}{\partial v} 
&=  \int_{\mathcal{H}_v}d^{d-2}x \, \sqrt{h} \int_v^{\infty} dv' \, (K_{ij} + X_{ij})(K^{ij} + X^{ij}) + \mathcal{O}(\epsilon^3) \\ 
& \ge 0 \, ,\quad \text{ignoring} \quad \mathcal{O}(\epsilon^2 \, l^N) \quad \text{corrections}.
\end{split}
\end{equation}
Therefore, we get the second law up to quadratic order in the amplitude of the fluctuations but in an EFT sense where $\mathcal{O}(l^N)$ terms are ignored. In deriving eq.\eqref{deftheta2}, we have used the following facts: firstly, $\nabla_i \mathcal{J}^i$ has boost weight $1$, and hence it vanishes when evaluated on equilibrium configurations (at the two end points of the $v$ integration in eq.\eqref{deftheta1}). Hence $\nabla_i \mathcal{J}^i$ drops out on the RHS of eq.\eqref{deftheta1}. Secondly, for the spatial total derivative term $\nabla_i \mathcal{Y}^i$, we note that $\mathcal{Y}^{i} \sim \mathcal{O}(\epsilon^2)$. Since we are also working to the same $\mathcal{O}(\epsilon^2)$, we can interchange the order of integration \footnote{We take $\sqrt{h}$ to be in the background metric eq.\eqref{eqlbmMetric} which becomes independent of $v$ on $\mathcal{H}_v$. Thus, we can interchange the $x^i$ integral and the $v$ integral up to terms that are $\mathcal{O}(\epsilon^3)$ which arise from the fluctuating part of $h_{ij}$ in eq.\eqref{defnoneqlbm}.} and get rid of the total derivative piece upon doing the spatial integration. We assume that $\mathcal{H}_v$ is compact at this step. 

The main point that we want to highlight here is that in deriving eq.\eqref{deftheta2}, the foremost step was to argue for the off-shell structure of $E_{vv}$ as in eq.\eqref{eom2}. Our aim, in \S \ref{nlCS2+1}, would be to establish that for CS theory in $(2+1)$ dimensions, one can manipulate its EoM in a brute-force manner such that it can be expressed as the RHS of eq.\eqref{eom2}. The following steps to arrive at eq.\eqref{deftheta2} do not depend on the theory under study. So the second law to quadratic order under EFT assumptions will follow straightforwardly. 

\subsection{Off-shell equations of motion in terms of Noether charge in Chern-Simons theories} 
\label{sec:cscovphase}
In \S\ref{sec:basicsetup} eq.\eqref{RCeqn}, we have shown how the component of the EoM $E_{vv}$ is crucial for the second law. Following \cite{Bhattacharyya:2021jhr, Biswas:2022grc}, we first relate the off-shell structure of equations of motion (EoM) to the Noether charge under diffeomorphism. As stressed above, the main difference between the Lagrangian densities considered here and the Lagrangian densities considered in \cite{Bhattacharyya:2021jhr, Biswas:2022grc} is that they are diffeomorphism invariant up to total derivatives only. Thus, one cannot directly borrow the covariant phase space approach of \cite{LeeWald,wald1993, Iyer:1994ys}. The modification of the covariant phase formalism for Chern-Simons theories has been worked out in \cite{Tachikawa:2006sz, Bonora:2011gz, Azeyanagi:2014sna}. We will review their modification here.

We consider a generic Chern-Simons (CS) Lagrangian of the form \cite{Azeyanagi:2014sna}
\begin{equation}\label{eq:cslagrangian}
    L = L(g_{\mu\nu}, \Gamma^{\alpha}_{\beta\mu},R^{\alpha}_{~\beta \mu\nu}, A_{\mu},F_{\mu\nu}) \, .
\end{equation}
The variation of the CS Lagrangian is given by
\begin{equation}\label{eq:varL}
    \delta (\sqrt{-g} L) = \sqrt{-g} E^{\mu\nu} \delta g_{\mu\nu} + \sqrt{-g} G^{\mu} \delta A_{\mu} + \sqrt{-g} D_{\mu} \Theta^{\mu}[g_{\mu\nu}, \Gamma^{\alpha}_{\beta\mu},A_{\mu},\delta g_{\mu\nu},\delta \Gamma^{\alpha}_{\beta\mu},\delta A_{\mu}] \, .
\end{equation}
$E^{\mu\nu}$ is the gravitational EoM and $G^{\mu}$ is the gauge EoM. $\Theta^{\mu}$ is the total derivative term generated in the variation of the Lagrangian. Here we have been explicit in pointing out the dependence of $\Theta^{\mu}$ on non-covariant quantities like $\Gamma^{\alpha}_{\beta\mu}$ and $A_{\mu}$. However, we will not split the covariant and non-covariant contributions separately as in \cite{Bonora:2011gz}. 

We now consider the variation $\delta g_{\mu\nu}$ due to a diffeomorphism $x^{\mu} \to x^{\mu} + \zeta^{\mu}$ and a $U(1)$ gauge transformation $A_\mu \rightarrow A_\mu + D_\mu \Lambda$, \cite{Biswas:2022grc},
\begin{equation}\label{eq:diffeo}
    \begin{split}
        \delta g_{\mu\nu} &= \mathcal{L}_{\zeta}g_{\mu\nu} =D_{\mu} \zeta_{\nu} + D_{\nu} \zeta_{\mu} \, , \\
        \delta A_{\mu} &= \mathcal{L}_{\zeta} A_{\mu} + D_{\mu} \Lambda = \zeta^{\alpha} F_{\alpha\mu} + D_{\mu} \left( A_{\alpha}\xi^{\alpha} + \Lambda \right) \, .
    \end{split}
\end{equation}
Now comes the crucial difference between the Lagrangian eq.\eqref{eq:cslagrangian} and the Lagrangians considered in \cite{Bhattacharyya:2021jhr, Biswas:2022grc}. The variation of the CS Lagrangian of the form eq.\eqref{eq:cslagrangian} under a diffeomorphism and $U(1)$ gauge transformation eq.\eqref{eq:diffeo} is given by
\begin{equation}\label{eq:diffvarl}
    \delta [\sqrt{-g} L] = \sqrt{-g} D_{\mu} ( \zeta^{\mu} L ) + \sqrt{-g} D_{\mu} \Xi^{\mu} \, .
\end{equation}
This additional variation due to $\Xi^{\mu}$ means that though the action remains diffeomophism/$U(1)$ invariant if we consider compact space-times, the Lagrangian is not diffeomorphism/$U(1)$ invariant. For quantities built out of the metric, the variation in eq.\eqref{eq:diffvarl} acts as a Lie derivative, and one should consider the Lie derivative acting on non-tensorial quantities like $\Gamma^{\alpha}_{\beta\mu}$ as if their indices were tensorial indices.

Substituting eq.\eqref{eq:diffvarl} and eq.\eqref{eq:diffeo} in eq.\eqref{eq:varL}, we have the following expression after some integration by parts manipulation:
\begin{equation}\label{eq:varlinter}
    \begin{split}
        D_{\mu}[ \zeta^{\mu} L + \Xi^{\mu} - 2 \zeta_{\nu} E^{\mu\nu} - G^{\mu}(A_{\nu}\zeta^{\nu} + \Lambda) - \Theta^{\mu} ] = - 2 \zeta_{\nu} D_{\mu} E^{\mu\nu} - (A_{\nu}\zeta^{\nu} + \Lambda) D_{\mu} G^{\mu} + G^{\mu}\zeta^{\nu} F_{\nu\mu} \, .
    \end{split}
\end{equation}
Following \cite{LeeWald,Bhattacharyya:2021jhr,Biswas:2022grc}, we can derive the following Bianchi identities
\begin{equation}\label{eq:bianchigen}
    - 2 D_{\mu} E^{\mu\nu} + G_{\mu} F^{\nu\mu} = 0 \, , ~~~ \text{and} ~~~ D_{\mu} G^{\mu} = 0 \, .
\end{equation}
This is derived by integrating both sides of eq.\eqref{eq:varlinter} over the entire space-time and by choosing $\zeta^{\mu}$ such that it is non-zero only in a small region $\mathcal{R}$. The $U(1)$ gauge transformation parameter, $\Lambda$, should also be restricted similarly. The LHS of eq.\eqref{eq:varlinter} would vanish since it would integrate to a pure boundary term at infinity where $\zeta$ vanishes. Thus, we obtain
\begin{equation}
    \int_{\text{full space-time}} \left[\zeta_{\nu}\left( - 2 D_{\mu} E^{\mu\nu} + G_{\mu} F^{\nu\mu} \right) - (A_{\nu}\zeta^{\nu} + \Lambda) D_{\mu} G^{\mu}  \right] \, .
\end{equation}
This is true as long as $\zeta$ is non-zero for a finite region in space-time. Thus eq.\eqref{eq:bianchigen} holds identically because $\zeta$ and $A_{\nu}$ are independent. Substituting eq.\eqref{eq:bianchigen} in eq.\eqref{eq:varlinter}, we see that we get an identically conserved vector from the LHS of eq.\eqref{eq:varlinter}. This can be written as a divergence of an anti-symmetric object $Q^{\mu\nu}$ called the Noether charge:
\begin{equation}\label{eq:csnoethercharge}
    \Theta^{\mu} - \Xi^{\mu} - \zeta^{\mu} L = - 2 E^{\mu\nu}\zeta_{\nu} - G^{\mu} (A^{\nu}\zeta_{\nu} + \Lambda) + D_{\nu} Q^{\mu\nu} \, .
\end{equation}
We now choose $\zeta$ as the Killing vector $\xi = v \partial_v - r \partial_r$ given by eq.\eqref{KillVec}. Contracting the above equation with $\xi$, we get
\begin{equation}
    \Theta^{\mu}\xi_{\mu} - \Xi^{\mu}\xi_{\mu} - \xi^{\mu}\xi_{\mu} L = - 2 E^{\mu\nu}\xi_{\mu}\xi_{\nu} - G^{\mu}\xi_{\mu} (A^{\nu}\xi_{\nu} + \Lambda) + \xi_{\mu} D_{\rho} Q^{\mu\rho} \, .
\end{equation}
Evaluating this resulting expression at $r=0$, we get
\begin{equation}\label{eq:main}
    2 v \, E_{vv} + G_v (v A_v + \Lambda) = (-\Theta^r + \Xi^r + D_{\rho} Q^{r\rho})|_{r=0} \, .
\end{equation}
One should note that this $Q^{\mu\nu}$ inherently has non-covariant pieces since it is derived from a non-covariant CS Lagrangian eq.\eqref{eq:cslagrangian}. 

To make further progress using boost weight analysis, we must work out $\Theta^r$, $\Xi^r$ and $Q^{r\rho}$ for the most general CS Lagrangian of the form eq.\eqref{eq:cslagrangian}. We will see that $G_v (v A_v + \Lambda) \sim \mathcal{O}(\epsilon^2)$ using eq.\eqref{eqlbmA1}. We don't want to set this at the level of eq.\eqref{eq:main} since, as we will see later, $\Xi^r$ depends on $\Lambda$. Let us highlight that eq.\eqref{eq:main} represents the main equation that we work with to establish that the $E_{vv}$ has the structure of eq.\eqref{eom1} and thus proving a linearized second law for CS Lagrangians of the form eq.\eqref{eq:cslagrangian}. The crucial difference when compared to the analysis of \cite{Bhattacharyya:2021jhr, Biswas:2022grc} is the explicit appearance of the non-covariant $\Xi^r$ in eq.\eqref{eq:main}. We will see below in \S \ref{sec:xistruchorizon} that this crucial piece is essential to arrive at an entropy current structure of eq.\eqref{eom1}. 

\subsection*{Tachikawa's proposal for black hole entropy and first law:} 

Let us now recap the main statements in \cite{Tachikawa:2006sz, Bonora:2011gz, Azeyanagi:2014sna} concerning the first law for CS theories. Eq.\eqref{eq:diffvarl} describes the main change to the covariant phase space methods of \cite{LeeWald, wald1993, Iyer:1994ys}. Tachikawa \cite{Tachikawa:2006sz} proposed a definition of the black hole entropy for CS theories based on this modification. His definition was to incorporate this $\Xi^{\mu}$ and define the Noether charge through eq.\eqref{eq:csnoethercharge}. This results in a $(d-2)$-form $\mathbf{Q}_{\xi}$. The pullback of the $(d-2)$-form $\mathbf{Q}_{\xi}$ to the bifurcation surface $\Sigma$ of the black hole results in a generalization of the black hole entropy to CS theories. This is given by
\begin{equation}\label{eq:tachikawaform}
    S_{IWT} = -2 \pi \int_{\Sigma} \mathbf{Q}_{\xi} \, |_{\xi \to 0, \, D_{\alpha} \xi_{\beta} \to \kappa \epsilon_{\alpha\beta}} \, .
\end{equation}
Here $\kappa$ is the surface gravity of the horizon, and the $\xi$ is the Killing vector that generates the horizon. By definition, this formula eq.\eqref{eq:tachikawaform} is not covariant under diffeomorphism. This issue of non-covariance was studied in \cite{Bonora:2011gz} and later significantly refined in \cite{Azeyanagi:2014sna}. In a special Kruskal-type coordinate system of the black hole, eq.\eqref{eq:tachikawaform} can be expressed in terms of the Lagrangian \cite{Bonora:2011gz}. For our gauge eq.\eqref{nhmetric}, we can get the following formula as \footnote{Here $\tilde{\epsilon}^{\alpha}_{~\beta}$ is the binormal to the Bifurcation surface $\Sigma$ and in our gauge $\Sigma$ is given by $\xi$ of eq.\eqref{KillVec} set to zero. Thus, the only non-trivial components are $\tilde{\epsilon}^{r}_{~r} = -1$ and $\tilde{\epsilon}^{v}_{~v} = 1$.}
\begin{equation}\label{eq:firstlawS1}
    S_{IWT} = -4 \pi \int_{\Sigma} d^{d-2}x \, \sqrt{h} \, \tilde{\epsilon}^{\alpha}_{~\beta} \pdv{L}{R^{\alpha}_{~\beta r v}} \, . 
\end{equation}
In \cite{Bonora:2011gz}, this formula was derived in a gauge which set $\Xi^{\mu}$ on the bifurcation surface $\Sigma$ to zero. However, as pointed out in \cite{Azeyanagi:2014sna} \footnote{The derivation in \cite{Azeyanagi:2014sna} uses the anomaly polynomial. One must do appropriate manipulations to express it in terms of the CS Lagrangian .}, one can derive eq.\eqref{eq:firstlawS1} covariantly without resorting to set $\Xi^{\mu}$ to zero in special gauges. In fact, we will show that $\Xi^r$ contributes to the Iyer-Wald-Tachikawa entropy and the full answer, including the contributions of $Q^{\mu\nu}$ give eq.\eqref{eq:firstlawS1}. One can expand the binormal to obtain the final result as \footnote{It should be noted that the convention for the Lagrangian defined this way has an overall factor of $\frac{1}{2\pi}$.}
\begin{equation}\label{eq:siwtfinal}
    S_{IWT} = -2 \int_{\Sigma} d^{d-2}x \, \sqrt{h} \, \left( \pdv{L}{R^{v}_{~v r v}} - \pdv{L}{R^{r}_{~r r v}} \right) = \int_{\Sigma} d^{d-2}x \, \sqrt{h} \, s_{IWT} \, .
\end{equation}
In explicit examples, we will show that $\mathcal{J}^v$ of eq.\eqref{eom1} reduces to the integrand of eq.\eqref{eq:siwtfinal} in the equilibrium limit. This is analogous to the diffeomorphism invariant case eq.\eqref{Jveq}.

\section{Brute-force calculation of entropy current for specific Chern-Simons theories}
\label{sec:bruteforce}

In this section, we will consider specific examples of CS theories, both purely gravitational and mixed gauge gravity ones, in various dimensions. We aim to obtain the expressions of the EoMs in all such theories and work them out in a brute-force way in the chosen coordinate system and gauge eq.\eqref{nhmetric}. In this process, we will distinguish possible terms according to their boost weights and figure out if they contribute to $\mathcal{O}(\epsilon)$ or higher, leading to establishing eq.\eqref{eom1} for linearized analysis and eq.\eqref{eom2} for analysis to quadratic order. We will also obtain the components of entropy current ($\mathcal{J}^v$ and $\mathcal{J}^i$) for all such example theories.

Before we proceed, let us list our conventions for the calculations that follow. 

\begin{itemize}
    \item The signature for the metric is mostly plus $(-++\dots+)$.
    \item The lower case Greek indices: $\{ \alpha,\beta,\dots,\mu,\nu,\dots, \alpha_1 , \alpha_2 , \dots \}$ are used for the full space-time coordinates denoted by $x^{\mu}$.
    \item The covariant derivative associated with the full space-time metric $g_{\mu\nu}$ is denoted by $D_{\mu}$. 
    \item The horizon is a co-dimension one surface $\mathcal{H}$ located at $r=0$. The constant $v$ slices are co-dimension two surfaces denoted by $\mathcal{H}_v$. 
    \item Lower case latin indices: $\{i,j,k,l,\dots, i_1,i_2,\dots\}$ are used for the spatial coordinates $x^i$ on $\mathcal{H}_v$.
    \item The intrinsic metric on $\mathcal{H}$ is $h_{ij}$ and the covariant derivative associated with it is $\nabla_i$.
    \item The Levi-Civita tensor is denoted by $\epsilon^{\alpha_1 \dots \alpha_k}$ and it is related to the totally anti-symmetric Levi-Civita symbol $\varepsilon^{\alpha_1 \dots \alpha_k}$ as
    $$\epsilon^{\alpha_1 \dots \alpha_k} = - \dfrac{1}{\sqrt{h}} \varepsilon^{\alpha_1 \dots \alpha_k} \, .$$
    Since we focus on $2+1$ and $4+1$ examples, in our horizon adapted coordinates eq.\eqref{nhmetric}, the independent components of the Levi-Civita tensors are given by
    $$\epsilon^{vrx}= -\dfrac{1}{\sqrt{h}} \, , ~~~~~~ \epsilon^{vrxyz} = - \dfrac{1}{\sqrt{h}} \, .$$
    \item The required Christoffel symbols and curvature components in our gauge eq.\eqref{nhmetric} are given in Appendix \ref{ap:christoffel}.
\end{itemize}

\subsection{Gravitational Chern-Simons theory in $(2+1)$-dimensions}
In $(2+1)$-dimensions, the action and Lagrangian for purely gravitational CS theory are given by 
\begin{equation} \label{lag2+1}
 I = \int d^3x \sqrt{-g} \, \mathcal{L} \, , \quad \text{with} \quad   \mathcal{L} = \tensor{\epsilon}{^\lambda^\mu^\nu} \,  \tensor{\Gamma}{^\rho_\lambda_\sigma} \left(\tensor{\partial}{_\mu} \tensor{\Gamma}{^\sigma_\rho_\nu} + \frac{2}{3}  \, \tensor{\Gamma}{^\sigma_\mu_\tau}  \, \tensor{\Gamma}{^\tau_\rho_\nu} \right) \, .
\end{equation}
One can obtain the EoM from here as
\begin{equation}
    \tensor{E}{^\mu^\nu} = - \left(\tensor{\epsilon}{^\nu^\rho^\sigma}  \, D{_\rho} \tensor{R}{_\sigma^\mu} + \tensor{\epsilon}{^\mu^\rho^\sigma} \, D{_\rho} \tensor{R}{_\sigma^\nu} \right) \, .
\end{equation}
Our chosen metric gauge and coordinate system in $(2+1)$-dimensions become
\begin{equation} \label{nhmet2+1}
ds^2 = 2 \, dv \, dr -r^2 X(r, \, v, \, x) \, dv^2 + 2 \, r \, \omega(r, \, v, \, x)\, dv \, dx + h(r, \, v, \, x) \, dx^2 \, ,
\end{equation} 
The $vv$-component of $\tensor{E}{^\mu^\nu}$ can be worked out as
\begin{equation}
    \begin{split}
        2 \tensor{E}{_v_v} &= 2 \tensor{E}{^r^r}
    = - 4\tensor{\epsilon}{^r^\rho^\sigma} D{_\rho} \tensor{R}{_\sigma^r}
    = 4 \tensor{\epsilon}{^r^\sigma^\rho} \left(\tensor{\partial}{_\rho} \tensor{R}{_\sigma_v} - \tensor{\Gamma}{^\alpha_\rho_\sigma} \tensor{R}{_\alpha_v} - \tensor{\Gamma}{^\alpha_\rho_v} \tensor{R}{_\sigma_\alpha} \right) \\
    &= 4\tensor{\epsilon}{^r^v^x} \left(\tensor{\partial}{_x} \tensor{R}{_v_v} - \tensor{\partial}{_v} \tensor{R}{_x_v} \right) + \tensor{\epsilon}{^r^v^x} \left(- \frac{\omega}{h} \tensor{\partial}{^2_v} h + \frac{\omega}{h^2} \left(\tensor{\partial}{_v} h \right)^2 + \frac{1}{h} \left(\tensor{\partial}{_v} h \right) \left(\tensor{\partial}{_v} \omega \right) \right) \, .
    \end{split}
\end{equation}
Thus,
\begin{equation}
    E_{vv} = \tensor{\epsilon}{^r^v^x} \left(\tensor{\partial}{^2_v} \omega + \frac{1}{h} \left(\tensor{\partial}{_v} \omega \right) \left(\tensor{\partial}{_v} h \right) - \frac{1}{h} \tensor{\partial}{^2_v} \tensor{\partial}{_x} h + \frac{1}{h} \tensor{\partial}{_v} \left(\frac{1}{h} \left(\tensor{\partial}{_v} h \right) \left(\tensor{\partial}{_x} h \right) \right) \right) \, .
\end{equation}
Each term on the RHS above is in the form $A \tensor{\partial}{^2_v} B$, which we can write as $A \tensor{\partial}{^2_v} B = \tensor{\partial}{_v} \left(A \tensor{\partial}{_v} B \right) + \mathcal{O}\left(\epsilon^2 \right)$. Thus, neglecting $\mathcal{O}\left(\epsilon^2 \right)$ terms we get
\begin{equation} \label{Evv2+1lin}
    \begin{split}
    \tensor{E}{_v_v} &= \tensor{\partial}{_v} \left(\tensor{\epsilon}{^r^v^x} \tensor{\partial}{_v} \omega - \frac{1}{h} \tensor{\epsilon}{^r^v^x} \tensor{\partial}{_v} \tensor{\partial}{_x} h + \frac{1}{h^2} \tensor{\epsilon}{^r^v^x} \left(\tensor{\partial}{_v} h \right) \left(\tensor{\partial}{_x} h \right) \right) + \mathcal{O}\left(\epsilon^2 \right) \\
    &= \tensor{\partial}{_v} \left(\tensor{\epsilon}{^r^v^x} \tensor{\partial}{_v} \omega - \tensor{\epsilon}{^r^v^x} \tensor{\partial}{_x} \left(\frac{1}{h} \tensor{\partial}{_v} h \right) \right) + \mathcal{O}\left(\epsilon^2 \right) \\
    &= \tensor{\partial}{_v} \left(\frac{1}{\sqrt{h}} \tensor{\partial}{_v} \left(\sqrt{h} \tensor{\epsilon}{^r^v^x} \omega \right) - \frac{1}{\sqrt{h}} \tensor{\partial}{_x} \left(\sqrt{h} \tensor{\epsilon}{^r^v^x} \frac{1}{h} \tensor{\partial}{_v} h \right) \right) + \mathcal{O}\left(\epsilon^2 \right) \, ,
\end{split}
\end{equation}
which indeed is of the form given in eq.\eqref{eom1}. Hence, the entropy current is
\begin{equation} \label{ec2+1lin}
    \tensor{\mathcal{J}}{^v} = \tensor{\epsilon}{^r^v^x}\omega  \, ,\qquad \tensor{\mathcal{J}}{^i} = - \tensor{\epsilon}{^r^v^x} \left(\frac{1}{h} \tensor{\partial}{_v} h \right) \, ,
\end{equation}
and from here, we also conclude that the linearized second law holds in this case. 

\subsection{Non-linear analysis of Chern-Simons theory in $(2+1)$-dimensions as an EFT} \label{nlCS2+1}
In this subsection, we aim to explore the second law beyond linear order in the $\epsilon$-expansion for CS theory in $(2+1)$-dimensions with the Lagrangian given in eq.\eqref{lag2+1}. In this case, the Lagrangian, and hence, the EoMs, will have at most three derivatives. Therefore, the $E_{vv}$ can have at most $\mathcal{O}\left(\epsilon^2 \right)$ terms. This is because $E_{vv}$ has boost-weight $2$, so any term in $E_{vv}$ must have two $\partial_v$ derivatives, while the third one has to be $\partial_x$. In that case, this term becomes $\mathcal{O}\left(\epsilon^2 \right)$ \footnote{One must note that a term in $E_{vv}$ cannot have all three $\partial_v$ derivatives, because that would make it a term with boost weight $3$. So, there is no possibility of a $\mathcal{O}\left(\epsilon^3 \right)$ contribution in $E_{vv}$.}. Keeping track of possible $\mathcal{O}\left(\epsilon^2 \right)$ terms in $E_{vv}$, eq.\eqref{Evv2+1lin} can be written more exactly as follows
\begin{equation} \label{Evv2+1qud}
    \begin{split}
    \tensor{E}{_v_v} &= \tensor{\partial}{_v} \left(\tensor{\epsilon}{^r^v^x} \tensor{\partial}{_v} \omega - \tensor{\epsilon}{^r^v^x} \tensor{\partial}{_x} \left(\frac{1}{h} \tensor{\partial}{_v} h \right) \right) - \frac{3}{4} \tensor{\epsilon}{^r^v^x} \tensor{\partial}{_x} \left(\frac{1}{h^2} \left(\tensor{\partial}{_v} h \right)^2 \right) + \frac{3}{2} \tensor{\epsilon}{^r^v^x} \left(\tensor{\partial}{_v} \omega \right) \left(\frac{1}{h} \tensor{\partial}{_v} h \right)
    \\
    &= \tensor{\partial}{_v} \left(\frac{1}{\sqrt{h}} \tensor{\partial}{_v} \left(\sqrt{h} \tensor{\epsilon}{^r^v^x} \omega \right) - \frac{1}{\sqrt{h}} \tensor{\partial}{_x} \left(\sqrt{h} \tensor{\epsilon}{^r^v^x} \frac{1}{h} \tensor{\partial}{_v} h \right) \right) \\
    &\quad - \frac{3}{4} \frac{1}{\sqrt{h}} \tensor{\partial}{_x} \left(\sqrt{h} \tensor{\epsilon}{^r^v^x} \frac{1}{h^2} \left(\tensor{\partial}{_v} h \right)^2 \right) + \frac{3}{2} \tensor{\epsilon}{^r^v^x} \left(\tensor{\partial}{_v} \omega \right) \left(\frac{1}{h} \tensor{\partial}{_v} h \right) \, .
\end{split}
\end{equation}
Note that the nonlinear terms in $E_{vv}$ are not of the form of a square plus a total derivative as explained in eq.\eqref{eom2}, which implies that the second law may be violated in the second order. 

Following our discussion in \S\ref{sec:basicsetup}, we would now consider pure $(2+1)$-dimensional gravitational CS theory as an EFT. The leading curvature contribution to the gravitational sector of the action must be the Einstein-Hilbert action, and pure $(2+1)$-dimensional gravitational CS theory is a sub-leading correction term. If we include the Einstein-Hilbert action, we get the Lagrangian of the theory as \footnote{Note that the CS part of the Lagrangian is written differently compared to eq.\eqref{lag2+1}. However, one can use the properties of $\epsilon^{\lambda\mu\nu}$ to show that they are identical.}
\begin{equation} \label{lagEFT2+1}
    \mathcal{L} = R + l \, \tensor{\epsilon}{^\lambda^\mu^\nu} \, \tensor{\Gamma}{^\rho_\lambda_\sigma} \left(\frac{1}{2} \tensor{R}{^\sigma_\rho_\mu_\nu} - \frac{1}{3} \,  \tensor{\Gamma}{^\sigma_\mu_\tau}  \, \tensor{\Gamma}{^\tau_\rho_\nu} \right) \, .
\end{equation}
It should be noted that the CS part of the Lagrangian has three derivatives acting on $g_{\mu\nu}$. Therefore, on dimensional grounds, this sub-leading term must be multiplied by a dimensionful coupling constant $l$ \footnote{This is also consistent with the following statement made in \S\ref{sec:basicsetup} that in an EFT expansion a term with $(k+2)$ number of derivatives in the Lagrangian should be accompanied by a coefficient $l^k$, see also \cite{Hollands:2022fkn}.}. The validity of EFT field theory relies on the approximation that $l$ can be treated small compared to any other scale present in the system, i.e., $L$, associated with the dynamical configuration described by the metric in eq.\eqref{nhmet2+1}. Thus, we will use $l \ll L$ and choose units as $L=1$. To be more precise, we will be neglecting terms of $\mathcal{O}\left(l^2 \right)$. This is justified since we have truncated the Lagrangian of the low energy effective theory only to $\mathcal{O}\left(l \right)$ in eq.\eqref{lagEFT2+1}. Consequently, we should be concerned with proving the second law to the same order in EFT expansion, i.e., to  $\mathcal{O}\left(l \right)$, up to $\mathcal{O}\left(l^2 \right)$ corrections. Obviously, we need to keep $\mathcal{O}\left(\epsilon^2 \right)$ terms as in eq.\eqref{Evv2+1qud} but we can ignore $\mathcal{O}(\epsilon^2 l^2)$ terms.

Our next job would be to look into the equations of motion that follow from eq.\eqref{lagEFT2+1},
\begin{equation} \label{eomEFT2+1}
    \tensor{E}{^\mu^\nu} = \frac{1}{2} R   \, \tensor{g}{^\mu^\nu} - \tensor{R}{^\mu^\nu} - l  \,  \left(\tensor{\epsilon}{^\nu^\rho^\sigma}   \, D{_\rho} \tensor{R}{^\mu_\sigma} + \tensor{\epsilon}{^\mu^\rho^\sigma}  \,  D{_\rho} \tensor{R}{^\nu_\sigma} \right) \, .
\end{equation}
From this, we can compute $\tensor{E}{_v_v}$ at the horizon
\begin{equation} \label{EvvEFT1}
    \begin{split}
    \tensor{E}{_v_v} &= - \tensor{R}{_v_v} + l \tensor{\epsilon}{^r^v^x} \left(\tensor{\partial}{^2_v} \omega + \frac{1}{h} \left(\tensor{\partial}{_v} \omega \right) \left(\tensor{\partial}{_v} h \right) - \frac{1}{h} \tensor{\partial}{^2_v} \tensor{\partial}{_x} h + \frac{1}{h} \tensor{\partial}{_v} \left(\frac{1}{h} \left(\tensor{\partial}{_v} h \right) \left(\tensor{\partial}{_x} h \right) \right) \right) \, .
\end{split}
\end{equation}
Keeping in mind that $\mathcal{O}\left(l^2 \right)$ pieces can be ignored but $\mathcal{O}\left(\epsilon^2 \right)$ pieces should be retained, we can manipulate it further to obtain
\begin{equation} \label{EvvEFT}
    \begin{split}
    \tensor{E}{_v_v} &= - \tensor{R}{_v_v} + l \tensor{\epsilon}{^r^v^x} \left(\tensor{\partial}{^2_v} \omega + \frac{1}{h} \left(\tensor{\partial}{_v} \omega \right) \left(\tensor{\partial}{_v} h \right) - \frac{1}{h} \tensor{\partial}{^2_v} \tensor{\partial}{_x} h + \frac{1}{h} \tensor{\partial}{_v} \left(\frac{1}{h} \left(\tensor{\partial}{_v} h \right) \left(\tensor{\partial}{_x} h \right) \right) \right) 
    \\
    &= \tensor{\partial}{_v} \left(\frac{1}{2} \tensor{h}{^i^j} \tensor{\partial}{_v} \tensor{h}{_i_j} + l \tensor{\epsilon}{^r^v^x} \tensor{\partial}{_v} \omega - l \tensor{\epsilon}{^r^v^x} \tensor{\partial}{_x} \left(\frac{1}{h} \tensor{\partial}{_v} h \right) \right) + \frac{1}{4} \left(\frac{1}{h} \tensor{\partial}{_v} h \right)^2 \\
    &\quad - \frac{3}{4} l \tensor{\epsilon}{^r^v^x} \tensor{\partial}{_x} \left(\frac{1}{h^2} \left(\tensor{\partial}{_v} h \right)^2 \right) + \frac{3}{2} l \tensor{\epsilon}{^r^v^x} \left(\tensor{\partial}{_v} \omega \right) \left(\frac{1}{h} \tensor{\partial}{_v} h \right) 
    \\
    &= \tensor{\partial}{_v} \left(\frac{1}{2} \tensor{h}{^i^j} \tensor{\partial}{_v} \tensor{h}{_i_j} + l \tensor{\epsilon}{^r^v^x} \tensor{\partial}{_v} \omega - l \tensor{\epsilon}{^r^v^x} \tensor{\partial}{_x} \left(\frac{1}{h} \tensor{\partial}{_v} h \right) \right) - \frac{3}{4} l \tensor{\epsilon}{^r^v^x} \tensor{\partial}{_x} \left(\frac{1}{h^2} \left(\tensor{\partial}{_v} h \right)^2 \right) \\
    &\quad + \left(\frac{1}{2h} \tensor{\partial}{_v} h + \frac{3}{2} l \tensor{\epsilon}{^r^v^x} \tensor{\partial}{_v} \omega \right)^2 - l^2 \left(\frac{9}{4} \tensor{\epsilon}{^r^v^x} \tensor{\partial}{_v} \omega \right)^2 
    \\
    &= \tensor{\partial}{_v} \left(\frac{1}{2} \tensor{h}{^i^j} \tensor{\partial}{_v} \tensor{h}{_i_j} + l \tensor{\epsilon}{^r^v^x} \tensor{\partial}{_v} \omega - l \tensor{\epsilon}{^r^v^x} \tensor{\partial}{_x} \left(\frac{1}{h} \tensor{\partial}{_v} h \right) \right) - \frac{3}{4} l \tensor{\epsilon}{^r^v^x} \tensor{\partial}{_x} \left(\frac{1}{h^2} \left(\tensor{\partial}{_v} h \right)^2 \right) \\
    &\quad + \left(\frac{1}{2h} \tensor{\partial}{_v} h + \frac{3}{2} l \tensor{\epsilon}{^r^v^x} \tensor{\partial}{_v} \omega \right)^2 + \mathcal{O}\left(l^2 \right) \, .
\end{split}
\end{equation}
It should be clear from the final expression above in eq.\eqref{EvvEFT} that we have indeed succeeded in casting the $E_{vv}$ as eq.\eqref{eom2}. We obtain the entropy current 
\begin{equation} \label{JvJiEFT}
    \begin{split}
    \tensor{\mathcal{J}}{^v} &= 1+ l \, \tensor{\epsilon}{^r^v^x} \, \omega  \, ,\quad \quad \tensor{\mathcal{J}}{^i} = - l \,  \tensor{\epsilon}{^r^v^x} \, \left(\frac{1}{h} \tensor{\partial}{_v} h \right) \,,
\end{split}
\end{equation}
and also identify the following
\begin{equation} 
    \begin{split}
    (K_{ij} + X_{ij}) &\sim \left(\frac{1}{2h} \tensor{\partial}{_v} h + \frac{3}{2} l \tensor{\epsilon}{^r^v^x} \tensor{\partial}{_v} \omega \right) \, , \quad \nabla_i \mathcal{Y}^i \sim - \frac{3}{4} l \dfrac{1}{\sqrt{h}} \tensor{\partial}{_x} \left( \tensor{\sqrt{h}\,\epsilon}{^r^v^x}\frac{1}{h^2} \left(\tensor{\partial}{_v} h \right)^2 \right) \, .
\end{split}
\end{equation}    
Another thing worth highlighting is that the whole square term on the RHS of eq.\eqref{EvvEFT} has appeared with a positive sign, which is crucial for demonstrating the second law. The entropy current piece is of $\mathcal{O}\left(\epsilon \right)$, which is also expected from the linearized analysis. The sign of the $\nabla_i \mathcal{Y}^i$ term is unimportant as this will not contribute to the entropy production - this term will drop out when integrated over a compact $\mathcal{H}_v$. From here, we can follow the same steps that were worked out in \S\ref{sec:basicsetup} to derive eq.\eqref{deftheta2} starting from eq.\eqref{eom2}  to obtain $\partial_v S_{tot} \ge 0$, where $S_{tot}$ is the total integrated entropy for the compact horizon slice.

Thus, finally, we see that the second law holds up to quadratic orders in the dynamical fluctuations for CS theories in $(2+1)$-dimensions in the low energy effective field theory sense ignoring $\mathcal{O}\left(l^2 \right)$ contributions. 
\subsection{Mixed gauge gravity Chern-Simons theories in $(4+1)$-dimensions} \label{BFCS4+1}
In $(4+1)$-dimensions, the mixed gauge gravity CS theory has the following Lagrangian 
\begin{equation} \label{lag4+1a}
 I = \int d^5x \sqrt{-g} \, \mathcal{L} \, , \quad \text{with} \quad   \mathcal{L} = 2 \tensor{\epsilon}{^\mu^\nu^\lambda^\rho^\sigma} \tensor{F}{_\mu_\nu} \tensor{\Gamma}{^\alpha_\lambda_\beta} \left(\frac{1}{2} \tensor{R}{^\beta_\alpha_\rho_\sigma} - \frac{1}{3} \tensor{\Gamma}{^\beta_\rho_\tau} \tensor{\Gamma}{^\tau_\alpha_\sigma} \right) \, ,
\end{equation}
where we can see that $\mathcal{L}$ explicitly depends on non-tensorial Christoffel symbols $\tensor{\Gamma}{^\beta_\rho_\tau}$, but depends on gauge invariant field strength tensor $\tensor{F}{_\mu_\nu}$. Alternatively, an equivalent Lagrangian, say $\tilde{\mathcal{L}}$, can be written down that differs from $\mathcal{L}$ up to the addition of a total derivative piece. One can get   
\begin{equation} \label{lag4+1b}
 \tilde{\mathcal{L}} = \tensor{\epsilon}{^\mu^\nu^\rho^\sigma^\delta} \tensor{A}{_\mu} \tensor{R}{^\alpha_\beta_\nu_\rho} \tensor{R}{^\beta_\alpha_\sigma_\delta} \, ,
\end{equation}
such that 
\begin{equation}
\begin{split}
 \mathcal{L} &= \mathcal{\tilde{L}} + D{_\mu} \tensor{B}{^\mu} \, , \\
 \text{where} \quad \tensor{B}{^\mu} &= 4 \tensor{\epsilon}{^\mu^\nu^\lambda^\rho^\sigma} \tensor{A}{_\nu} \tensor{\Gamma}{^\alpha_\lambda_\beta} \left(\frac{1}{2} \tensor{R}{^\beta_\alpha_\rho_\sigma} - \frac{1}{3} \tensor{\Gamma}{^\beta_\rho_\tau} \tensor{\Gamma}{^\tau_\alpha_\sigma} \right) \, .
 \end{split}
\end{equation}
Both $\mathcal{L}$ and $\mathcal{\tilde{L}}$ will lead to the same EoM
\begin{equation} \label{eom4+1a}
\begin{split}
 \tensor{E}{^\mu^\nu} = \tensor{\epsilon}{^\mu^\beta^\rho^\sigma^\delta} D{_\alpha} \left(\tensor{R}{^\nu^\alpha_\beta_\rho} \tensor{F}{_\sigma_\delta} \right) + \tensor{\epsilon}{^\nu^\beta^\rho^\sigma^\delta} D{_\alpha} \left(\tensor{R}{^\mu^\alpha_\beta_\rho} \tensor{F}{_\sigma_\delta} \right) \, .
 \end{split}
\end{equation}
The near-horizon metric for the dynamical black hole is given by eq.\eqref{nhmetric} where the co-dimension $2$ horizon slice is now spanned by $3$ coordinates $x^i$ for $i=1,2,3$. For this example, we aim to study the second law at the linearized level. 

At the horizon, $E_{vv}$ becomes
\begin{equation} \label{Evv4+1a}
\begin{split}
  \tensor{E}{_v_v} =  \tensor{E}{^r^r} &= 2 \tensor{\epsilon}{^r^\beta^\rho^\sigma^\delta} D{_\alpha} \left(\tensor{R}{^r^\alpha_\beta_\rho} \tensor{F}{_\sigma_\delta} \right) = 2 \tensor{\epsilon}{^r^\beta^\rho^\sigma^\delta} \left(\tensor{F}{_\sigma_\delta} \left(D{_\rho} \tensor{R}{_v_\beta} - D{_\beta} \tensor{R}{_v_\rho} \right) + \tensor{R}{_v^\alpha_\beta_\rho} D{_\alpha} \tensor{F}{_\sigma_\delta} \right) \\
    &= 2 \tensor{\epsilon}{^r^\beta^\rho^\sigma^\delta} \left(2 \tensor{F}{_\sigma_\delta} \left(\tensor{\partial}{_\rho} \tensor{R}{_v_\beta} - \tensor{\Gamma}{^\lambda_\rho_v} \tensor{R}{_\beta_\lambda} \right) + \tensor{R}{_v^\alpha_\beta_\rho} \left(\tensor{\partial}{_\alpha} \tensor{F}{_\sigma_\delta} - 2 \tensor{\Gamma}{^\lambda_\alpha_\sigma} \tensor{F}{_\lambda_\delta} \right) \right) \, .
    \end{split}
\end{equation}
Let us evaluate each term one by one. We have
\begin{equation} \label{dtls4+1a}
\begin{split}
    \tensor{\epsilon}{^r^\beta^\rho^\sigma^\delta} \tensor{R}{_v^\alpha_\beta_\rho} \tensor{\Gamma}{^\lambda_\alpha_\sigma} \tensor{F}{_\lambda_\delta} &= 2 \tensor{\epsilon}{^r^v^i^j^k} \tensor{R}{_v_n_v_i} \tensor{h}{^m^n} \tensor{\Gamma}{^l_m_j} \tensor{F}{_l_k} + \mathcal{O}\left(\epsilon^2 \right) \\
    & = \tensor{\epsilon}{^r^v^i^j^k} \tensor{h}{^m^n} \tensor{\hat{\Gamma}}{^l_m_j} \tensor{F}{_k_l} \tensor{\partial}{^2_v} \tensor{h}{_n_i} + \mathcal{O}\left(\epsilon^2 \right)
    \\
    \tensor{\epsilon}{^r^\beta^\rho^\sigma^\delta} \tensor{R}{_v^\alpha_\beta_\rho} \tensor{\partial}{_\alpha} \tensor{F}{_\sigma_\delta} &= 2 \tensor{\epsilon}{^r^v^i^j^k} \left(\tensor{R}{_v_n_v_i} \tensor{h}{^m^n} \tensor{\partial}{_m} \tensor{F}{_j_k} + \tensor{R}{_v_r_j_k} \tensor{\partial}{_v} \tensor{F}{_v_i} \right) + \mathcal{O}\left(\epsilon^2 \right) \\
    &= \tensor{\epsilon}{^r^v^i^j^k} \left[- \tensor{h}{^m^n} \left(\tensor{\partial}{_m} \tensor{F}{_j_k} \right) \left(\tensor{\partial}{^2_v} \tensor{h}{_n_i} \right) + 2 \left(\tensor{\partial}{_j} \tensor{\omega}{_k} \right) \left(\tensor{\partial}{_v} \tensor{F}{_v_i} \right) \right] + \mathcal{O}\left(\epsilon^2 \right)
    \\
    \tensor{\epsilon}{^r^\beta^\rho^\sigma^\delta} \tensor{F}{_\sigma_\delta} \tensor{\Gamma}{^\lambda_\rho_v} \tensor{R}{_\beta_\lambda} &= \tensor{\epsilon}{^r^v^i^j^k} \tensor{F}{_j_k} \tensor{\Gamma}{^v_i_v} \tensor{R}{_v_v} + \mathcal{O}\left(\epsilon^2 \right) \\
    &= \frac{1}{4} \tensor{\epsilon}{^r^v^i^j^k} \tensor{F}{_j_k} \tensor{\omega}{_i} \tensor{h}{^m^n} \tensor{\partial}{^2_v} \tensor{h}{_m_n} + \mathcal{O}\left(\epsilon^2 \right)
    \\
    \tensor{\epsilon}{^r^\beta^\rho^\sigma^\delta} \tensor{F}{_\sigma_\delta} \tensor{\partial}{_\rho} \tensor{R}{_v_\beta} &= \tensor{\epsilon}{^r^v^i^j^k} \tensor{F}{_j_k} \left(\tensor{\partial}{_i} \tensor{R}{_v_v} - \tensor{\partial}{_v} \tensor{R}{_v_i} \right) + \mathcal{O}\left(\epsilon^2 \right) \\
    &= \frac{1}{4} \tensor{\epsilon}{^r^v^i^j^k} \tensor{F}{_j_k} \left(2 \tensor{\partial}{^2_v} \tensor{\omega}{_i} + \tensor{\omega}{_i} \tensor{h}{^m^n} \tensor{\partial}{^2_v} \tensor{h}{_m_n} \right) \\
    &  - \frac{1}{2} \tensor{\epsilon}{^r^v^i^j^k} \tensor{F}{_j_k} \tensor{h}{^n^m} \left(\tensor{\partial}{_m} \tensor{\partial}{^2_v} \tensor{h}{_n_i} - \tensor{\hat{\Gamma}}{^l_m_n} \tensor{\partial}{^2_v} \tensor{h}{_i_l} - \tensor{\hat{\Gamma}}{^l_m_i} \tensor{\partial}{^2_v} \tensor{h}{_n_l} \right) + \mathcal{O}\left(\epsilon^2 \right) \, .
    \end{split}
\end{equation}
Thus, substituting eq.\eqref{dtls4+1a} in eq.\eqref{Evv4+1a} we have the following expression for $E_{vv}$, up to $\mathcal{O}(\epsilon^2)$ corrections, 
\begin{equation} \label{Evv4+1b}
\begin{split}
    \tensor{E}{_v_v} &= 2 \tensor{\epsilon}{^r^v^i^j^k} \left[2 \left(\tensor{\partial}{_j} \tensor{\omega}{_k} \right) \left(\tensor{\partial}{_v} \tensor{F}{_v_i} \right) - \tensor{h}{^m^n} \tensor{\partial}{_m} \left(\tensor{F}{_j_k} \tensor{\partial}{^2_v} \tensor{h}{_n_i} \right) - 2 \tensor{h}{^m^n} \tensor{\hat{\Gamma}}{^l_m_j} \tensor{F}{_k_l} \tensor{\partial}{^2_v} \tensor{h}{_n_i} \right] \\
    &\quad + 2 \tensor{\epsilon}{^r^v^i^j^k} \tensor{F}{_j_k} \left[\tensor{\partial}{^2_v} \tensor{\omega}{_i} +  \tensor{h}{^n^m} \left(\tensor{\hat{\Gamma}}{^l_m_n} \tensor{\partial}{^2_v} \tensor{h}{_i_l} + \tensor{\hat{\Gamma}}{^l_m_i} \tensor{\partial}{^2_v} \tensor{h}{_n_l} \right) \right] + \mathcal{O}\left(\epsilon^2 \right) 
    \\
    &= 2 \tensor{\partial}{_v} \left[2 \tensor{\epsilon}{^r^v^i^j^k} \left(\tensor{\partial}{_j} \tensor{\omega}{_k} \right) \tensor{F}{_v_i} - \tensor{\epsilon}{^r^v^i^j^k} \tensor{h}{^m^n} \tensor{\partial}{_m} \left(\tensor{F}{_j_k} \tensor{\partial}{_v} \tensor{h}{_n_i} \right) - 2 \tensor{\epsilon}{^r^v^i^j^k} \tensor{h}{^m^n} \tensor{\hat{\Gamma}}{^l_m_j} \tensor{F}{_k_l} \tensor{\partial}{_v} \tensor{h}{_n_i} \right] \\
    &\quad + 2 \tensor{\partial}{_v} \left[\tensor{\epsilon}{^r^v^i^j^k} \tensor{F}{_j_k} \tensor{\partial}{_v} \tensor{\omega}{_i} + \tensor{\epsilon}{^r^v^i^j^k} \tensor{F}{_j_k} \tensor{h}{^n^m} \left(\tensor{\hat{\Gamma}}{^l_m_n} \tensor{\partial}{_v} \tensor{h}{_i_l} + \tensor{\hat{\Gamma}}{^l_m_i} \tensor{\partial}{_v} \tensor{h}{_n_l} \right) \right] + \mathcal{O}\left(\epsilon^2 \right) \, .
    \end{split}
\end{equation}
To bring the $\omega$-dependent terms into the entropy current structure, we do some further manipulations involving Bianchi identities, as follows
\begin{equation} \label{dtls4+1b}
\begin{split}
    \tensor{\epsilon}{^r^v^i^j^k} \tensor{F}{_j_k} \tensor{\partial}{_v} \tensor{\omega}{_i} &= \tensor{\epsilon}{^r^v^i^j^k} \tensor{\partial}{_v} \left(\tensor{F}{_j_k} \tensor{\omega}{_i} \right) - \tensor{\epsilon}{^r^v^i^j^k} \tensor{\omega}{_i} \tensor{\partial}{_v} \tensor{F}{_j_k} \\
    &= \tensor{\epsilon}{^r^v^i^j^k} \tensor{\partial}{_v} \left(\tensor{F}{_j_k} \tensor{\omega}{_i} \right) + \tensor{\epsilon}{^r^v^i^j^k} \tensor{\omega}{_i} \left(\tensor{\partial}{_j} \tensor{F}{_k_v} + \tensor{\partial}{_k} \tensor{F}{_v_j} \right) \\
    &= \tensor{\epsilon}{^r^v^i^j^k} \left[\tensor{\partial}{_v} \left(\tensor{F}{_j_k} \tensor{\omega}{_i} \right) + 2 \tensor{\partial}{_i} \left(\tensor{\omega}{_j} \tensor{F}{_v_k} \right) - 2 \tensor{F}{_v_i} \tensor{\partial}{_j} \tensor{\omega}{_k} \right] \, .
    \end{split}
\end{equation}
Additionally, we can rewrite the $\tensor{F}{_k_l}$ term as
\begin{equation} \label{dtls4+1c}
\begin{split}
    \tensor{\epsilon}{^r^v^i^j^k} \tensor{\hat{\Gamma}}{^l_m_j} \tensor{F}{_k_l} \tensor{\partial}{_v} \tensor{h}{_n_i} &= \left(\tensor{\epsilon}{^r^v^a^j^k} \tensor{\delta}{^i_l} + \tensor{\epsilon}{^r^v^i^a^k} \tensor{\delta}{^j_l} + \tensor{\epsilon}{^r^v^i^j^a} \tensor{\delta}{^k_l} \right) \tensor{\hat{\Gamma}}{^l_m_j} \tensor{F}{_k_a} \tensor{\partial}{_v} \tensor{h}{_n_i} \\
    &= \tensor{\epsilon}{^r^v^i^j^k} \left(\tensor{\hat{\Gamma}}{^l_m_j} \tensor{F}{_k_i} \tensor{\partial}{_v} \tensor{h}{_n_l} + \tensor{\hat{\Gamma}}{^l_m_l} \tensor{F}{_k_j} \tensor{\partial}{_v} \tensor{h}{_n_i} + \tensor{\hat{\Gamma}}{^l_m_j} \tensor{F}{_l_k} \tensor{\partial}{_v} \tensor{h}{_n_i} \right) \, ,\\
    \text{hence}, \quad 2 \tensor{\epsilon}{^r^v^i^j^k} \tensor{\hat{\Gamma}}{^l_m_j} \tensor{F}{_k_l} \tensor{\partial}{_v} \tensor{h}{_n_i} &= \tensor{\epsilon}{^r^v^i^j^k} \tensor{F}{_j_k} \left(\tensor{\hat{\Gamma}}{^l_m_i} \tensor{\partial}{_v} \tensor{h}{_n_l} - \tensor{\hat{\Gamma}}{^l_m_l} \tensor{\partial}{_v} \tensor{h}{_n_i} \right) \, .
    \end{split}
\end{equation}
Thus, substituting eq.\eqref{dtls4+1b} and eq.\eqref{dtls4+1c} in eq.\eqref{Evv4+1b} we get
\begin{equation} \label{Evv4+1c}
\begin{split}
    \tensor{E}{_v_v} &= 2 \tensor{\partial}{_v} \left[\tensor{\epsilon}{^r^v^i^j^k} \tensor{\partial}{_v} \left(\tensor{F}{_j_k} \tensor{\omega}{_i} \right) + 2 \tensor{\epsilon}{^r^v^i^j^k} \tensor{\partial}{_i} \left(\tensor{\omega}{_j} \tensor{F}{_v_k} \right) - \tensor{\epsilon}{^r^v^i^j^k} \tensor{h}{^m^n} \tensor{\partial}{_m} \left(\tensor{F}{_j_k} \tensor{\partial}{_v} \tensor{h}{_n_i} \right) \right] \\
    &\quad + 2 \tensor{\partial}{_v} \left[\tensor{\epsilon}{^r^v^i^j^k} \tensor{F}{_j_k} \left(\tensor{h}{^l^m} \tensor{\hat{\Gamma}}{^n_m_l} + \tensor{h}{^n^m} \tensor{\hat{\Gamma}}{^l_m_l} \right) \left(\tensor{\partial}{_v} \tensor{h}{_i_n} \right) \right] + \mathcal{O}\left(\epsilon^2 \right) \, .
    \end{split}
\end{equation}
Let us evaluate the term in the second line further
\begin{equation}
\begin{split}
    \tensor{h}{^l^m} \tensor{\hat{\Gamma}}{^n_m_l} + \tensor{h}{^n^m} \tensor{\hat{\Gamma}}{^l_m_l} &= \frac{1}{2} \tensor{h}{^l^m} \tensor{h}{^n^o} \left(\tensor{\partial}{_m} \tensor{h}{_o_l} + \tensor{\partial}{_l} \tensor{h}{_o_m} - \tensor{\partial}{_o} \tensor{h}{_m_l} \right) + \frac{1}{2} \tensor{h}{^n^m} \tensor{h}{^l^o} \tensor{\partial}{_m} \tensor{h}{_o_l} \\
    &= \tensor{h}{^l^m} \tensor{h}{^n^o} \tensor{\partial}{_m} \tensor{h}{_o_l} = - \tensor{\partial}{_m} \tensor{h}{^n^m} \, .
    \end{split}
\end{equation}
Finally, we get for $\tensor{E}{_v_v}$
\begin{equation} \label{Evv4+1d}
\begin{split}
    \tensor{E}{_v_v} &= 2 \tensor{\partial}{_v} \left[\tensor{\epsilon}{^r^v^i^j^k} \tensor{\partial}{_v} \left(\tensor{F}{_j_k} \tensor{\omega}{_i} \right) + 2 \tensor{\epsilon}{^r^v^i^j^k}\tensor{\partial}{_i} \left(\tensor{\omega}{_j} \tensor{F}{_v_k} \right) - \tensor{\epsilon}{^r^v^i^j^k} \tensor{\partial}{_m} \left(\tensor{h}{^n^m} \tensor{F}{_j_k} \tensor{\partial}{_v} \tensor{h}{_n_i} \right) \right] + \mathcal{O}\left(\epsilon^2 \right) \\
    &= 2 \tensor{\partial}{_v} \left[\frac{1}{\sqrt{h}} \tensor{\partial}{_v} \left(\sqrt{h} \tensor{\epsilon}{^r^v^i^j^k} \tensor{F}{_j_k} \tensor{\omega}{_i} \right) + \tensor{\nabla}{_i} \left(2 \tensor{\epsilon}{^r^v^i^j^k} \tensor{\omega}{_j} \tensor{F}{_v_k} - \tensor{\epsilon}{^r^v^m^j^k} \tensor{h}{^n^i} \tensor{F}{_j_k} \tensor{\partial}{_v} \tensor{h}{_n_m} \right) \right] + \mathcal{O}\left(\epsilon^2 \right) \, .
    \end{split}
\end{equation}
Thus, it is obvious that the RHS of eq.\eqref{Evv4+1d} has been cast in the form of eq.\eqref{eom1}. Also, we can obtain the following components of the entropy current \footnote{The expression of $\mathcal{J}^v$ in eq.\eqref{JvJi4+1} for $(4+1)$-dimensional CS theory is schematically of the form $\mathcal{J}^v \sim \omega \wedge dA$. A similar prediction was made in \cite{Hollands:2022fkn} (see the remark after Proposition 1) regarding a possible term in $\mathcal{J}^v$ in $(4+1)$-dimensions.}
\begin{equation} \label{JvJi4+1}
    \tensor{\mathcal{J}}{^v} = 2 \tensor{\epsilon}{^r^v^i^j^k} \tensor{F}{_j_k} \tensor{\omega}{_i} \, ,  \qquad\qquad \tensor{\mathcal{J}}{^i} = 4 \tensor{\epsilon}{^r^v^i^j^k} \tensor{\omega}{_j} \tensor{F}{_v_k} - 2 \tensor{\epsilon}{^r^v^m^j^k} \tensor{h}{^n^i} \tensor{F}{_j_k} \tensor{\partial}{_v} \tensor{h}{_n_m} \, .
\end{equation}
Thus the linearized second law for $(4+1)$-dimensional CS theory is also established.

\section{Reparametrization covariance of entropy production}
\label{sec:reparam2+1d}

In this section, we will examine the covariance of entropy production under reparametrization of the horizon slicing for the specific examples of CS theories studied in the previous section. In \S\ref{repara2+1}, we look at gravitational CS theory in $(2+1)$-dimensions as a low energy EFT with the Lagrangian given in eq.\eqref{lagEFT2+1}. For this case, in \S\ref{nlCS2+1}, we have proved the second law to quadratic order in dynamical perturbations within the EFT approximations. Next, in \S\ref{repara4+1}, we will turn our attention to mixed gauge gravity CS theory in $(4+1)$-dimensions, for which in \S\ref{BFCS4+1} we have proved the second law, but only to the linear order of dynamical fluctuations. At linearized order, instead of entropy production, we can, at the most, check that no entropy is destroyed. However, since for $(2+1)$-dimensions, we are working to the non-linear order in perturbations, we will be checking the reparametrization covariance of actual entropy production.

As described in \S\ref{sec:basicsetup}, the reparametrization of horizon slicing is implemented by coordinate transformations, eq.\eqref{reslice}, which keep our choice of gauge for the near horizon metric eq.\eqref{nhmetric} the same. We rewrite the coordinate change for going from $v, r, x^i$ to $\tau, \rho, y^i$ here again for convenience 
 \begin{equation} \label{reslice1}
     v = e^{\zeta(\vec{y})} \tau + \mathcal{O} (\rho) \, , \quad r = e^{-\zeta(\vec{y})} \rho + \mathcal{O} (\rho^2) \, , \quad x^a = y^a + \mathcal{O} (\rho) \, .
 \end{equation}
We have also explained that the information of entropy production is seeded in the off-shell structure of $E_{vv}$ written in terms of the divergence of the entropy current, i.e., in eq.\eqref{eom2}. We will check that both sides of eq.\eqref{Evvlin2+1a} transform identically under eq.\eqref{reslice1}.
\subsection{Gravitational Chern-Simons theory in $(2+1)$-dimensions as EFT to $\mathcal{O}(\epsilon^2)$} \label{repara2+1}
In \S\ref{nlCS2+1}, we have obtained the following expressions from our brute-force analysis; see eq.\eqref{EvvEFT} and eq.\eqref{JvJiEFT}, 
\begin{equation} \label{Evvlin2+1a}
\begin{split}
 E_{vv} \, |_{r=0} &= \partial_v \left[ \frac{1}{\sqrt{h}}  \partial_v \left( \sqrt{h} \, \mathcal{J}^v\right) + \nabla_i \mathcal{J}^i\right]  \\
    &\quad - \frac{3}{4} l \tensor{\epsilon}{^r^v^x} \tensor{\partial}{_x} \left(\frac{1}{h^2} \left(\tensor{\partial}{_v} h \right)^2 \right)+ \left(\frac{1}{2h} \tensor{\partial}{_v} h + \frac{3}{2} l \tensor{\epsilon}{^r^v^x} \tensor{\partial}{_v} \omega \right)^2 + \mathcal{O}\left(l^2 \right) \, , 
 \end{split}
\end{equation}
with 
\begin{equation} \label{JvJi2+1a}
\begin{split}
\tensor{\mathcal{J}}{^v} &= 1+ l \, \tensor{\epsilon}{^r^v^x} \, \omega  \, ,\qquad \tensor{\mathcal{J}}{^i} = - l \, \tensor{\epsilon}{^r^v^x} \left(\frac{1}{h} \tensor{\partial}{_v} h \right) \, .
 \end{split}
\end{equation}
The transformation of $E_{vv}$ in the LHS of eq.\eqref{Evvlin2+1a} can be obtained straightforwardly as follows
\begin{equation} \label{Evvtrans}
\begin{split}
 E_{vv} = \left(\frac{\partial \tau}{\partial v}\right)^2 E_{\tau \tau} \, , \quad \text{implying} \quad E_{vv} = e^{-2 \zeta} E_{\tau \tau} \, .
 \end{split}
\end{equation}

Next, for the RHS of eq.\eqref{Evvlin2+1a}, individually, each of the terms will transform non-trivially, but they must combine together to make the transformation of the full RHS of eq.\eqref{Evvlin2+1a} consistent with eq.\eqref{Evvtrans}. Before further investigating that, we need to set up the rules with which various ingredients transform under eq.\eqref{reslice1}. Let us note the transformation rule for various metric coefficients under eq.\eqref{reslice1}, 
\begin{equation} \label{metrepara1}
\begin{split}
ds^2 &= 2 \, d\tau \, d\rho -\rho^2 \widetilde{X}(\rho, \, \tau, \, y) \, d\tau^2 + 2 \, \rho \, \widetilde{\omega}(\rho, \, \tau, \, y) d\tau dy + \widetilde{h}(\rho, \, \tau, \, y) \, dy^2 \, ,
 \end{split}
\end{equation}
where the required transformation rules are that of $\omega$ and $h$ given in eq.\eqref{reslice}
\begin{equation} \label{metrepara2}
\begin{split} 
\omega = \widetilde{\omega} + 2 (\partial_y \zeta) - \tau (\partial_y \zeta) \frac{1}{\widetilde{h}} \partial_\tau \widetilde{h} + \mathcal{O}(\rho) \, , \quad 
h  = \widetilde{h} + \mathcal{O}(\rho) \, .
 \end{split}
\end{equation}
Here we follow a convention that the quantities with a tilde are evaluated in the transformed coordinates. From here, we also note down how the derivatives transform, 
\begin{equation} \label{derrepara}
\begin{split}
 \partial_v = e^{-\zeta} \partial_\tau +  \mathcal{O}(\rho) \, , \quad \partial_x = \partial_y - \tau (\partial_y \zeta) \partial_\tau +  \mathcal{O}(\rho) \, .
 \end{split} 
\end{equation}
With the knowledge of eq.\eqref{metrepara2} and eq.\eqref{derrepara}, we first derive the following result: Let us assume $B$ is a quantity in terms of $(v, r, x)$ coordinates, and $\widetilde{B}$ be the same quantity transformed to $(\tau, \rho, y)$ coordinates such that  $B$ transform as
\begin{equation}
    B = e^{- \phi(y)} \tilde{B}\, .
\end{equation}
Then, we may show how $\tensor{\partial}{_x} B$ would transform 
\begin{equation} \label{rule1}
    \tensor{\partial}{_x} B = \tensor{\partial}{_y} \left(e^{- \phi} \tilde{B} \right) - \tau \pdv{\zeta}{y} \tensor{\partial}{_\tau} \left(e^{- \phi} \tilde{B} \right)  = e^{- \phi} \left(\tensor{\partial}{_y} \tilde{B} - \tensor{\partial}{_\tau} \left(\tau \pdv{\zeta}{y} \tilde{B} \right) - \pdv{\left(\phi - \zeta \right)}{y} \tilde{B} \right)\, .
\end{equation}
Using eq.\eqref{rule1}, we can derive the following relations
\begin{equation} \label{rule2}
\begin{split}
    \tensor{\partial}{_x} \left(\frac{1}{h} \tensor{\partial}{_v} h \right) &= e^{- \zeta} \left[\tensor{\partial}{_y} \left(\frac{1}{\tilde{h}} \tensor{\partial}{_\tau} \tilde{h} \right) - \tensor{\partial}{_\tau} \left(\tau \pdv{\zeta}{y} \frac{1}{\tilde{h}} \tensor{\partial}{_\tau} \tilde{h} \right) \right] \, ,
    \\
    \tensor{\partial}{_x} \left(\frac{1}{h^2} \left(\tensor{\partial}{_v} h \right)^2 \right) &= e^{- 2 \zeta} \left[\tensor{\partial}{_y} \left(\frac{1}{\tilde{h}^2} \left(\tensor{\partial}{_\tau} \tilde{h} \right)^2 \right) - \tensor{\partial}{_\tau} \left(\tau \pdv{\zeta}{y} \frac{1}{\tilde{h}^2} \left(\tensor{\partial}{_\tau} \tilde{h} \right)^2 \right) - \pdv{\zeta}{\tensor{y}{^i}} \frac{1}{\tilde{h}^2} \left(\tensor{\partial}{_\tau} \tilde{h} \right)^2  \right] \, .
\end{split}
\end{equation}

Let us now return to the RHS of eq.\eqref{Evvlin2+1a}, and first, we focus on the $\mathcal{O} (\epsilon)$ pieces involving $\mathcal{J}^v$ and $\mathcal{J}^i$. Using eq.\eqref{metrepara2}, eq.\eqref{derrepara}, eq.\eqref{rule2} we get for $\mathcal{J}^v$ in eq.\eqref{JvJi2+1a}
\begin{equation} \label{reparaJv}
\begin{split}
 \tensor{\mathcal{J}}{^v} = 1+ l \, \tensor{\epsilon}{^r^v^x} \, \omega = 1+ l \, \tensor{\tilde{\epsilon}}{^\rho^\tau^y} \, \left(\widetilde{\omega} + 2 (\partial_y \zeta) - \tau (\partial_y \zeta) \frac{1}{\widetilde{h}} \partial_\tau \widetilde{h}\right) \, ,
 \end{split}
\end{equation}
and similarly, for $\mathcal{J}^i$ in eq.\eqref{JvJi2+1a} we get
\begin{equation} \label{reparaJi}
\begin{split}
\tensor{\mathcal{J}}{^i} = - l \, \tensor{\epsilon}{^r^v^x} \left(\frac{1}{h} \tensor{\partial}{_v} h \right) = - e^{-\zeta} \, l \, \tensor{\tilde{\epsilon}}{^\rho^\tau^y} \,\left(\frac{1}{\widetilde{h}} \tensor{\partial}{_\tau} \widetilde{h} \right)  \, .
 \end{split}
\end{equation}
Using eq.\eqref{reparaJv} and eq.\eqref{reparaJi} we can immediately check that 
\begin{equation} \label{reparaEvvOe}
\begin{split}
&\partial_v \left[ \frac{1}{\sqrt{h}}  \partial_v \left( \sqrt{h} \, \mathcal{J}^v\right) + \nabla_i \mathcal{J}^i\right] =  \tensor{\partial}{_v} \left(\frac{1}{2}\frac{\tensor{\partial}{_v} h}{h} +l \, \tensor{\epsilon}{^r^v^x} \tensor{\partial}{_v} \omega - l \, \tensor{\epsilon}{^r^v^x} \tensor{\partial}{_x} \left(\frac{\tensor{\partial}{_v} h}{h}\right) \right) \\
    &= e^{- 2 \zeta} \tensor{\partial}{_\tau} \left[\frac{1}{2}\frac{\tensor{\partial}{_\tau} \widetilde{h}}{\widetilde{h}} +l \, \tensor{\tilde{\epsilon}}{^\rho^\tau^y} \left\{ \tensor{\partial}{_\tau} \left(\tilde{\omega} + 2 \pdv{\zeta}{y} - \tau \pdv{\zeta}{y} \frac{\tensor{\partial}{_\tau} \widetilde{h}}{\widetilde{h}} \right) -  \tensor{\partial}{_y} \left(\frac{\tensor{\partial}{_\tau} \widetilde{h}}{\widetilde{h}}\right) + \tensor{\partial}{_\tau} \left(\tau \pdv{\zeta}{y} \frac{\tensor{\partial}{_\tau} \widetilde{h}}{\widetilde{h}} \right) \right\} \right] \\
    &= e^{- 2 \zeta} \tensor{\partial}{_\tau} \left(\frac{1}{2}\frac{\tensor{\partial}{_\tau} \widetilde{h}}{\widetilde{h}} +l \, \tensor{\tilde{\epsilon}}{^\rho^\tau^y} \tensor{\partial}{_\tau} \tilde{\omega} -l \,  \tensor{\tilde{\epsilon}}{^\rho^\tau^y} \tensor{\partial}{_y} \left(\frac{1}{\tilde{h}} \tensor{\partial}{_\tau} \tilde{h} \right) \right)\, .
 \end{split}
\end{equation}
From eq.\eqref{reparaEvvOe}, what we essentially get is the following 
\begin{equation} \label{reparaEvvOe1}
\begin{split}
\partial_v \left[ \frac{1}{\sqrt{h}}  \partial_v \left( \sqrt{h} \, \mathcal{J}^v\right) + \nabla_i \mathcal{J}^i\right] = e^{- 2 \zeta} \, \partial_\tau \left[ \frac{1}{\sqrt{\widetilde{h}}}  \partial_\tau \left( \sqrt{\widetilde{h}} \, \widetilde{\mathcal{J}}^\tau \right) + \widetilde{\nabla}_i \widetilde{\mathcal{J}}^i\right] \, ,
 \end{split}
\end{equation}
such that 
\begin{equation} \label{JvJi2+1c}
\begin{split}
\tensor{\widetilde{\mathcal{J}}}{^\tau} &= 1+ l \, \tensor{\tilde{\epsilon}}{^\rho^\tau^y} \, \left( \widetilde{\omega} + 2 \pdv{\zeta}{y} \right)  \, ,\qquad \tensor{\widetilde{\mathcal{J}}}{^i} = - l \, \tensor{\tilde{\epsilon}}{^\rho^\tau^y} \left(\frac{\tensor{\partial}{_v} \widetilde{h}}{\widetilde{h}}  \right) \, ,
 \end{split}
\end{equation}
are the components of the entropy current in the transformed coordinates. We can clearly see that eq.\eqref{reparaEvvOe1} is consistent with eq.\eqref{Evvtrans}. Therefore, we also learn that the $\mathcal{O}\left(\epsilon \right)$ terms on the RHS of eq.\eqref{Evvlin2+1a} reparameterize among themselves to give the reparameterized currents. While $\mathcal{J}^i$ has the same functional form in any parametrization, the form of the $\mathcal{J}^v$ differs by
\begin{equation}\label{eq:Jv2+1extra}
    \widetilde{\mathcal{J}}^{\tau}_{ex} = 2 l \, \tilde{\epsilon}^{\rho \tau y} \pdv{\zeta}{y} \, ,
\end{equation}
which can be seen by comparing eq.\eqref{JvJi2+1a} with eq.\eqref{JvJi2+1c}. The difference in the form of $\mathcal{J}^v$ for different choice of parametrization of the null generators will not make any difference to the entropy production because
\begin{equation}
	\partial_{\tau} \left( \sqrt{h}  \widetilde{\mathcal{J}}^{\tau}_{ex} \right) = 2 \, l \, \tilde{\varepsilon}^{\rho \tau y}  \, \partial_{\tau} \partial_y \zeta = 0 \, ,
\end{equation}
Thus, the part of $\widetilde{\mathcal{J}}^{\tau}_{ex}$ in $\widetilde{\mathcal{J}}^{\tau}$, present on the RHS of eq.\eqref{JvJi2+1c}, does not make any contribution to the RHS of eq.\eqref{reparaEvvOe1}. This is an ambiguity in resolving $E_{\tau\tau}$ as $\widetilde{\mathcal{J}}^{\tau}$ and $\widetilde{\mathcal{J}}^i$ because $E_{\tau\tau}$ is unchanged even in the presence of $\widetilde{\mathcal{J}}^{\tau}_{ex}$. We will also see a similar issue for the $(4+1)$-dimensional mixed CS theory. Interestingly, one can show that the additional term in the transformation of $\mathcal{J}^v$ in eq.\eqref{eq:Jv2+1extra} will not contribute to the full Iyer-Wald-Tachikawa entropy as
\begin{equation}
    \int dy \, \sqrt{h} \,  \widetilde{\mathcal{J}}^{\tau}_{ex} = \int dy \, 2l \, \Tilde{\varepsilon}^{\rho \tau y} \pdv{\zeta}{y} = 0 \, .
\end{equation}
In the final step, we have used the property of compact horizons (in this case, $S^1$). An analysis analogous to the above for the $(4+1)$-dimensional case is given below eq.\eqref{eq:newJvJi4+1}.

Next, we focus on the quadratic terms, i.e., terms in the second line on the RHS of eq.\eqref{Evvlin2+1a}, and ignore the $\mathcal{O}\left(l^2 \right)$ terms
\begin{equation} \label{RepaOe2a}
\begin{split}
 &- \frac{3}{4} l \tensor{\epsilon}{^r^v^x} \tensor{\partial}{_x} \left(\frac{1}{h^2} \left(\tensor{\partial}{_v} h \right)^2 \right)+ \left(\frac{1}{2h} \tensor{\partial}{_v} h + \frac{3}{2} l \tensor{\epsilon}{^r^v^x} \tensor{\partial}{_v} \omega \right)^2 + \mathcal{O}\left(l^2 \right) \\
 &= 
\frac{1}{4} \left(\frac{1}{h} \tensor{\partial}{_v} h \right)^2 - \frac{3}{4} l \tensor{\epsilon}{^r^v^x} \tensor{\partial}{_x} \left(\frac{1}{h^2} \left(\tensor{\partial}{_v} h \right)^2 \right) + \frac{3}{2} l \tensor{\epsilon}{^r^v^x} \left(\tensor{\partial}{_v} \omega \right) \left(\frac{1}{h} \tensor{\partial}{_v} h \right)+ \mathcal{O}\left(l^2 \right) \, .
\end{split}
\end{equation}
Following a similar approach to the $\mathcal{O}\left(\epsilon \right)$ terms, we study the RHS of eq.\eqref{RepaOe2a}. The first term is easy to handle, as it transforms as 
\begin{equation} \label{RepaOe2b}
\begin{split}
\frac{1}{4} \left(\frac{1}{h} \tensor{\partial}{_v} h \right)^2 =e^{- 2 \zeta} \frac{1}{4}\left(\frac{1}{\tilde{h}^2} \left(\tensor{\tilde{\partial}}{_\tau} \tilde{h} \right)^2 \right) \, .
 \end{split}
\end{equation}
For the last two terms on the RHS of eq.\eqref{RepaOe2a}, we get 
\begin{equation} \label{RepaOe2c}
\begin{split}
    &\quad\ \tensor{\epsilon}{^r^v^x} \left(\tensor{\partial}{_v} \omega \right) \left(\frac{1}{h} \tensor{\partial}{_v} h \right) - \frac{1}{2} \tensor{\epsilon}{^r^v^x} \tensor{\partial}{_x} \left(\frac{1}{h^2} \left(\tensor{\partial}{_v} h \right)^2 \right) 
    \\
    &= e^{- 2 \zeta} \tensor{\tilde{\epsilon}}{^\rho^\tau^y} \left[\tensor{\tilde{\partial}}{_\tau} \left(\tilde{\omega} + 2 \pdv{\zeta}{y} - \tau \pdv{\zeta}{y} \frac{1}{\tilde{h}} \tensor{\tilde{\partial}}{_\tau} \tilde{h} \right) \left(\frac{1}{\tilde{h}} \tensor{\tilde{\partial}}{_\tau} \tilde{h} \right) \right] \\
    &\quad - \frac{1}{2} e^{- 2 \zeta} \tensor{\tilde{\epsilon}}{^\rho^\tau^y} \left[\tensor{\tilde{\partial}}{_y} \left(\frac{1}{\tilde{h}^2} \left(\tensor{\tilde{\partial}}{_\tau} \tilde{h} \right)^2 \right) - \tensor{\tilde{\partial}}{_\tau} \left(\tau \pdv{\zeta}{y} \frac{1}{\tilde{h}^2} \left(\tensor{\tilde{\partial}}{_\tau} \tilde{h} \right)^2 \right) - \pdv{\zeta}{\tensor{y}{^i}} \frac{1}{\tilde{h}^2} \left(\tensor{\tilde{\partial}}{_\tau} \tilde{h} \right)^2 \right] 
    \\
    &= e^{- 2 \zeta} \tensor{\tilde{\epsilon}}{^\rho^\tau^y} \left[\left(\tensor{\tilde{\partial}}{_\tau} \tilde{\omega} \right) \left(\frac{1}{\tilde{h}} \tensor{\tilde{\partial}}{_\tau} \tilde{h} \right) - \frac{1}{2} \tensor{\tilde{\partial}}{_y} \left(\frac{1}{\tilde{h}^2} \left(\tensor{\tilde{\partial}}{_\tau} \tilde{h} \right)^2 \right) \right] \, .
 \end{split}
\end{equation}
Therefore, combining eq.\eqref{RepaOe2b} and eq.\eqref{RepaOe2c} together, we see that even the $\mathcal{O}\left(\epsilon^2 \right)$ terms get reparameterized in the desired covariant way, but up to $ \mathcal{O}\left(l^2 \right)$ corrections, consistent with the EFT expansion
\begin{equation} \label{RepaOe2d}
\begin{split}
 &- \frac{3}{4} l \,  \tensor{\epsilon}{^r^v^x} \tensor{\partial}{_x} \left(\frac{1}{h^2} \left(\tensor{\partial}{_v} h \right)^2 \right)+ \left(\frac{1}{2h} \tensor{\partial}{_v} h + \frac{3}{2} l \tensor{\epsilon}{^r^v^x} \tensor{\partial}{_v} \omega \right)^2 \\
 &= e^{- 2 \zeta} \left[-\frac{3}{4} l \, \tensor{\tilde{\epsilon}}{^\rho^\tau^y} \tensor{\partial}{_y} \left(\frac{1}{\widetilde{h}^2} \left(\tensor{\partial}{_\tau} \widetilde{h} \right)^2 \right)+ \left(\frac{1}{2\widetilde{h}} \tensor{\partial}{_\tau} \widetilde{h} + \frac{3}{2} l \tensor{\tilde{\epsilon}}{^\rho^\tau^y} \tensor{\partial}{_\tau} \widetilde{\omega} \right)^2 \right]
+ \mathcal{O}\left(l^2 \right) \, .
\end{split}
\end{equation}
Therefore, we see that the RHS of eq.\eqref{Evvlin2+1a} indeed transforms covariantly to be consistent with eq.\eqref{Evvtrans}. 


Hence, we prove that for CS theory in $(2+1)$-dimensions as low energy EFT, the second law holds under reparameterizations of horizon slicing up to quadratic orders in the dynamical fluctuations.  
\subsection{Mixed gauge gravity Chern-Simons theory in $(4+1)$-dimensions to $\mathcal{O}(\epsilon)$} \label{repara4+1}
In this sub-section, we study the reparametrization covariance of the $(4+1)$-dimensional mixed Chern-Simons theory given by eq.\eqref{lag4+1a}. The analysis is similar to \S\ref{repara2+1}. We want to study how the $\mathcal{J}^v$ and  $\mathcal{J}^i$ of eq.\eqref{JvJi4+1} transform under eq.\eqref{reslice1}. Since, for this example, our brute-force analysis in \S\ref{BFCS4+1} to derive eq.\eqref{Evv4+1d} involved only $\mathcal{O}(\epsilon)$ terms, we should focus on how the RHS of eq.\eqref{eom1} transforms. For this, the transformation rule for $\mathcal{J}^v$ and  $\mathcal{J}^i$ is sufficient.

Firstly, we would like to have a result similar to eq.\eqref{rule1}, but a higher dimensional generalization of that. If we have a term $C$ which transforms under reparameterization as $C = e^{- \phi} \tilde{C}$, then we have
\begin{equation}\label{eq:rulehigher}
    \begin{split}
        \tensor{\partial}{_i} C &= \tensor{\tilde{\partial}}{_i} \left(e^{- \phi} \tilde{C} \right) - \tau \pdv{\zeta}{\tensor{y}{^i}} \tensor{\tilde{\partial}}{_\tau} \left(e^{- \phi} \tilde{C} \right) = e^{- \phi} \left[\tensor{\tilde{\partial}}{_i} \tilde{C} - \tensor{\tilde{\partial}}{_\tau} \left(\tau \pdv{\zeta}{\tensor{y}{^i}} \tilde{C} \right) - \pdv{\left(\phi - \zeta \right)}{\tensor{y}{^i}} \tilde{C} \right] \, .
    \end{split}
\end{equation}
We can now see how the RHS of $E_{vv}$ in eq.\eqref{Evv4+1d} transforms. This should be given by eq.\eqref{Evvtrans}.
Using eq.\eqref{eq:rulehigher}, we get the following relations
\begin{equation}\label{eq:4+1rep1}
    \begin{split}
        \tensor{\partial}{_i}& \left(\tensor{\omega}{_j} \tensor{F}{_v_k} \right) = e^{- \zeta} \tensor{\tilde{\partial}}{_i} \left(\tensor{\omega}{_j} \tensor{\tilde{F}}{_\tau_k} \right) - e^{- \zeta} \tensor{\tilde{\partial}}{_\tau} \left(\pdv{\zeta}{\tensor{y}{^i}} \tau \tensor{\omega}{_j} \tensor{\tilde{F}}{_\tau_k} \right) 
    \\
    &= e^{- \zeta} \tensor{\tilde{\partial}}{_i} \left(\left(\tensor{\tilde{\omega}}{_j} + 2 \pdv{\zeta}{\tensor{y}{^j}} \right) \tensor{\tilde{F}}{_\tau_k} \right) - e^{- \zeta} \tensor{\tilde{\partial}}{_\tau} \left(\tau \pdv{\zeta}{\tensor{y}{^i}} \left(\tensor{\tilde{\omega}}{_j} + 2 \pdv{\zeta}{\tensor{y}{^j}} \right) \tensor{\tilde{F}}{_\tau_k} \right) + \mathcal{O}\left(\epsilon^2 \right)
    \\
    \tensor{\partial}{_m} &\left(\tensor{h}{^n^m} \tensor{F}{_j_k} \tensor{\partial}{_v} \tensor{h}{_n_i} \right) = e^{- \zeta} \tensor{\tilde{\partial}}{_m} \left(\tensor{\tilde{h}}{^n^m} \tensor{F}{_j_k} \tensor{\partial}{_\tau} \tensor{\tilde{h}}{_n_i} \right) - e^{- \zeta} \tensor{\tilde{\partial}}{_\tau} \left(\tau \pdv{\zeta}{\tensor{y}{^m}} \tensor{\tilde{h}}{^n^m} \tensor{F}{_j_k} \tensor{\partial}{_\tau} \tensor{\tilde{h}}{_n_i} \right) 
    \\
    &= e^{- \zeta} \tensor{\tilde{\partial}}{_m} \left(\tensor{\tilde{h}}{^n^m} \tensor{\tilde{F}}{_j_k} \tensor{\partial}{_\tau} \tensor{\tilde{h}}{_n_i} \right) - e^{- \zeta} \tensor{\tilde{\partial}}{_\tau} \left(\tau \pdv{\zeta}{\tensor{y}{^m}} \tensor{\tilde{h}}{^n^m} \tensor{\tilde{F}}{_j_k} \tensor{\partial}{_\tau} \tensor{\tilde{h}}{_n_i} \right) + \mathcal{O}\left(\epsilon^2 \right) \, .
    \end{split}
\end{equation}
We also have,
\begin{equation}\label{eq:4+1rep2}
    \begin{split}
        & \tensor{\partial}{_v} \left(\tensor{F}{_j_k} \tensor{\omega}{_i} \right)= e^{- \zeta} \tensor{\tilde{\partial}}{_\tau} \left[\left(\tau \pdv{\zeta}{\tensor{y}{^k}} \tensor{\tilde{F}}{_\tau_j} - \tau \pdv{\zeta}{\tensor{y}{^j}} \tensor{\tilde{F}}{_\tau_k} + \tensor{\tilde{F}}{_j_k} \right) \left(\tensor{\tilde{\omega}}{_i} + 2 \pdv{\zeta}{\tensor{y}{^i}} - \tau \tensor{\tilde{h}}{^m^n} \pdv{\zeta}{\tensor{y}{^n}} \tensor{\partial}{_\tau} \tensor{\tilde{h}}{_i_m} \right) \right]
    \\
    &= e^{- \zeta} \tensor{\tilde{\partial}}{_\tau} \left[\left(\tau \pdv{\zeta}{\tensor{y}{^k}} \tensor{\tilde{F}}{_\tau_j} - \tau \pdv{\zeta}{\tensor{y}{^j}} \tensor{\tilde{F}}{_\tau_k} + \tensor{\tilde{F}}{_j_k} \right) \left(\tensor{\tilde{\omega}}{_i} + 2 \pdv{\zeta}{\tensor{y}{^i}} \right) - \tau \tensor{\tilde{h}}{^m^n} \tensor{\tilde{F}}{_j_k} \pdv{\zeta}{\tensor{y}{^n}} \tensor{\partial}{_\tau} \tensor{\tilde{h}}{_i_m} \right] + \mathcal{O}\left(\epsilon^2 \right) \, .
    \end{split}
\end{equation}
Combining eq.\eqref{eq:4+1rep1} and eq.\eqref{eq:4+1rep2}, we get
\begin{equation}
    \begin{split}
        &\quad\ \tensor{\epsilon}{^r^v^i^j^k} \tensor{\partial}{_v} \left(\tensor{F}{_j_k} \tensor{\omega}{_i} \right) + 2 \tensor{\epsilon}{^r^v^i^j^k} \tensor{\partial}{_i} \left(\tensor{\omega}{_j} \tensor{F}{_v_k} \right) - \tensor{\epsilon}{^r^v^i^j^k} \tensor{\partial}{_m} \left(\tensor{h}{^n^m} \tensor{F}{_j_k} \tensor{\partial}{_v} \tensor{h}{_n_i} \right) 
    \\
    &= e^{- \zeta} \tensor{\tilde{\epsilon}}{^\rho^\tau^i^j^k} \tensor{\tilde{\partial}}{_\tau} \left[\left(\tau \pdv{\zeta}{\tensor{y}{^i}} \tensor{\tilde{F}}{_\tau_k} - \tau \pdv{\zeta}{\tensor{y}{^k}} \tensor{\tilde{F}}{_\tau_i} + \tensor{\tilde{F}}{_k_i} \right) \left(\tensor{\tilde{\omega}}{_j} + 2 \pdv{\zeta}{\tensor{y}{^j}} \right) \right] \\ &- e^{- \zeta} \tensor{\tilde{\epsilon}}{^\rho^\tau^i^j^k} \tensor{\tilde{\partial}}{_\tau} \left[\tau \pdv{\zeta}{\tensor{y}{^m}} \tensor{\tilde{h}}{^m^n} \tensor{\tilde{F}}{_j_k} \tensor{\partial}{_\tau} \tensor{\tilde{h}}{_i_n} \right] \\
    &\quad + 2 e^{- \zeta} \tensor{\tilde{\epsilon}}{^\rho^\tau^i^j^k} \tensor{\tilde{\partial}}{_i} \left(\left(\tensor{\tilde{\omega}}{_j} + 2 \pdv{\zeta}{\tensor{y}{^j}} \right) \tensor{\tilde{F}}{_\tau_k} \right) - 2 e^{- \zeta} \tensor{\tilde{\epsilon}}{^\rho^\tau^i^j^k} \tensor{\tilde{\partial}}{_\tau} \left(\tau \pdv{\zeta}{\tensor{y}{^i}} \left(\tensor{\tilde{\omega}}{_j} + 2 \pdv{\zeta}{\tensor{y}{^j}} \right)  \tensor{\tilde{F}}{_\tau_k} \right) \\
    &\quad - e^{- \zeta} \tensor{\tilde{\epsilon}}{^\rho^\tau^i^j^k} \tensor{\tilde{\partial}}{_m} \left(\tensor{\tilde{h}}{^n^m} \tensor{\tilde{F}}{_j_k} \tensor{\partial}{_\tau} \tensor{\tilde{h}}{_n_i} \right) + e^{- \zeta} \tensor{\tilde{\epsilon}}{^\rho^\tau^i^j^k} \tensor{\tilde{\partial}}{_\tau} \left(\tau \pdv{\zeta}{\tensor{y}{^m}} \tensor{\tilde{h}}{^n^m} \tensor{\tilde{F}}{_j_k} \tensor{\partial}{_\tau} \tensor{\tilde{h}}{_n_i} \right) + \mathcal{O}\left(\epsilon^2 \right)
    \\
    &= e^{- \zeta} \tensor{\tilde{\epsilon}}{^\rho^\tau^i^j^k} \left(\tensor{\tilde{\partial}}{_\tau} \left[\tensor{\tilde{F}}{_k_i} \left(\tensor{\tilde{\omega}}{_j} + 2 \pdv{\zeta}{\tensor{y}{^j}} \right) \right] + 2 \tensor{\tilde{\partial}}{_i} \left[\left(\tensor{\tilde{\omega}}{_j} + 2 \pdv{\zeta}{\tensor{y}{^j}} \right) \tensor{\tilde{F}}{_\tau_k} \right] - \tensor{\tilde{\partial}}{_m} \left[\tensor{\tilde{h}}{^n^m} \tensor{\tilde{F}}{_j_k} \tensor{\partial}{_\tau} \tensor{\tilde{h}}{_n_i} \right] \right) + \mathcal{O}\left(\epsilon^2 \right) \, .
    \end{split}
\end{equation}
Thus, the $E_{vv}$ of eq.\eqref{Evv4+1d} transforms as
\begin{equation}\label{eq:Evv4+1reparam}
    \begin{split}
        \tensor{E}{_v_v} &= 2 \tensor{\partial}{_v} \left[\tensor{\epsilon}{^r^v^i^j^k} \tensor{\partial}{_v} \left(\tensor{F}{_j_k} \tensor{\omega}{_i} \right) + 2 \tensor{\epsilon}{^r^v^i^j^k} \tensor{\partial}{_i} \left(\tensor{\omega}{_j} \tensor{F}{_v_k} \right) - \tensor{\epsilon}{^r^v^i^j^k} \tensor{\partial}{_m} \left(\tensor{h}{^n^m} \tensor{F}{_j_k} \tensor{\partial}{_v} \tensor{h}{_n_i} \right) \right] + \mathcal{O}\left(\epsilon^2 \right)
    \\
    &= 2 e^{- 2 \zeta} \tensor{\tilde{\partial}}{_\tau} \left(\tensor{\tilde{\epsilon}}{^\rho^\tau^i^j^k} \tensor{\tilde{\partial}}{_\tau} \left[\tensor{\tilde{F}}{_j_k} \left(\tensor{\tilde{\omega}}{_i} + 2 \pdv{\zeta}{\tensor{y}{^i}} \right) \right] + 2 \tensor{\tilde{\epsilon}}{^\rho^\tau^i^j^k} \tensor{\tilde{\partial}}{_i} \left[\left(\tensor{\tilde{\omega}}{_j} + 2 \pdv{\zeta}{\tensor{y}{^j}} \right) \tensor{\tilde{F}}{_\tau_k} \right] \right) \\
    &\quad - 2 e^{- 2 \zeta} \tensor{\tilde{\partial}}{_\tau} \left(\tensor{\tilde{\epsilon}}{^\rho^\tau^i^j^k} \tensor{\tilde{\partial}}{_m} \left[\tensor{\tilde{h}}{^n^m} \tensor{\tilde{F}}{_j_k} \tensor{\partial}{_\tau} \tensor{\tilde{h}}{_n_i} \right] \right) + \mathcal{O}\left(\epsilon^2 \right) \, .
    \end{split}
\end{equation}
From eq.\eqref{eq:Evv4+1reparam}, consistency of eq.\eqref{Evvtrans} is confirmed, $\tensor{E}{_v_v} = e^{- 2 \zeta} \tensor{E}{_\tau_\tau}$. Consequently, in the transformed coordinates, we still get the off-shell structure of $E_{\tau\tau}$ as in eq.\eqref{eom1}, but now in terms of the transformed $\tensor{\widetilde{\mathcal{J}}}{^\tau}$ and $\tensor{\widetilde{\mathcal{J}}}{^i}$
\begin{equation} \label{eomtrns}
     E_{\tau\tau} \, |_{\rho=0} = \partial_\tau \left[ \frac{1}{\sqrt{\widetilde{h}}}  \partial_\tau \left( \sqrt{\widetilde{h}} \, \widetilde{\mathcal{J}}^\tau \right) + \widetilde{\nabla}_i \widetilde{\mathcal{J}}^i\right] + \mathcal{O}(\epsilon^2) \, ,
\end{equation}
and we can readily obtain the transformed entropy current components in the new coordinate system as
\begin{equation}\label{eq:newJvJi4+1}
    \tensor{\widetilde{\mathcal{J}}}{^\tau} = 2 \tensor{\tilde{\epsilon}}{^\rho^\tau^i^j^k} \tensor{\tilde{F}}{_j_k} \left(\tensor{\tilde{\omega}}{_i} + 2 \pdv{\zeta}{\tensor{y}{^i}} \right) \, , \quad \tensor{\widetilde{\mathcal{J}}}{^i} = 4 \tensor{\tilde{\epsilon}}{^\rho^\tau^i^j^k} \left(\tensor{\tilde{\omega}}{_j} + 2 \pdv{\zeta}{\tensor{y}{^j}} \right) \tensor{\tilde{F}}{_\tau_k} - 2 \tensor{\tilde{\epsilon}}{^\rho^\tau^m^j^k} \tensor{\tilde{h}}{^n^i} \tensor{\tilde{F}}{_j_k} \tensor{\partial}{_\tau} \tensor{\tilde{h}}{_n_m} \, . 
\end{equation}

The expressions of $\tensor{\widetilde{\mathcal{J}}}{^\tau}$ and $\tensor{\widetilde{\mathcal{J}}}{^i}$ written in eq.\eqref{eq:newJvJi4+1} exemplifies an important aspect of the entropy current construction in general which we would like to highlight. If we compare eq.\eqref{eq:newJvJi4+1} with eq.\eqref{JvJi4+1}, we see that there are additional terms in eq.\eqref{eq:newJvJi4+1} which we denote by 
\begin{equation}\label{eq:4+1repamb}
    \tensor{\widetilde{\mathcal{J}}}{^\tau}_{ex} = 4\tensor{\tilde{\epsilon}}{^\rho^\tau^i^j^k} \tensor{\tilde{F}}{_j_k}  \pdv{\zeta}{\tensor{y}{^i}}  \, , \qquad \tensor{\widetilde{\mathcal{J}}}{^i}_{ex} = 8 \tensor{\tilde{\epsilon}}{^\rho^\tau^i^j^k}  \pdv{\zeta}{\tensor{y}{^j}} \tensor{\tilde{F}}{_\tau_k} \, .
\end{equation}
This additional term in $\widetilde{\mathcal{J}}^{\tau}$ is similar to the additional term in the $(2+1)$-case eq.\eqref{JvJi2+1c}. One can similarly show that  \footnote{Here $\tensor{\tilde{\varepsilon}}{^\rho^\tau^i^j^k}$ denotes the $(4+1)$-dimensional Levi Civita symbol.}
\begin{equation}
    \begin{split}
        \partial_{\tau} \left(\sqrt{h}\tensor{\widetilde{\mathcal{J}}}{^\tau}_{ex} \right) &= 4 \tensor{\tilde{\varepsilon}}{^\rho^\tau^i^j^k} \partial_{\tau} \tilde{F}_{jk}  \pdv{\zeta}{\tensor{y}{^i}} = 4 \tensor{\tilde{\varepsilon}}{^\rho^\tau^i^j^k} \partial_{[\tau} \tilde{F}_{jk]}  \pdv{\zeta}{\tensor{y}{^i}} =0 \, , \\
        \partial_i \left(\sqrt{h}\tensor{\widetilde{\mathcal{J}}}{^i}_{ex} \right) &= 8 \tensor{\tilde{\varepsilon}}{^\rho^\tau^i^j^k} \partial_{[i}\tilde{F}_{\tau k]} \pdv{\zeta}{\tensor{y}{^j}} + 8 \tensor{\tilde{\varepsilon}}{^\rho^\tau^i^j^k} \tensor{\tilde{F}}{_\tau_k} \pdv{\zeta}{y^{[i}y^{j]}} = 0 \, , 
    \end{split}
\end{equation}
and hence 
\begin{equation}
    \begin{split}
         \frac{1}{\sqrt{\widetilde{h}}}  \partial_\tau \left( \sqrt{\widetilde{h}} \, \widetilde{\mathcal{J}}^\tau_{ex}  \right) + \widetilde{\nabla}_i \widetilde{\mathcal{J}}^i_{ex} = 0 \, . 
    \end{split}
\end{equation}
This means that the extra terms of eq.\eqref{eq:4+1repamb} (and that of eq.\eqref{JvJi2+1c}) drop out from the $E_{vv}$ combination in eq.\eqref{eq:Evv4+1reparam} (in eq.\eqref{reparaEvvOe1}). In other words, although in entropy current components there are these additional contributions, they do not contribute to the off-shell structure of $E_{\tau\tau}$, in the RHS of eq.\eqref{eomtrns}. Thus, these additional terms of eq.\eqref{eq:4+1repamb} should be thought of as ambiguities that contribute to the components of the entropy current but do not contribute to $E_{vv}$. These ambiguities have appeared before, e.g., in \cite{Bhattacharya:2019qal} in the context of $(3+1)$-dimensional Gauss-Bonnet gravity \footnote{In $(3+1)$-dimensions the Gauss-Bonnet term does not contribute to the EoMs, but the local entropy current receives contributions from them. See the recent work \cite{Gadioux:2023pmw} where this has been looked at carefully.}. In that context, they existed irrespective of what horizon slicing we used. However, here for our example in $(4+1)$-dimensional CS theory, we find an instance where we started with a coordinate system without these ambiguities. Still, they are generated due to a change of horizon slicing \footnote{Note that for a particular type of coordinate reparametrization with constant $\zeta$, both $\tensor{\widetilde{\mathcal{J}}}{^\tau}_{ex}$ and $\tensor{\widetilde{\mathcal{J}}}{^i}_{ex}$ vanishes.}. 

It is also important to note that the $\tensor{\widetilde{\mathcal{J}}}{^\tau}_{ex}$ obtained in eq.\eqref{eq:4+1repamb} is boost-invariant and it survives in the equilibrium limit, implying that it should get captured in the equilibrium definition of entropy eq.\eqref{Jveq}. As the entropy of the black holes defined through the Iyer-Wald-Tachikawa Noether charge eq.\eqref{eq:siwtfinal} is non-covariant, we expect such non-covariant pieces under a coordinate transformation. However, one can see that this extra term doesn't contribute to the total entropy eq.\eqref{stotdyn}. This is because
\begin{equation}
    \begin{split}
        \int d^3 y \, \sqrt{\tilde{h}} \, \widetilde{J}^{\tau}_{ex} &= \int d^3y \, 4 \tilde{\varepsilon}^{\rho \tau i j k} \Tilde{F}_{jk} \partial_i \zeta = \int d^3 y \, \left[ 4 \partial_i \left( \tilde{\varepsilon}^{\rho \tau i j k} \Tilde{F}_{jk} \zeta \right) + 4 \tilde{\varepsilon}^{\rho \tau i j k} \partial_i \Tilde{F}_{jk} \zeta  \right] \\
        &= \int d^3y \, \sqrt{\Tilde{h}} \nabla_i \left[ 4 \tilde{\epsilon}^{\rho \tau i j k} \Tilde{F}_{jk} \zeta \right] = 0 \, ,
    \end{split}
\end{equation}
where to get to the final step, we used $\partial_{[i} \Tilde{F}_{jk]} = 0$ and we integrated the remaining total derivative term on the assumption of compact horizons \footnote{It should be noted that recently, in \cite{Hollands:2022fkn}, the transformation of the entropy current components was studied under coordinate reparameterization. They focus only on diffeomorphism invariant theories of gravity. The final conclusion of our result for the $(2+1)$-dimensional pure gravity CS theory and the $4+1$-dimensional mixed CS theory is nevertheless consistent with their transformation of $\mathcal{J}^v$ on $v=\tau=0$: $\widetilde{\mathcal{J}}^{\tau} = \mathcal{J}^v + \nabla_i B^i$. Thus, eq.\eqref{eq:Jv2+1extra} and eq.\eqref{eq:4+1repamb} are clear cases that demonstrate Proposition 4.1 of \cite{Hollands:2022fkn} for particular CS theories. It would be interesting to see if we can generalize Proposition 4.1 for generic CS theories.}. This is what one should expect physically as well. The total entropy of the stationary black hole should be independent of the choice of coordinates. 


\section{Proof of constructing entropy current in generic Chern-Simons theories} \label{sec:absprf}

We will now use all the ingredients of \S\ref{sec:basicsetup} and \S \ref{sec:cscovphase} to construct an entropy current for dynamical black holes in arbitrary CS theories of gravity of the form given by Lagrangian eq.\eqref{eq:cslagrangian}. We will follow the strategy outlined in \cite{Bhattacharyya:2021jhr, Biswas:2022grc}. Our starting point would be the main equation eq.\eqref{eq:main}:
\begin{equation}
    2 v \, E_{vv} + G_v (v A_v + \Lambda) = (-\Theta^r + \Xi^r + D_{\rho} Q^{r\rho})|_{r=0} \, .
\end{equation}
We will now use the boost weight analysis to establish that $\Theta^r$, $\Xi^r$ and $Q^{r\rho}$ take a particular form that will allow for an entropy current structure of eq.\eqref{eom1} for $E_{vv}$. We will first briefly review the proof of the existence of entropy current for diffeomorphism and $U(1)$ invariant theories of gravity. We will then impose constraints that restrict the form of the Lagrangian for Chern-Simons theories of gravity. Once we have the Lagrangian, we work out all the associated structures. Finally, we evaluate the structures in our gauge to establish the existence of entropy current for arbitrary Chern-Simons theories of gravity.

We first quickly recap the idea behind the proof of the existence of an entropy current for diffeomorphism invariant theories of gravity \cite{Bhattacharyya:2021jhr, Biswas:2022grc}. Since we have chosen a particular horizon adapted coordinates in eq.\eqref{nhmetric}, the stationary configuration has a particular Killing vector given by eq.\eqref{KillVec}. This ultimately results in a factor of $v$ multiplying $E_{vv}$ in eq.\eqref{eq:main}. If we consider the presence of gauge fields as in \cite{Biswas:2022grc} and the Lagrangian is gauge invariant, then the boost weight analysis together with eq.\eqref{eqlbmA1} would imply that $G_v (v A_v + \Lambda) \sim \mathcal{O}(\epsilon^2)$. Thus, the main task of proving the entropy current structure of eq.\eqref{eom1} from eq.\eqref{eq:main} (with $\Xi^r$ set to zero since we are reviewing diffeomorphism invariant theories) would require us to track down the explicit factors of $v$ from the RHS of eq.\eqref{eq:main}. These factors of $v$ arise from $\Theta^r$ and $Q^{r\rho}$. From the general analysis of the construction of Noether charge for arbitrary diffeomorphism invariant theories of gravity \cite{Iyer:1994ys}, it is known that $\Theta^{\mu}$ and $Q^{\mu\nu}$ are linear in $\xi$. The off-shell structure of $E_{vv}$ is built out of the metric functions of eq.\eqref{nhmetric} $(X,\omega_i,h_{ij})$, and various derivatives $(\partial_v,\partial_r,\nabla_i)$ acting on them. We must also include gauge invariant $U(1)$ field strength tensor components $F_{vr}$, $F_{vi}$, $F_{ri}$ and $F_{ij}$ if we consider gauge invariant Lagrangians. These functions only have implicit dependence on $v$; thus, any explicit dependence on $v$ should come solely from $\xi$. This implies, on the horizon $r=0$, $\Theta^r$ and $Q^{r\rho}$ have the following structure \footnote{In $Q^{ri}$, we have chosen to represent the boost weight $+2$ quantity as $\partial_v \tilde{J}^i_{(1)}$. We can always do this up to $\mathcal{O}(\epsilon^2)$ terms.}
\begin{equation}\label{eq:thetaQinitial}
    \Theta^r |_{r=0} = \Theta_{(1)} + v \Theta_{(2)} \, , ~~~ Q^{rv}|_{r=0} = Q_{(0)} + v Q_{(1)} \, , ~~~ Q^{ri}|_{r=0} = J^i_{(1)} + v \partial_v \tilde{J}^i_{(1)} \, . 
\end{equation}
The subscripts in the RHS denote the boost-weight defined by eq.\eqref{defboostwt} of the various quantities. One can show that these quantities evaluated on the horizon are given by (see Appendix \ref{sec:reviewcurrent} for the details of this review)
\begin{equation}\label{eq:thetardiffgen}
    \Theta^r |_{r=0} = (1+ v \partial_v) \mathcal{A}_{(1)} + v \partial^2_v \mathcal{B}_{(0)} \, .
\end{equation}
Here $\mathcal{B}_{(0)}$ denotes the JKM ambiguity, and it is $\mathcal{O}(\epsilon)$ even though it has zero boost-weight. This is because the $\mathcal{B}_{(0)}$ takes the form of a product of two terms that are individually not boost-invariant:
\begin{equation}\label{eq:B0diffgen}
    \mathcal{B}_{(0)} \sim X_{(-k+m)} \partial^{k-m}_v Y_{(0)} \sim \mathcal{O}(\epsilon) \, .
\end{equation}
Similarly, we can get
\begin{equation}\label{eq:Qrnustrucdiffgen}
    Q^{r\mu} = \widetilde{Q}^{r\mu} + v \, W^{r\mu}_v \, .
\end{equation}
We can combine eq.\eqref{eq:thetardiffgen} and eq.\eqref{eq:Qrnustrucdiffgen} in eq.\eqref{eq:main} (with $\Xi^r=0$) by carefully tracking the factors of $v$ to get
\begin{equation}\label{eq:evvfinaldiffgen}
    2 E_{vv}|_{r=0} = - \partial_v \left( \dfrac{1}{\sqrt{h}} \partial_v \left[ \sqrt{h} \left( \widetilde{Q}^{rv} + \mathcal{B}_{(0)} \right) \right] + \nabla_i \left[ \widetilde{Q}^{ri} - J^i_{(1)} \right] \right) + \mathcal{O}(\epsilon^2) \, .
\end{equation}
Here $J^i_{(1)}$ is defined through $W^{ri}_v = \partial_v J^i_{(1)} + \mathcal{O}(\epsilon^2)$. This allows us to write the components of the entropy current to be \footnote{Notably, one can show that when $U(1)$ gauge invariant Lagrangians are considered, $\mathcal{J}^v$ and $\mathcal{J}^i$ are gauge invariant. This is because the structures $\Theta^r$ and $Q^{r\rho}$ are $U(1)$ gauge invariant.}
\begin{equation}\label{eq:jvjidifffinal}
    \mathcal{J}^v = - \dfrac{1}{2} \left( \widetilde{Q}^{rv} + \mathcal{B}_{(0)} \right) \, , ~~~ \text{and} ~~ \mathcal{J}^i = - \dfrac{1}{2} \left( \widetilde{Q}^{ri} - J^i_{(1)} \right) \, .
\end{equation}
We now proceed to focus on how to extend this analysis for CS theories of gravity. 

\subsection{Structure of the Lagrangian for Chern-Simons theories}

In this section, we elucidate the structure of the CS Lagrangian eq.\eqref{eq:cslagrangian}. The Lagrangian of eq.\eqref{eq:cslagrangian} has an explicit dependence of $A_{\mu}$. This makes it difficult to study the structures $\Theta^r$, $\Xi^r$, and $Q^{r\rho}$ because they too will carry the explicit dependence of $A_{\mu}$. $A_v$ will enter the boost weight analysis since it has a non-trivial boost weight and will mess up the structure of Result:1 given in eq.\eqref{eq:genboostwt}. If the structure of a generic covariant tensor with positive boost weight changes, then eq.\eqref{eq:thetardiffgen} and eq.\eqref{eq:Qrnustrucdiffgen} will change due to additional factors proportional to $A_v$. This will significantly affect the final structure of eq.\eqref{eq:evvfinaldiffgen} due to $A_v$ factors polluting the terms. Let us point this out by first computing the various structures: $\Theta^{\mu}$, $\Xi^{\mu}$ and $Q^{\mu\rho}$ for eq.\eqref{eq:cslagrangian}.

\subsubsection{Noether charge for Chern-Simons theories}
\label{sec:noetherchargecs}

The form of the Lagrangian in eq.\eqref{eq:cslagrangian} can in principle depend arbitrarily on $\Gamma^{\alpha}_{\beta\mu}$ and $A_{\mu}$. The crucial input of the Chern-Simons class of theories comes from eq.\eqref{eq:diffvarl}, i.e., under a combined action of the diffeomorphism and a $U(1)$ gauge transformation eq.\eqref{eq:diffeo}, the Lagrangian is constrained to vary as eq.\eqref{eq:diffvarl}. This imposes severe restrictions on the form of the Lagrangian eq.\eqref{eq:cslagrangian}. Before studying these restrictions, it is useful to compute the Noether charge $Q^{\mu\nu}$ for generic Chern-Simons theories. This has been studied before in \cite{Bonora:2011gz, Azeyanagi:2014sna, Copetti:2017cin}, but we work with the component notation, unlike those works which work in the differential form language. The standard variation of the Lagrangian eq.\eqref{eq:cslagrangian} gives
\begin{equation}\label{eq:deltaL}
    \delta L = \pdv{L}{\tensor{A}{_\mu}} \delta \tensor{A}{_\mu} + \pdv{L}{\tensor{F}{_\mu_\nu}} \delta \tensor{F}{_\mu_\nu} + \pdv{L}{\tensor{\Gamma}{^\lambda_\mu_\nu}} \delta \tensor{\Gamma}{^\lambda_\mu_\nu} + \pdv{L}{\tensor{R}{^\alpha_\beta_\mu_\nu}} \delta \tensor{R}{^\alpha_\beta_\mu_\nu} + \pdv{L}{\tensor{g}{^\mu^\nu}} \delta \tensor{g}{^\mu^\nu} \, .
\end{equation}
Now, eq.\eqref{eq:deltaL} gives eq.\eqref{eq:varL} with $E^{\mu\nu}$, $G^{\mu}$ and $\Theta^{\mu}$ given by
\begin{equation}\label{eq:cseom}
    \begin{split}
    \tensor{E}{^\mu^\nu} &= \frac{1}{2} L \tensor{g}{^\mu^\nu} - \tensor{g}{^\mu^\alpha} \tensor{g}{^\nu^\beta} \pdv{L}{\tensor{g}{^\alpha^\beta}} - D{_\alpha} \tensor{S}{^\alpha^\mu^\nu} \, , ~~~~ 
    \tensor{G}{^\mu} = \pdv{L}{\tensor{A}{_\mu}} + 2 D{_\nu} \pdv{L}{\tensor{F}{_\mu_\nu}} \, , \\
    \tensor{\Theta}{^\mu} &= 2 \pdv{L}{\tensor{F}{_\mu_\nu}} \delta \tensor{A}{_\nu} + 2 \pdv{L}{\tensor{R}{^\alpha_\beta_\mu_\nu}} \delta \tensor{\Gamma}{^\alpha_\beta_\nu} + \tensor{S}{^\mu^\alpha^\beta} \delta \tensor{g}{_\alpha_\beta} \, ,
\end{split}
\end{equation}
where 
\begin{equation}\label{eq:defS}
    \begin{split}
    \tensor{S}{^\mu^\alpha^\beta} &= \frac{1}{2} D{_\nu} \left[\tensor{g}{^\alpha^\lambda} \left(\pdv{L}{\tensor{R}{^\lambda_\beta_\mu_\nu}} + \pdv{L}{\tensor{R}{^\lambda_\mu_\beta_\nu}} \right) + \tensor{g}{^\beta^\lambda} \left(\pdv{L}{\tensor{R}{^\lambda_\alpha_\mu_\nu}} + \pdv{L}{\tensor{R}{^\lambda_\mu_\alpha_\nu}} \right) \right. \\
    &\quad \left. - \tensor{g}{^\mu^\lambda} \left(\pdv{L}{\tensor{R}{^\lambda_\beta_\alpha_\nu}} + \pdv{L}{\tensor{R}{^\lambda_\alpha_\beta_\nu}} \right) \right] + \frac{1}{2} \left(\tensor{g}{^\alpha^\lambda} \pdv{L}{\tensor{\Gamma}{^\lambda_\beta_\mu}} + \tensor{g}{^\beta^\lambda} \pdv{L}{\tensor{\Gamma}{^\lambda_\alpha_\mu}} - \tensor{g}{^\mu^\lambda} \pdv{L}{\tensor{\Gamma}{^\lambda_\beta_\alpha}} \right) \, .
    \end{split}
\end{equation}

Next, we consider the variation under eq.\eqref{eq:diffeo}. Under eq.\eqref{eq:diffeo} $A_{\mu},\Gamma^{\alpha}_{\beta\mu}$ transform as
\begin{equation} \label{delAchris}
\begin{split}
    \delta \tensor{A}{_\mu} &= \mathcal{L}_{\xi} \tensor{A}{_\mu} + D{_\mu} \Lambda \, ,
    \\
    \delta \tensor{\Gamma}{^\lambda_\mu_\nu} &= \mathcal{L}_{\xi} \tensor{\Gamma}{^\lambda_\mu_\nu} + \partial{_\mu} \partial_{\nu} \tensor{\xi}{^\lambda} \, .
\end{split}
\end{equation}
If we implement eq.\eqref{delAchris} in eq.\eqref{eq:deltaL}, we have
\begin{equation}\label{eq:varlxider}
    \begin{split}
    \delta L &= \pdv{L}{\tensor{A}{_\mu}} \mathcal{L}_{\xi} \tensor{A}{_\mu} + \pdv{L}{\tensor{F}{_\mu_\nu}} \mathcal{L}_{\xi} \tensor{F}{_\mu_\nu} + \pdv{L}{\tensor{\Gamma}{^\lambda_\mu_\nu}} \mathcal{L}_{\xi} \tensor{\Gamma}{^\lambda_\mu_\nu} + \pdv{L}{\tensor{R}{^\alpha_\beta_\mu_\nu}} \mathcal{L}_{\xi} \tensor{R}{^\alpha_\beta_\mu_\nu} +  \pdv{L}{\tensor{g}{^\mu^\nu}} \mathcal{L}_{\xi} \tensor{g}{^\mu^\nu} \\
    &\quad + \pdv{L}{\tensor{A}{_\mu}} D{_\mu} \Lambda + \pdv{L}{\tensor{\Gamma}{^\lambda_\mu_\nu}} \tensor{\partial}{_\mu}\partial_{\nu} \tensor{\xi}{^\lambda} = \mathcal{L}_{\xi} L + \pdv{L}{\tensor{A}{_\mu}} D{_\mu} \Lambda + \pdv{L}{\tensor{\Gamma}{^\lambda_\mu_\nu}} \tensor{\partial}{_\mu}\tensor{\partial}{_\nu} \tensor{\xi}{^\lambda} \, .
    \end{split}
\end{equation}
This clarifies that the Lagrangian in eq.\eqref{eq:cslagrangian} is not diffeomorphism/$U(1)$ covariant. 
After some integration by parts manipulations (see eq.\eqref{eq:partialLG} in Appendix \ref{ap:noethercs} for details), we get
\begin{equation}\label{eq:finaldiffvarL}
    \delta_{\xi}\left(\sqrt{-g} L \right) = \sqrt{-g} D{_\mu} \left(\tensor{\xi}{^\mu} L \right) + \sqrt{-g} D{_\mu} \tensor{\Xi}{^\mu} - \sqrt{-g} \Lambda D{_\mu} \pdv{L}{\tensor{A}{_\mu}} + \tensor{\xi}{^\lambda} \tensor{\partial}{_\mu}\tensor{\partial}{_\nu} \left(\sqrt{-g} \pdv{L}{\tensor{\Gamma}{^\lambda_\mu_\nu}} \right) \, ,
\end{equation}
where $\tensor{\Xi}{^\mu}$ is \footnote{\label{foot:xiamb}This structure of $\Xi^{\mu}$ is slightly different from the differential form $\Xi$ derived in \cite{Bonora:2011gz,Azeyanagi:2014sna,Copetti:2017cin}. It should be mentioned that we are treating the Christoffel symbol as being symmetric in its two lower indices. This differs from the differential form notation, where the Christoffel symbol is treated as a matrix-valued one form. This singles out the first index (the index of the one form) as special. Due to this ambiguity in treating the Christoffel symbol, the expression of $\Xi^{\mu}$ for specific examples derived from eq.\eqref{eq:defxi} differs from the brute force computations by a total derivative ambiguity. See eq.\eqref{eq:2+1gengravstruc} and eq.\eqref{eq:4+1genstruc} in \S \ref{sec:proofverf} for details.}
\begin{equation}\label{eq:defxi}
    \tensor{\Xi}{^\mu} = \pdv{L}{\tensor{A}{_\mu}} \Lambda + \pdv{L}{\tensor{\Gamma}{^\lambda_\mu_\nu}} \tensor{\partial}{_\nu} \tensor{\xi}{^\lambda} - \frac{1}{\sqrt{-g}} \tensor{\xi}{^\lambda} \tensor{\partial}{_\nu} \left(\sqrt{-g} \pdv{L}{\tensor{\Gamma}{^\lambda_\mu_\nu}} \right) \, .
\end{equation}
Thus, we see that in order for eq.\eqref{eq:finaldiffvarL} to match with the definition of a CS Lagrangian eq.\eqref{eq:diffvarl}, we have two separate identities that need to be satisfied \footnote{The symmetric treatment of the two lower indices of the Christoffel symbol is evident from the second Bianchi identity for the gravity sector. In the differential form notation \cite{Bonora:2011gz, Azeyanagi:2014sna, Copetti:2017cin}, the first index would be special, and we would have ended with just one derivative contracting the first index similar to the gauge sector Bianchi identity.}:
\begin{equation}\label{eq:bianchics}
    D{_\mu} \pdv{L}{\tensor{A}{_\mu}} = 0 \, , \qquad\qquad \tensor{\partial}{_\mu}\tensor{\partial}{_\nu} \left(\sqrt{-g} \pdv{L}{\tensor{\Gamma}{^\lambda_\mu_\nu}} \right) = 0 \, , 
\end{equation}
which are obtained by using the fact that $\xi^\lambda$ and $\Lambda$ are arbitrary in eq.\eqref{eq:finaldiffvarL} above.
These are additional Chern-Simons Bianchi identities apart from the general Bianchi identities eq.\eqref{eq:bianchigen}. We can now use eq.\eqref{eq:cseom} to derive $Q^{\mu\nu}$ from eq.\eqref{eq:csnoethercharge}. After a tedious calculation which is given in Appendix \ref{ap:noethercs} (steps following eq.\eqref{eq:Jmu}), we finally have
\begin{equation}\label{eq:defQ}
    \begin{split}
        \tensor{Q}{^\mu^\nu} = 2 \pdv{L}{\tensor{F}{_\mu_\nu}} \left(\tensor{A}{_\lambda} \tensor{\xi}{^\lambda} + \Lambda \right) + 2 \pdv{L}{\tensor{R}{^\alpha_\beta_\mu_\nu}} D{_\beta} \tensor{\xi}{^\alpha} + \tensor{\xi}{_\alpha} \left(\tensor{S}{^\mu^\nu^\alpha} - \tensor{S}{^\nu^\mu^\alpha} \right) + \tensor{\xi}{^\lambda} D{_\beta} \left[\pdv{L}{\tensor{R}{^\lambda_\mu_\nu_\beta}} - \pdv{L}{\tensor{R}{^\lambda_\nu_\mu_\beta}} \right] \, ,
    \end{split}
\end{equation}
where $S^{\mu\nu\alpha}$ is given by eq.\eqref{eq:defS}. 

\subsubsection{Gauge field dependence of the Chern-Simons Lagrangian }
\label{sec:gaugedepcslagrangian}

In order to establish the entropy current structure from eq.\eqref{eq:main}, we need to process further the CS Lagrangian eq.\eqref{eq:cslagrangian}. From eq.\eqref{eq:cseom}, eq.\eqref{eq:defxi} and eq.\eqref{eq:defQ}, we see that the structures generically depend on $A_{\mu}$ because of the explicit dependence of the Lagrangian eq.\eqref{eq:cslagrangian} on $A_{\mu}$. Consider an example of the $(4+1)$-dimensional mixed CS theory
\begin{equation}\label{eq:4+1mixGanomaly}
    L = \tensor{\epsilon}{^\mu^\nu^\rho^\sigma^\delta} \tensor{A}{_\mu} \tensor{R}{^\alpha_\beta_\nu_\rho} \tensor{R}{^\beta_\alpha_\sigma_\delta} \, .
\end{equation}
Differentiating the above Lagrangian with respect to $\tensor{R}{^\alpha_\beta_\nu_\rho}$ gives an expression that depends on $A_{\mu}$. This presents a problem because the gauge invariance of the entropy current would be murky. We should expect the entropy current to be $U(1)$ gauge invariant because the bulk EoM of CS theories is $U(1)$ gauge and diffeomorphism covariant. Another technical issue arises from eq.\eqref{eq:defxi}; we would have terms proportional to $\Lambda$. The boost weight analysis does not constrain $\Lambda$. The only constraint on expressions involving $\Lambda$ is given in eq.\eqref{eqlbmA1}. Thus, the analysis of mixed CS theories becomes tricky because it is difficult to argue that the $\Lambda$ term of eq.\eqref{eq:defxi} combines with $v A_v$ in such a way that it drops out (being of $\mathcal{O}(\epsilon^2)$) from eq.\eqref{eq:main}. To get around this, we will follow the approach of \cite{Copetti:2017cin}. Due to the Bianchi identities eq.\eqref{eq:bianchics}, mixed CS Lagrangians typically have the anomaly either in the $U(1)$ sector or the gravitational sector. One can push the anomaly from the $U(1)$ sector to the gravitational sector at the cost of a total derivative term \footnote{Let us try to elaborate on what we mean by putting the anomaly in the gauge or gravity sector. This terminology is actually borrowed from the holographic context, wherein an anomaly in the boundary field theory is equivalent to having a CS term in the dual bulk theory Lagrangian. In the boundary theory, the anomaly can be in the gauge sector or in the gravitational sector. In dual bulk language, which is what we are mainly concerned with in our work, it corresponds to having a CS term with the explicit appearance of the gauge field $A_\mu$ or the non-tensorial $\Gamma^{\mu}_{\alpha\beta}$. So, for us pushing the anomaly from one sector to the other actually means that we are either having a CS Lagrangian with an explicit $A_\mu$ or transferring to a Lagrangian that depends on $\Gamma^{\mu}_{\alpha\beta}$. This transformation is achieved by adding a total derivative piece to the Lagrangian; hence, the local dynamics remain unaltered.}, see for example in \cite{Copetti:2017cin}. If the Lagrangian eq.\eqref{eq:cslagrangian} has the anomaly in the gravitational sector, then it essentially becomes independent of $A_{\mu}$. This makes analyzing the entropy current structure easier if we can argue that the total derivative term arising from pushing the anomaly doesn't contribute to the entropy current. We will precisely do this by carefully using the Bianchi identities eq.\eqref{eq:bianchics} below. We will give the main steps and leave the details to Appendix \ref{ap:gaugedepcslagrangian}.

As $L = L \left(\tensor{g}{^\mu^\nu}, \tensor{\Gamma}{^\alpha_\beta_\mu}, \tensor{R}{^\alpha_\beta_\mu_\nu}, \tensor{A}{_\mu}, \tensor{F}{_\mu_\nu} \right)$ from eq.\eqref{eq:cslagrangian}, $L$ consists of sums of terms of the form
\begin{equation}\label{eq:lncs}
    \tensor{L}{_n} = \tensor{\tilde{\mathcal{L}}}{^{\nu_1}^\cdots^{\nu_n}} \prod_{i=1}^n \tensor{A}{_{\nu_i}} \, ,
\end{equation}
where $\tensor{\tilde{\mathcal{L}}}{^{\nu_1}^\cdots^{\nu_n}}$ is $U(1)$ gauge invariant and totally symmetric in all its indices (as $\prod_{i=1}^n \tensor{A}{_{\nu_i}}$ is totally symmetric). The Bianchi identity eq.\eqref{eq:bianchics} implies that 
\begin{equation}\label{eq:csgaugeconds1}
    D{_\mu} \pdv{\tensor{L}{_n}}{\tensor{A}{_\mu}} = 0 \implies n D{_\mu} \tensor{\tilde{\mathcal{L}}}{^\mu^{\nu_1}^\cdots^{\nu_{n-1}}} = 0 \quad \text{and} \quad n(n-1) \tensor{\tilde{\mathcal{L}}}{^\mu^\nu^{\nu_1}^\cdots^{\nu_{n-2}}} = 0 \, .
\end{equation}
The second condition of eq.\eqref{eq:csgaugeconds1} immediately implies that for $n \ge 2$, $\tensor{L}{_n} = 0$ in eq.\eqref{eq:lncs}. This necessarily means that a generic CS Lagrangian eq.\eqref{eq:cslagrangian} either has no factors of $A_{\mu}$ or it has a single factor of $A_{\mu}$ \footnote{\label{foot:nonabelian} This is consistent with the fact that if we are working with a $U(1)$ gauge field and treat it as a one form, one cannot construct a higher dimensional form by taking the wedge product of two $U(1)$ gauge field forms. This is why the analysis of this section is restricted to $U(1)$ gauge fields. Thus, we won't consider non-abelian gauge fields for which the Bianchi identity eq.\eqref{eq:bianchics} would be different.}, i.e.,
\begin{equation}
    L = \tilde{\mathcal{L}} + \tilde{\mathcal{L}}^{\mu} A_{\mu} \, .
\end{equation}
The first condition of eq.\eqref{eq:csgaugeconds1} implies that there are two different choices for $\tilde{\mathcal{L}}^{\mu}$:
\begin{equation}\label{eq:lmufinal1}
    \tilde{\mathcal{L}}^{\mu} = 
    \begin{cases}
        D{_\nu} \tensor{\mathcal{B}}{^\nu^\mu^{\rho_1}^{\sigma_1}^\cdots^{\rho_n}^{\sigma_n}} \prod_{i=1}^{n} F_{\rho_i \sigma_i} ~~~ \text{if} ~~ 2n+1 < D \\
        a_g \, \tensor{\epsilon}{^\mu^{\rho_1}^{\sigma_1}^\cdots^{\rho_n}^{\sigma_n}} \prod_{i=1}^n F_{\rho_i \sigma_i} ~~~ \text{if} ~~ 2n+1 = D
    \end{cases}
\end{equation}
Here $D$ is the dimension of the spacetime and $\tensor{\mathcal{B}}{^\mu^\nu^{\rho_1}^{\sigma_1}^\cdots^{\rho_n}^{\sigma_n}}$ is independent of the $U(1)$ gauge field and totally antisymmetric and $a_g$ is a constant.

We can combine eq.\eqref{eq:lncs} and eq.\eqref{eq:lmufinal1} to explicitly state that a generic Chern-Simons Lagrangian eq.\eqref{eq:cslagrangian} must be of the form \footnote{We have done an integration by parts manipulation on the  $\tensor{\mathcal{B}}{^\mu^\nu^{\rho_1}^{\sigma_1}^\cdots^{\rho_n}^{\sigma_n}}$ 
 term to express the final result by dumping terms into $\mathcal{L}$. See eq.\eqref{eq:lmuamu}.}
\begin{equation}\label{eq:finalcslagrangian}
    L = \mathcal{L} + D{_\mu} \left[\tensor{A}{_\nu} \sum_{n=0}^{N-1} \tensor{\mathcal{B}}{^\mu^\nu^{\rho_1}^{\sigma_1}^\cdots^{\rho_n}^{\sigma_n}} \left(\prod_{i=1}^n \tensor{F}{_{\rho_i}_{\sigma_i}} \right) \right] + a_g \, \tensor{\epsilon}{^\mu^{\rho_1}^{\sigma_1}^\cdots^{\rho_N}^{\sigma_N}} \tensor{A}{_\mu} \prod_{i=1}^N \tensor{F}{_{\rho_i}_{\sigma_i}} \, ,
\end{equation}
where $\mathcal{L}$ is $U(1)$ gauge invariant, and $a_g$ is a constant \footnote{This form of the Lagrangian is consistent with eq.(1.5) and eq.(1.6) of \cite{Bonora:2011gz} which gave a general form of the Lagrangian for mixed CS theories.}. 

As a demonstration of this construction, we consider the $(4+1)$-dimensional mixed CS theory of eq.\eqref{eq:4+1mixGanomaly}. We can push the anomaly from the $U(1)$ sector to the gravitational sector as follows:
\begin{equation}\label{eq:4+1example}
	L = \mathcal{L} + D_{\mu} (U^{\mu\nu}A_{\nu}) \, ,
\end{equation}
where 
\begin{equation}\label{eq:4+1mixGRanomaly}
	\mathcal{L} = 2 \tensor{\epsilon}{^\mu^\nu^\lambda^\rho^\sigma} \tensor{F}{_\mu_\nu} \tensor{\Gamma}{^\alpha_\lambda_\beta} \left(\frac{1}{2} \tensor{R}{^\beta_\alpha_\rho_\sigma} - \frac{1}{3} \tensor{\Gamma}{^\beta_\rho_\tau} \tensor{\Gamma}{^\tau_\alpha_\sigma} \right) \, ,
\end{equation}
is the $U(1)$ gauge invariant Lagrangian and $U^{\mu\nu}$ is given by
\begin{equation}\label{eq:Bterm}
	U^{\mu\nu} = -4 \tensor{\epsilon}{^\mu^\nu^\lambda^\rho^\sigma} \tensor{\Gamma}{^\alpha_\lambda_\beta} \left(\frac{1}{2} \tensor{R}{^\beta_\alpha_\rho_\sigma} - \frac{1}{3} \tensor{\Gamma}{^\beta_\rho_\tau} \tensor{\Gamma}{^\tau_\alpha_\sigma} \right) \, .
\end{equation}
One can see that eq.\eqref{eq:4+1example} along with eq.\eqref{eq:4+1mixGRanomaly} and eq.\eqref{eq:Bterm} is of the form eq.\eqref{eq:finalcslagrangian}. It should be noted that in $(4+1)$-dimensions, we have a pure gauge CS Lagrangian
\begin{equation}
    L = \epsilon^{\mu\nu\lambda\rho\sigma} A_{\mu} F_{\nu\lambda} F_{\rho\sigma} \, ,
\end{equation}
which is of the form of the last term in eq.\eqref{eq:finalcslagrangian}.

This final result for the Lagrangian is a direct consequence of the CS theory Bianchi identities eq.\eqref{eq:bianchics} \footnote{We note that we only needed to use the Bianchi identity of the gauge sector in eq.\eqref{eq:bianchics}. It turns out this is enough for us to argue for the entropy current structure of eq.\eqref{eom1}. Imposing the gravitational sector Bianchi identity of eq.\eqref{eq:bianchics} would impose further restrictions on $\mathcal{L}$ and $\mathcal{B}^{\mu\nu\rho_1 \sigma_1 \dots \rho_n \sigma_n}$. This also points to the fact that there are generally ``hidden" indices in $\mathcal{B}$ corresponding to the gravitational terms. For example in eq.\eqref{eq:Bterm}, the $\lambda,\rho,\sigma$ correspond to these hidden indices.}. This form of the Lagrangian is considerably easier to work with because the $A_{\mu}$ dependence has been absorbed into a total derivative term and a pure gauge term. Once we carefully deal with these terms, the analysis of $\mathcal{L}$ mostly follows that of \cite{Bhattacharyya:2021jhr, Biswas:2022grc} since it is $U(1)$ gauge invariant. The equivalent structures of eq.\eqref{eq:cseom} and eq.\eqref{eq:defQ} for the total derivative term in eq.\eqref{eq:finalcslagrangian} have been worked out in Appendix \ref{sec:totalterm}. We will evaluate those structures using our gauge eq.\eqref{nhmetric} on the horizon $r=0$ in \S\ref{sec:totaltermhorizon}. We now proceed to construct a proof of the entropy current from eq.\eqref{eq:finalcslagrangian} using eq.\eqref{eq:main}.

\subsection{Proof of the existence of the entropy current for Chern-Simons theories}

Now that we have carefully analyzed the structure of the Lagrangian in eq.\eqref{eq:cslagrangian}, we can construct the proof of the entropy current structure for CS theories. The final form of the Lagrangian is given by eq.\eqref{eq:finalcslagrangian}, which we rewrite here for the convenience of the reader:
\begin{equation}\label{eq:cslagform1}
    L = \underbrace{\mathcal{L}}_{\text{Gauge invariant}} + \underbrace{D{_\mu} \left(\tensor{U}{^\mu^\nu} \tensor{A}{_\nu} \right)}_{\text{total derivative}} + \underbrace{a_g\, \tensor{\epsilon}{^\mu^{\rho_1}^{\sigma_1}^\cdots^{\rho_N}^{\sigma_N}} \tensor{A}{_\mu} \prod_{i=1}^N \tensor{F}{_{\rho_i}_{\sigma_i}} }_{\text{pure gauge}}\, ,
\end{equation}
where $\tensor{U}{^\mu^\nu}$ is $U(1)$ gauge invariant and antisymmetric in $(\mu, \nu)$, $\mathcal{L}$ is $U(1)$ gauge invariant, and $a_g$ is a constant. The proof that eq.\eqref{eq:main} given by
\begin{equation}\label{eq:main1}
    2 v \, E_{vv} + G_v (v A_v + \Lambda) = (-\Theta^r + \Xi^r + D_{\rho} Q^{r\rho})|_{r=0} \, ,
\end{equation}
has the structure of eq.\eqref{eom1} follows after we analyze the three terms in eq.\eqref{eq:cslagform1} separately. We will show below that the total derivative and pure $U(1)$ gauge terms do not contribute to the entropy current of eq.\eqref{eom1}. Thus, the entropy current solely receives contributions from $\mathcal{L}$, the CS Lagrangian, where the anomaly is in the gravitational sector. The non-trivial part of the proof is the structure of $\Xi^{\mu}$ that is non-zero for CS theories through eq.\eqref{eq:defxi}. This is the new element compared to the proofs constructed in \cite{Bhattacharyya:2021jhr, Biswas:2022grc}. We will show that this additional term will not spoil the entropy current structure of eq.\eqref{eom1}.

\subsubsection{Analysis of the total derivative term in eq.(\ref{eq:cslagform1})}
\label{sec:totaltermhorizon}

The total derivative term in eq.\eqref{eq:cslagform1} has the form $\mathcal{L}^{\mu} = U^{\mu\nu} A_{\nu}$ where $U^{\mu\nu}$ is $U(1)$ gauge invariant and antisymmetric. This is a special case of the general analysis in \S\ref{sec:totalterm}. We evaluate the various structures in eq.\eqref{eq:main1} in our gauge eq.\eqref{nhmetric} on the horizon $r=0$. The Noether charge is given by eq.\eqref{eq:totalQ}:
\begin{equation}\label{eq:totalQfinal}
    \begin{split}
        Q^{r\rho}_t|_{r=0} &= \tensor{U}{^r^\lambda} \tensor{A}{_\lambda} \tensor{\xi}{^\rho} = v \tensor{U}{^r^\lambda} \tensor{A}{_\lambda} \tensor{\delta}{^\rho_v} \\
    \implies \frac{1}{\sqrt{h}} \tensor{\partial}{_\rho} \left(\sqrt{h} Q^{r\rho}_t \right)|_{r=0} &= \frac{1}{\sqrt{h}} \left(1 + v \tensor{\partial}{_v} \right) \left(\sqrt{h} \tensor{U}{^r^\rho} \tensor{A}{_\rho} \right) \\ &= \left(1 + v \tensor{\partial}{_v} \right) \left(\tensor{U}{^r^\rho} \tensor{A}{_\rho} \right) + \frac{1}{2} v \tensor{A}{_v} \tensor{U}{^r^v} \tensor{h}{^m^n} \tensor{\partial}{_v} \tensor{h}{_m_n} + \mathcal{O}(\epsilon^2) \, .
    \end{split}
\end{equation}
The $\Theta^{\mu}$ and $\Xi^{\mu}$ are given by eq.\eqref{eq:totaltheta} and eq.\eqref{eq:totalxi} respectively. Thus, we have
\begin{equation}\label{eq:totalfinal}
    \begin{split}
        \Xi^r_t|_{r=0} &= \tensor{U}{^r^\rho} D{_\rho} \Lambda \, ,
    \\
    \Theta^r_t |_{r=0} &= A_v \mathcal{L}_{\xi} U^{rv} + A_i \mathcal{L}_{\xi} U^{ri} + U^{rv} \mathcal{L}_{\xi} A_v + U^{ri} \mathcal{L}_{\xi} A_i + \dfrac{1}{2}U^{rv} v A_v h^{mn} \partial_v h_{mn} + \mathcal{O}(\epsilon^2) \\
    &= (1+ v \partial_v)(U^{rv}A_v + U^{ri}A_i) + U^{r\rho} \partial_{\rho}\Lambda +  \dfrac{1}{2}U^{rv} v A_v h^{mn} \partial_v h_{mn} \, .
    \end{split}
\end{equation}
One can use eq.\eqref{eq:totalQfinal} and eq.\eqref{eq:totalfinal} to show that \footnote{This equation is exact to all orders in $\epsilon$ since we use the LHS of eq.\eqref{eq:twowaysthetatotal}. If we use the equivalent RHS of eq.\eqref{eq:twowaysthetatotal}, we would get a result up to $\mathcal{O}(\epsilon^2)$ only.}
\begin{equation}\label{eq:totalfinal2}
    \begin{split}
        -\Theta^r_t |_{r=0} + \Xi^r_t |_{r=0} + \frac{1}{\sqrt{h}} \tensor{\partial}{_\rho} \left(\sqrt{h} Q^{r\rho}_t \right) |_{r=0} = 0 \, .
    \end{split}
\end{equation}
Thus we see that the total derivative term in eq.\eqref{eq:cslagform1} doesn't contribute to the entropy current structure and drops out entirely from $E_{vv}$ of eq.\eqref{eq:main1}. Notice how in eq.\eqref{eq:totalfinal} and eq.\eqref{eq:totalQfinal}, there are terms explicitly proportional to $A_v$ and $A_i$. This is due to the $\delta A_{\nu}$ term in eq.\eqref{eq:totaltheta} \footnote{The $\mathcal{G}^{\mu\nu}$ of eq.\eqref{eq:totaltheta} is not anti-symmetric and thus if we use eq.\eqref{eq:diffeo}, the $\delta A_{\nu}$ term in $\Theta^{\mu}$ doesn't cancel out the corresponding contribution in $Q^{\mu\nu}$. This leads to a split of $vA_v$ and $\Lambda$ terms in eq.\eqref{eq:main1}. }. Even if the $v A_v$ terms and $\Lambda$ terms appear independent, they cancel out in the combination of eq.\eqref{eq:main1}. This points to the advantage of the structure of the Lagrangian in eq.\eqref{eq:cslagform1}. If we had worked with the original Lagrangian eq.\eqref{eq:cslagrangian}, then we would have had to track these appearances of $A_v$ and $A_i$ carefully. Arguing the entropy current structure of eq.\eqref{eom1} would have been a formidable task.

\subsubsection{Analysis of the pure gauge term in eq.(\ref{eq:cslagform1})}
\label{sec:puregaugetermhorizon}

We now consider the pure gauge term of eq.\eqref{eq:cslagform1} given by
\begin{equation}\label{eq:lgterm}
    L_g = a_g \, \tensor{\epsilon}{^\mu^{\rho_1}^{\sigma_1}^\cdots^{\rho_N}^{\sigma_N}} \tensor{A}{_\mu} \prod_{i=1}^N \tensor{F}{_{\rho_i}_{\sigma_i}} \, .
\end{equation}
For this $L_g$, we can use the general structures derived in eq.\eqref{eq:cseom}, eq.\eqref{eq:defxi} and eq.\eqref{eq:defQ} to evaluate the contribution to eq.\eqref{eq:main1}. We thus have (see eq.\eqref{eq:lgevvgenap})
\begin{equation}\label{eq:lgevvgen}
    \begin{split}
        \eval{\left[D{_\rho} Q^{r\rho}_g - \Theta^{r}_g + \Xi^{r}_g \right]}_{r=0} - G^{r}_g \left(\tensor{A}{_\rho} \tensor{\xi}{^\rho} + \Lambda \right) = - v \left(\pdv{L_g}{\tensor{A}{_r}} \tensor{A}{_v} + 2 \pdv{L_g}{\tensor{F}{_r_i}} \tensor{F}{_v_i} \right) \, .
    \end{split}
\end{equation}
Here the subscript $g$ denotes the contribution of $L_g$ in eq.\eqref{eq:lgterm} to eq.\eqref{eq:cseom}, eq.\eqref{eq:defxi} and eq.\eqref{eq:defQ}. Thus, we have (see eq.\eqref{eq:lgevvfinalap})
\begin{equation}\label{eq:lgevvfinal}
    \begin{split}
         \eval{\left[D{_\rho} Q^{r\rho}_g - \Theta^{r}_g + \Xi^{r}_g \right]}_{r=0} - G^{r}_g \left(\tensor{A}{_\rho} \tensor{\xi}{^\rho} + \Lambda \right) = 0 \, .
    \end{split}
\end{equation}
Thus $L_g$ of eq.\eqref{eq:lgterm} in eq.\eqref{eq:cslagform1} doesn't contribute to the entropy current at linear order. This is in line with the gauge invariant analysis of \cite{Biswas:2022grc}, where the pure gauge terms in the Lagrangian didn't contribute to the final entropy current. 

\subsubsection{Analysis of the structure of $\Xi^{\mu}$ coming from $\mathcal{L}$ in eq.(\ref{eq:cslagform1})}
\label{sec:xistruchorizon}

Since we have analyzed the total derivative term and the pure gauge term of eq.\eqref{eq:cslagform1} and showed that they are inconsequential, we can finally analyze the gauge invariant term $\mathcal{L}$ that truly contributes to the entropy current. Before making the final analysis of the entropy current contribution of the $U(1)$ gauge invariant $\mathcal{L}$, we analyze the non-trivial structure of $\Xi^{\mu}$ of eq.\eqref{eq:defxi}. To handle these terms, we must return to the derivation of eq.\eqref{eq:defxi}. During the derivation in eq.\eqref{eq:varlxider}, we obtained the relation (here $\frac{\partial \mathcal{L}}{\partial A_{\mu}} = 0$ because it is $U(1)$ gauge invariant)
\begin{equation}
    \delta \mathcal{L} = \mathcal{L}_{\xi} \mathcal{L} + \pdv{\mathcal{L}}{\tensor{\Gamma}{^\lambda_\mu_\nu}} \partial{_\mu} \partial_{\nu} \tensor{\xi}{^\lambda}
\end{equation}
which was rearranged to get the value of $\Xi^{\mu}_{\mathcal{L}}$ (see eq.\eqref{eq:partialLG})
\begin{equation}\label{eq:dxiinter}
    \pdv{\mathcal{L}}{\tensor{\Gamma}{^\lambda_\mu_\nu}} \partial{_\mu} \partial_{\nu} \tensor{\xi}{^\lambda} = D{_\mu} \Xi^{\mu}_{\mathcal{L}} + \frac{\tensor{\xi}{^\lambda}}{\sqrt{-g}} \partial{_\mu} \partial_{\nu} \left(\sqrt{-g} \pdv{\mathcal{L}}{\tensor{\Gamma}{^\lambda_\mu_\nu}} \right) \, .
\end{equation}
Here the subscript $\mathcal{L}$ denotes that the $\Xi^{\mu}$ only receives contributions from $\mathcal{L}$ of eq.\eqref{eq:cslagform1}. We now use the Bianchi identity of eq.\eqref{eq:bianchics} to set the last term in RHS of eq.\eqref{eq:dxiinter} to $0$. So, we are left with
\begin{equation}\label{eq:xidefbianchi}
    D{_\mu} \Xi^{\mu}_{\mathcal{L}} = \pdv{\mathcal{L}}{\tensor{\Gamma}{^\lambda_\mu_\nu}} \partial{_\mu} \partial_{\nu} \tensor{\xi}{^\lambda} \, .
\end{equation}

Now comes the crucial point. When we go to the horizon $r=0$ in our gauge eq.\eqref{nhmetric}, the RHS of eq.\eqref{eq:xidefbianchi} is zero because the Killing vector $\xi$ of eq.\eqref{KillVec} is linear in the coordinates $v$ and $r$. Thus eq.\eqref{eq:xidefbianchi} in our gauge becomes
\begin{equation*}
    D{_\mu} \Xi^{\mu}_{\mathcal{L}} \Bigg|_{r=0} = 0 \, .
\end{equation*}
Hence, in our gauge eq.\eqref{nhmetric}, we must have for some antisymmetric $2$-symbol $q^{\mu\nu}$ such that
\begin{equation}\label{eq:xitoq}
    \Xi^{\mu}_{\mathcal{L}} = D{_\nu} \tensor{q}{^\mu^\nu} \, .
\end{equation}
This looks similar to the Noether charge eq.\eqref{eq:csnoethercharge}. But, to repeat the conclusions of  \S \ref{sec:reviewcurrent}, in particular eq.\eqref{eq:Qrnustrucdiffgen} and eq.\eqref{eq:divQdiffgen}, we need to establish two more things for $\tensor{q}{^\mu^\nu}$ - linearity in $v$ and gauge invariance. We can show that $q^{\mu\nu}$ is $U(1)$ gauge invariant up to a total derivative. The details of the proof are given in Appendix \ref{ap:xistruc}. This allows us to express the $q^{r\rho}$ on the horizon as
\begin{equation}\label{eq:xiqdefmain}
\tensor{q}{^r^\rho}| = \tensor{\tilde{q}}{^r^\rho} + v \,\tensor{w}{_v^r^\rho} \, ,
\end{equation}
where $\tilde{q}, w$ have no explicit factors of $v$. We also treat them to be $U(1)$ gauge invariant because the inconsequential total derivative term of eq.\eqref{eq:qmunufinal} drops out in eq.\eqref{eq:xitoq}.

\subsubsection*{Final result on the horizon:}

We can now combine eq.\eqref{eq:xitoq} and eq.\eqref{eq:xiqdefmain} to express the result of $\Xi^r_{\mathcal{L}}$ on the horizon $r=0$. The final result is thus given by
\begin{equation}\label{eq:xirfinal}
    \tensor{\Xi}{^r}_{\mathcal{L}} = \frac{1}{\sqrt{h}} \tensor{\partial}{_\rho} \left(\sqrt{h} \tensor{\tilde{q}}{^r^\rho} \right) + \left(1 + v \tensor{\partial}{_v} \right) \tensor{w}{_v^r^v} + v \tensor{\partial}{_v} \left(\frac{1}{\sqrt{h}} \tensor{\partial}{_i} \left(\sqrt{h} \tensor{j}{^i_{(1)}} \right) \right) + \mathcal{O}\left(\epsilon^2 \right) \, .
\end{equation}
Here $\Tilde{q}^{r\rho}$, $\tensor{j}{^i_{(1)}}$ and $\tensor{w}{_v^r^v}$ are $U(1)$ gauge invariant. This is entirely analogous to eq.\eqref{eq:divQdiffgen} (also eq.(3.66) of \cite{Bhattacharyya:2021jhr}).

\subsubsection{Entropy current from the gauge invariant term $\mathcal{L}$ in eq.(\ref{eq:cslagform1})}
\label{sec:finalentropycurrent}

As we have handled two of the terms in eq.\eqref{eq:cslagform1} (the total derivative term in \S \ref{sec:totaltermhorizon} and the pure gauge term in \S \ref{sec:puregaugetermhorizon}), we can finally focus on the contribution of $\mathcal{L}$. This term is $U(1)$ gauge invariant by construction (eq.\eqref{eq:finalcslagrangian}) and has a structure of $\Xi^r$ on $r=0$ given by eq.\eqref{eq:xirfinal}. We now consider the contribution of $\Theta^r$, $Q^{r\rho}$ and $G_v$ in eq.\eqref{eq:main1}. We first recall, the general structures for $G^{\mu}$, $\Theta^{\mu}$ and $Q^{\mu\nu}$ from eq.\eqref{eq:cseom}, eq.\eqref{eq:defS} and eq.\eqref{eq:defQ} respectively evaluated for $\mathcal{L}$: 
\begin{equation}\label{eq:gmuL}
     G^{\mu}_{\mathcal{L}} = \pdv{\mathcal{L}}{\tensor{A}{_\mu}} + 2 D{_\nu} \pdv{\mathcal{L}}{\tensor{F}{_\mu_\nu}} \, ,
\end{equation}
\begin{equation}\label{eq:thetamuL}
    \Theta^{\mu}_{\mathcal{L}} = 2 \pdv{\mathcal{L}}{\tensor{F}{_\mu_\nu}} \delta \tensor{A}{_\nu} + 2 \pdv{\mathcal{L}}{\tensor{R}{^\alpha_\beta_\mu_\nu}} \delta \tensor{\Gamma}{^\alpha_\beta_\nu} + S^{\mu\alpha\beta}_{\mathcal{L}} \delta \tensor{g}{_\alpha_\beta} \, ,
\end{equation}
\begin{equation}\label{eq:QmunuL}
    \begin{split}
        Q^{\mu\nu}_{\mathcal{L}} &= 2 \pdv{\mathcal{L}}{\tensor{F}{_\mu_\nu}} \left(\tensor{A}{_\lambda} \tensor{\xi}{^\lambda} + \Lambda \right) + 2 \pdv{\mathcal{L}}{\tensor{R}{^\alpha_\beta_\mu_\nu}} D{_\beta} \tensor{\xi}{^\alpha} + \tensor{\xi}{_\alpha} \left(S^{\mu\nu\alpha}_{\mathcal{L}} - S^{\nu\mu\alpha}_{\mathcal{L}} \right) + \tensor{\xi}{^\lambda} D{_\beta} \left[\pdv{\mathcal{L}}{\tensor{R}{^\lambda_\mu_\nu_\beta}} - \pdv{\mathcal{L}}{\tensor{R}{^\lambda_\nu_\mu_\beta}} \right] \, ,
    \end{split}
\end{equation}
where
\begin{equation}
    \begin{split}
        S^{\mu\alpha\beta}_{\mathcal{L}} = &\frac{1}{2} D{_\nu} \left[\tensor{g}{^\alpha^\lambda} \left(\pdv{\mathcal{L}}{\tensor{R}{^\lambda_\beta_\mu_\nu}} + \pdv{\mathcal{L}}{\tensor{R}{^\lambda_\mu_\beta_\nu}} \right) + \tensor{g}{^\beta^\lambda} \left(\pdv{\mathcal{L}}{\tensor{R}{^\lambda_\alpha_\mu_\nu}} + \pdv{\mathcal{L}}{\tensor{R}{^\lambda_\mu_\alpha_\nu}} \right) \right. \\
    &\quad \left. - \tensor{g}{^\mu^\lambda} \left(\pdv{\mathcal{L}}{\tensor{R}{^\lambda_\beta_\alpha_\nu}} + \pdv{\mathcal{L}}{\tensor{R}{^\lambda_\alpha_\beta_\nu}} \right) \right] + \frac{1}{2} \left(\tensor{g}{^\alpha^\lambda} \pdv{\mathcal{L}}{\tensor{\Gamma}{^\lambda_\beta_\mu}} + \tensor{g}{^\beta^\lambda} \pdv{\mathcal{L}}{\tensor{\Gamma}{^\lambda_\alpha_\mu}} - \tensor{g}{^\mu^\lambda} \pdv{\mathcal{L}}{\tensor{\Gamma}{^\lambda_\beta_\alpha}} \right) \, .
    \end{split}
\end{equation}

For convenience, we break apart the $\Theta^{\mu}_{\mathcal{L}}$ and $Q^{\mu\nu}_{\mathcal{L}}$ of eq.\eqref{eq:thetamuL} and eq.\eqref{eq:QmunuL} respectively into their ``gauge" parts and ``gravity" parts. 
\begin{equation}\label{eq:thetamuLbr}
    \Theta^{\mu}_{\mathcal{L}} = \Theta^{\mu}_{\mathcal{L}} \Bigg|_{\text{gauge}} + \Theta^{\mu}_{\mathcal{L}} \Bigg|_{\text{gravity}} \, , 
\end{equation}
where
\begin{equation}
    \begin{split}
        \Theta^{\mu}_{\mathcal{L}} \Bigg|_{\text{gauge}} = 2 \pdv{\mathcal{L}}{\tensor{F}{_\mu_\nu}} \delta \tensor{A}{_\nu} \, , ~~~~
        \Theta^{\mu}_{\mathcal{L}} \Bigg|_{\text{gravity}} &= 2 \pdv{\mathcal{L}}{\tensor{R}{^\alpha_\beta_\mu_\nu}} \delta \tensor{\Gamma}{^\alpha_\beta_\nu} + S^{\mu\alpha\beta}_{\mathcal{L}} \delta \tensor{g}{_\alpha_\beta} \, .
    \end{split}
\end{equation}
\begin{equation}\label{eq:QmunuLbr}
    Q^{\mu\nu}_{\mathcal{L}} = Q^{\mu\nu}_{\mathcal{L}} \Bigg|_{\text{gauge}} + Q^{\mu\nu}_{\mathcal{L}} \Bigg|_{\text{gravity}} \, ,
\end{equation}
where
\begin{equation}
    \begin{split}
        Q^{\mu\nu}_{\mathcal{L}} \Bigg|_{\text{gauge}} &= 2 \pdv{\mathcal{L}}{\tensor{F}{_\mu_\nu}} \left(\tensor{A}{_\lambda} \tensor{\xi}{^\lambda} + \Lambda \right) \, , \\
        Q^{\mu\nu}_{\mathcal{L}} \Bigg|_{\text{gravity}} &= 2 \pdv{\mathcal{L}}{\tensor{R}{^\alpha_\beta_\mu_\nu}} D{_\beta} \tensor{\xi}{^\alpha} + \tensor{\xi}{_\alpha} \left(S^{\mu\nu\alpha}_{\mathcal{L}} - S^{\nu\mu\alpha}_{\mathcal{L}} \right) + \tensor{\xi}{^\lambda} D{_\beta} \left[\pdv{\mathcal{L}}{\tensor{R}{^\lambda_\mu_\nu_\beta}} - \pdv{\mathcal{L}}{\tensor{R}{^\lambda_\nu_\mu_\beta}} \right] \, .
    \end{split}
\end{equation}
It should be noted that the break up of terms in eq.\eqref{eq:thetamuLbr} and eq.\eqref{eq:QmunuLbr} is done to treat the terms separately, i.e., the terms proportional to $\delta A_{\nu}$ and $\delta g_{\mu\nu}$. It does not mean that the gauge field and metric contributions factor out. $\mathcal{L}$ in eq.\eqref{eq:cslagform1} depends on both $F_{\mu\nu}$ and $g_{\mu\nu}$ in a mixed way, so the ``gauge" and "gravity" parts of eq.\eqref{eq:thetamuL} and eq.\eqref{eq:QmunuL} in general have mixing between the two fields. 

\subsubsection*{Contribution of the ``gauge" terms on the horizon:}

The ``gauge" terms of eq.\eqref{eq:thetamuLbr} and eq.\eqref{eq:QmunuLbr} along with eq.\eqref{eq:gmuL} make the following contribution to eq.\eqref{eq:main} on the horizon $r=0$ (see eq.\eqref{eq:gaugeLcontap})
\begin{equation}\label{eq:gaugeLcont}
    \begin{split}
        \eval{\left[D{_\rho} Q^{r\rho}_{\mathcal{L}} - \Theta^r_{\mathcal{L}} \right]}_{\text{gauge}} - G^r_{\mathcal{L}} \left(\tensor{A}{_\rho} \tensor{\xi}{^\rho} + \Lambda \right)  = \mathcal{O}(\epsilon^2) \, .
    \end{split}
\end{equation}
Thus, the ``gauge" terms of eq.\eqref{eq:thetamuLbr} and eq.\eqref{eq:QmunuLbr} do not contribute to the entropy current in eq.\eqref{eq:main1}. 

\subsubsection*{Noether charge of ``gravity" terms on the horizon:}

From eq.\eqref{eq:QmunuLbr}, the expression of $Q^{r\rho}$ on the horizon $r=0$ evaluates to
\begin{equation}
    \begin{split}
        Q^{r\rho}_{\mathcal{L}} \Bigg|_{\text{gravity}} &= 2 \pdv{\mathcal{L}}{\tensor{R}{^\alpha_\beta_r_\rho}} D{_\beta} \tensor{\xi}{^\alpha} + v \left( S^{r\rho r}_{\mathcal{L}} - S^{\rho rr}_{\mathcal{L}} \right) + v D{_\beta} \left[\pdv{\mathcal{L}}{\tensor{R}{^v_r_\rho_\beta}} - \pdv{\mathcal{L}}{\tensor{R}{^v_\rho_r_\beta}} \right]
    \\
    &= 2 \pdv{\mathcal{L}}{\tensor{R}{^v_v_r_\rho}} - 2 \pdv{\mathcal{L}}{\tensor{R}{^r_r_r_\rho}} + v \pdv{\mathcal{L}}{\tensor{R}{^m_r_r_\rho}} \tensor{\omega}{^m} - v \pdv{\mathcal{L}}{\tensor{R}{^v_m_r_\rho}} \tensor{\omega}{_m} + v \pdv{\mathcal{L}}{\tensor{R}{^n_m_r_\rho}} \tensor{h}{^n^l} \tensor{\partial}{_v} \tensor{h}{_m_l} \\
    &\quad + v \left(\tensor{g}{^\rho^\lambda} \pdv{\mathcal{L}}{\tensor{\Gamma}{^\lambda_r_r}} - \pdv{\mathcal{L}}{\tensor{\Gamma}{^v_r_\rho}} \right) + 2 v D{_\beta} \left[\tensor{g}{^\rho^\lambda} \pdv{\mathcal{L}}{\tensor{R}{^\lambda_r_r_\beta}} - \pdv{\mathcal{L}}{\tensor{R}{^v_\rho_r_\beta}} \right] \, .
    \end{split}
\end{equation}
Thus, we have on the horizon $r=0$
\begin{equation}\label{eq:Qmunustruc}
    Q^{r\rho}_{\mathcal{L}} \Bigg|_{\text{gravity}} = \tensor{\widetilde{Q}}{^r^\rho} + v \tensor{W}{_v^r^\rho} \, ,
\end{equation}
where $\widetilde{Q}, W$ have no explicit factors of $v$. This structure of $Q^{r\rho}$ is identical to the structure of $Q^{\mu\nu}$ for diffeomorphism invariant theories of gravity \cite{Bhattacharyya:2021jhr,Biswas:2022grc} derived in eq.\eqref{eq:Qrnustrucdiffgen} (see eq.(3.60) of \cite{Bhattacharyya:2021jhr}). As $\tensor{W}{_v^r^i}$ is gauge invariant and boost weight $2$, we have
\begin{equation}
    \tensor{W}{_v^r^i} = \tensor{\partial}{_v} \tensor{J}{^i_{(1)}} + \mathcal{O}(\epsilon^2) \, .
\end{equation}
This implies that
\begin{equation}
    \begin{split}
         \frac{1}{\sqrt{h}} \tensor{\partial}{_v} \left(\sqrt{h} v \tensor{W}{_v^r^v} \right) &= \left(1 + v \tensor{\partial}{_v} \right) \tensor{W}{_v^r^v} \, , \\
    \frac{1}{\sqrt{h}} \tensor{\partial}{_i} \left(\sqrt{h} v \tensor{W}{_v^r^i} \right) &= v \tensor{\partial}{_v} \left(\frac{1}{\sqrt{h}} \tensor{\partial}{_i} \left(\sqrt{h} \tensor{J}{^i_{(1)}} \right) \right) + \mathcal{O}\left(\epsilon^2 \right) \, .
    \end{split}
\end{equation}
Using these expressions, we can straightforwardly evaluate the divergence of the Noether charge on the horizon $r=0$ \cite{Bhattacharyya:2021jhr, Biswas:2022grc} as
\begin{equation}\label{eq:QmunuLcurrent}
    \begin{split}
        D_{\rho} Q^{r\rho}_{\mathcal{L}}\Bigg|_{\text{gravity}} = \eval{\frac{1}{\sqrt{h}} \tensor{\partial}{_\rho} \left(\sqrt{h} Q^{r\rho}_{\mathcal{L}}\right)}_{\text{gravity}} &= \frac{1}{\sqrt{h}} \tensor{\partial}{_\rho} \left(\sqrt{h} \tensor{\widetilde{Q}}{^r^\rho} \right) + \left(1 + v \tensor{\partial}{_v} \right) \tensor{W}{_v^r^v} \\ & \quad + v \tensor{\partial}{_v} \left(\frac{1}{\sqrt{h}} \tensor{\partial}{_i} \left(\sqrt{h} \tensor{J}{^i_{(1)}} \right) \right) + \mathcal{O}\left(\epsilon^2 \right) \, ,
    \end{split}
\end{equation}
where $\widetilde{Q}^{r\rho}$, $\tensor{W}{_v^r^v}$ and $\tensor{J}{^i_{(1)}}$ are all $U(1)$ gauge invariant. This is identical to the diffeomorphism invariant case in eq.\eqref{eq:divQdiffgen} (see eq.(3.66) of \cite{Bhattacharyya:2021jhr}).

\subsubsection*{$\Theta^r$ of ``gravity" terms on the horizon:}

We now analyze the structure of $\Theta^r$ in eq.\eqref{eq:thetamuLbr} from the contributions of the ``gravity" terms. We first define
\begin{equation}\label{eq:ELcs}
    \begin{split}
        E^{r\alpha\beta \nu}_{\mathcal{L}} &= \frac{1}{2} \left(\tensor{g}{^\alpha^\lambda} \pdv{\mathcal{L}}{\tensor{R}{^\lambda_\beta_r_\nu}} + \tensor{g}{^\alpha^\lambda} \pdv{\mathcal{L}}{\tensor{R}{^\lambda_\nu_r_\beta}} + \tensor{g}{^\beta^\lambda} \pdv{\mathcal{L}}{\tensor{R}{^\lambda_\alpha_r_\nu}}   + \tensor{g}{^\beta^\lambda} \pdv{\mathcal{L}}{\tensor{R}{^\lambda_\nu_r_\alpha}} - \tensor{g}{^\nu^\lambda} \pdv{\mathcal{L}}{\tensor{R}{^\lambda_\beta_r_\alpha}} - \tensor{g}{^\nu^\lambda} \pdv{\mathcal{L}}{\tensor{R}{^\lambda_\alpha_r_\beta}} \right) 
    \end{split}
\end{equation}
This is the CS theory equivalent of eq.\eqref{eq:ERdiff}. Using the definition of eq.\eqref{eq:ELcs} and the steps following eq.\eqref{eq:thetarLinter} to eq.\eqref{eq:thetarfinalap}, $\Theta^r$ from the ``gravity" contributions of eq.\eqref{eq:thetamuLbr} take the following form on the horizon:
\begin{equation}\label{eq:thetarfinal}
    \begin{split}
        \eval{\Theta^r_{\mathcal{L}}}_{\text{gravity}} &= - \left(1 + v \tensor{\partial}{_v} \right) \left(E^{rmnr}_{\mathcal{L}} \tensor{\partial}{_r} \tensor{h}{_m_n} \right) + v \tensor{\partial}{^2_v} \left(\tensor{J}{_{(1)}^m^n} \tensor{\partial}{_r} \tensor{h}{_m_n} \right) + \mathcal{O}\left(\epsilon^2 \right) \\
         &= \left(1 + v \tensor{\partial}{_v} \right) \tensor{\mathcal{A}}{_{(1)}} + v \tensor{\partial}{^2_v} \tensor{\mathcal{B}}{_{(0)}} + \mathcal{O} \left(\epsilon^2 \right) \, .
    \end{split}
\end{equation}
Here $\tensor{\mathcal{A}}{_{(1)}}$ and $\tensor{\mathcal{B}}{_{(0)}}$ are $U(1)$ gauge invariant and $\tensor{\mathcal{B}}{_{(0)}}$ is $\mathcal{O}(\epsilon)$ because it is of the form eq.\eqref{eq:B0diffgen}.

This structure of $\Theta^r$ is identical to the structure of $\Theta^r$ of \cite{Bhattacharyya:2021jhr, Biswas:2022grc} derived in eq.\eqref{eq:thetardiffgen} (see eq.(3.56) of \cite{Bhattacharyya:2021jhr}). Thus, the $\mathcal{B}_{(0)}$ denotes the JKM ambiguity, which is analogous to the JKM ambiguities appearing in \cite{Bhattacharyya:2021jhr, Biswas:2022grc}, i.e., eq.\eqref{eq:B0diffgen}. It is clear from eq.\eqref{eq:dPexpr} that the $\tensor{\mathcal{B}}{_{(0)}}$ term is zero only if $E^{rmnr}_{\mathcal{L}} = \mathcal{O}\left(\epsilon^2 \right)$ where
\begin{equation}\label{eq:B0termcont}
    E^{rmnr}_{\mathcal{L}} = \frac{1}{2} \left(\tensor{h}{^m^l} \pdv{\mathcal{L}}{\tensor{R}{^l_r_r_n}} + \tensor{h}{^n^l} \pdv{\mathcal{L}}{\tensor{R}{^l_r_r_m}} - \pdv{\mathcal{L}}{\tensor{R}{^v_n_r_m}} - \pdv{\mathcal{L}}{\tensor{R}{^v_m_r_n}} \right) \, .
\end{equation}
For the specific examples we study in \S\ref{sec:proofverf}, the JKM ambiguity $\mathcal{B}_{(0)}$ is zero.

\subsubsection*{Final structure of the entropy current:}

We have shown that the structures of $\Theta^r_{\mathcal{L}}$ in eq.\eqref{eq:thetarfinal} and $Q^{r\rho}_{\mathcal{L}}$ in eq.\eqref{eq:Qmunustruc} retain the same form as that of \cite{Bhattacharyya:2021jhr,Biswas:2022grc}, i.e., eq.\eqref{eq:thetardiffgen} and eq.\eqref{eq:Qrnustrucdiffgen} respectively. The new non-trivial element was the structure of $\Xi^r$, which is non-zero for CS theories, and it has been analyzed in eq.\eqref{eq:xirfinal}. This $\Xi^r$ is of the form of eq.\eqref{eq:divQdiffgen}. The analysis now basically follows that of \cite{Bhattacharyya:2021jhr,Biswas:2022grc} (\S\ref{sec:reviewcurrent}). Substituting eq.\eqref{eq:xirfinal}, eq.\eqref{eq:QmunuLcurrent}, eq.\eqref{eq:thetarfinal} and eq.\eqref{eq:gaugeLcont} in eq.\eqref{eq:main1}, we have
\begin{equation}
    \begin{split}
        2v \tensor{E}{_v_v} &= \frac{1}{\sqrt{h}} \tensor{\partial}{_\rho} \left(\sqrt{h} Q^{r\rho}_{\mathcal{L}} \right) - \Theta^r_{\mathcal{L}} + \Xi^r_{\mathcal{L}} - G^r_{\mathcal{L}} \left(\tensor{A}{_\rho} \tensor{\xi}{^\rho} + \Lambda \right) \\
    &= v \tensor{\partial}{_v} \left(\tensor{W}{_v^r^v} + \tensor{w}{_v^r^v} - \tensor{\mathcal{A}}{_{(1)}} + \frac{1}{\sqrt{h}} \tensor{\partial}{_i} \left(\sqrt{h} \left(\tensor{J}{^i_{(1)}} + \tensor{j}{^i_{(1)}} \right) \right) - \frac{1}{\sqrt{h}} \tensor{\partial}{_v} \left(\sqrt{h} \tensor{\mathcal{B}}{_{(0)}} \right) \right) \\
    &\quad + \frac{1}{\sqrt{h}} \tensor{\partial}{_\rho} \left(\sqrt{h} \left(\tensor{\widetilde{Q}}{^r^\rho} + \tensor{\tilde{q}}{^r^\rho} \right) \right) + \tensor{W}{_v^r^v} + \tensor{w}{_v^r^v} - \tensor{\mathcal{A}}{_{(1)}} + \mathcal{O}\left(\epsilon^2 \right) \, .
    \end{split}
\end{equation}
As the LHS is proportional to $v$, the powers of $v$ on both sides should match \cite{Bhattacharyya:2021jhr,Biswas:2022grc} analogous to eq.\eqref{eq:matchvidentity} and thus,
\begin{equation}
    \begin{split}
        \tensor{\mathcal{A}}{_{(1)}} &= \tensor{W}{_v^r^v} + \tensor{w}{_v^r^v} + \frac{1}{\sqrt{h}} \tensor{\partial}{_v} \left(\sqrt{h} \left(\tensor{\widetilde{Q}}{^r^v} + \tensor{\tilde{q}}{^r^v} \right) \right) + \frac{1}{\sqrt{h}} \tensor{\partial}{_i} \left(\sqrt{h} \left(\tensor{\widetilde{Q}}{^r^i} + \tensor{\tilde{q}}{^r^i} \right) \right) + \mathcal{O}\left(\epsilon^2 \right) \, ,
    \\
    2 \tensor{E}{_v_v} &= \tensor{\partial}{_v} \left(\tensor{W}{_v^r^v} + \tensor{w}{_v^r^v} - \tensor{\mathcal{A}}{_{(1)}} + \frac{1}{\sqrt{h}} \tensor{\partial}{_i} \left(\sqrt{h} \left(\tensor{J}{^i_{(1)}} + \tensor{j}{^i_{(1)}} \right) \right) - \frac{1}{\sqrt{h}} \tensor{\partial}{_v} \left(\sqrt{h} \tensor{\mathcal{B}}{_{(0)}} \right) \right) + \mathcal{O}\left(\epsilon^2 \right) \, .
    \end{split}
\end{equation}
Substituting $\tensor{\mathcal{A}}{_{(1)}}$ in the expression for $\tensor{E}{_v_v}$, we finally get
\begin{equation}\label{eq:evvfinal}
    2 \tensor{E}{_v_v} = - \tensor{\partial}{_v} \left(\frac{1}{\sqrt{h}} \tensor{\partial}{_v} \left(\sqrt{h} \left(\tensor{\widetilde{Q}}{^r^v} + \tensor{\tilde{q}}{^r^v} + \tensor{\mathcal{B}}{_{(0)}} \right) \right) + \frac{1}{\sqrt{h}} \tensor{\partial}{_i} \left(\sqrt{h} \left(\tensor{\widetilde{Q}}{^r^i} + \tensor{\tilde{q}}{^r^i} - \tensor{J}{^i_{(1)}} - \tensor{j}{^i_{(1)}} \right) \right) \right) + \mathcal{O}\left(\epsilon^2 \right)
\end{equation}
Thus, we have recast $E_{vv}$ of eq.\eqref{eq:main1} in the form eq.\eqref{eq:evvfinal} which is the same as eq.\eqref{eom1} with the components of the entropy current given by  
\begin{equation}\label{eq:finalcurrents}
	\begin{split}
		\mathcal{J}^v = -\dfrac{1}{2} \left( \widetilde{Q}^{rv} + \tilde{q}^{rv} + \mathcal{B}_{(0)} \right) \, , ~~~~~
		\mathcal{J}^i = -\dfrac{1}{2} \left( \widetilde{Q}^{ri} + \tilde{q}^{ri} - J^i_{(1)} - j^i_{(1)} \right) \, .
	\end{split}
\end{equation}
This is what we originally set out to prove. Most importantly, since we derived these currents from $\mathcal{L}$ which is $U(1)$ gauge invariant, and we know that $q^{r\rho}$ of eq.\eqref{eq:xiqdefmain} is $U(1)$ gauge invariant, the final currents of eq.\eqref{eq:finalcurrents} are $U(1)$ gauge invariant.

Let us summarize the major steps of our proof. We started off with a generic form of the CS theory Lagrangian given by eq.\eqref{eq:cslagrangian}. This form, as such, was difficult to work with because of the explicit dependence of the Lagrangian on $A_{\mu}$. We employed the Bianchi identities of the CS theory given by eq.\eqref{eq:bianchics} to recast the Lagrangian in a form given by eq.\eqref{eq:finalcslagrangian}. The explicit dependence of $A_{\mu}$ was relegated to a total derivative term and a pure gauge term, both of which do not contribute to the $E_{vv}$ of eq.\eqref{eq:main1}. The leftover part of the CS Lagrangian denoted by $\mathcal{L}$ is $U(1)$ gauge invariant but crucially not diffeomorphism invariant. Thus, the $\Xi^{\mu}$ term of eq.\eqref{eq:diffvarl} was non-zero. This new structure wasn't analyzed in \cite{Bhattacharyya:2021jhr, Biswas:2022grc} because the authors were looking at diffeomorphism invariant theories of gravity. The final structure of $\Xi^r_{\mathcal{L}}$ was worked out in eq.\eqref{eq:xirfinal}. The structures of $\Theta^r_{\mathcal{L}}$ and $Q^{r\rho}_{\mathcal{L}}$ were worked out in eq.\eqref{eq:thetarfinal} and eq.\eqref{eq:Qmunustruc} and they have the same structure as their diffeomorphism counterparts worked out in \cite{Bhattacharyya:2021jhr, Biswas:2022grc} (eq.\eqref{eq:thetardiffgen} and eq.\eqref{eq:Qrnustrucdiffgen} respectively). Putting all the structures together, we got eq.\eqref{eq:evvfinal}, which is what we wanted to prove. 

\subsubsection*{Entropy current for pure gravitational Chern-Simons term:}

We remark that if we had considered a pure gravitational Chern-Simons term, we have a significant simplification in the proof. The Lagrangian will already be of the form $\mathcal{L}$ with no total derivative term and pure gauge terms of eq.\eqref{eq:cslagform1}. Thus, we don't need the analysis of \S\ref{sec:totaltermhorizon} and \S\ref{sec:puregaugetermhorizon}. In the study of the structure of $\Xi^{\mu}$ in \S\ref{sec:xistruchorizon}, we only need to prove the linearity in $v$ in eq.\eqref{eq:xiqdefmain}. This is done in the first subsection of Appendix \ref{ap:xistruc}. Thus eq.\eqref{eq:xirfinal} follows. Since the Lagrangian is independent of the gauge field, we only need to consider the ``gravity" terms of eq.\eqref{eq:thetamuLbr} and eq.\eqref{eq:QmunuLbr}. The analysis of \S\ref{sec:finalentropycurrent} simplifies considerably, and we straightaway get to eq.\eqref{eq:evvfinal} resulting in the entropy current structure eq.\eqref{eom1}.

\section{Verification of the abstract proof with brute force results}
\label{sec:proofverf}

In this section, we will verify the components of the entropy current in eq.\eqref{eq:finalcurrents} by computing $\Theta^r$, $Q^{r\rho}$ and $\Xi^r$. We will also show that the final expressions match the brute force results of \S\ref{sec:bruteforce}. To do that, we first collect the relevant expressions we need to verify the abstract proof. From eq.\eqref{eq:thetarfinal}, eq.\eqref{eq:Qmunustruc} and eq.\eqref{eq:xirfinal}, we have
\begin{equation}
	\Theta^r = (1+v\partial_v) \mathcal{A}_{(1)} + v \partial^2_v \mathcal{B}_{(0)} + \mathcal{O}(\epsilon^2) \, .
\end{equation}
\begin{equation}
	\begin{split}
		Q^{rv} = \widetilde{Q}^{rv} + v W^{rv}_v \, , ~~~~~ 
		Q^{ri} = \widetilde{Q}^{ri} + v W^{ri}_v \, , ~~~~~
		W^{ri}_v = \partial_v J^i_{(1)} + \mathcal{O}(\epsilon^2) \, .
	\end{split}
\end{equation}
\begin{equation}
	\begin{split}
		\tensor{\Xi}{^r} &= \frac{1}{\sqrt{h}} \tensor{\partial}{_v} \left(\sqrt{h} \tensor{\tilde{q}}{^r^v} \right) +\frac{1}{\sqrt{h}} \tensor{\partial}{_i} \left(\sqrt{h} \tensor{\tilde{q}}{^r^i} \right) + \left(1 + v \tensor{\partial}{_v} \right) \tensor{w}{_v^r^v} + v \tensor{\partial}{_v} \left(\frac{1}{\sqrt{h}} \tensor{\partial}{_i} \left(\sqrt{h} \tensor{j}{^i_{(1)}} \right) \right) + \mathcal{O}\left(\epsilon^2 \right) \, .
	\end{split}
\end{equation}
The components of the entropy current are given by eq.\eqref{eq:finalcurrents}:
\begin{equation}
	\begin{split}
		\mathcal{J}^v = -\dfrac{1}{2} \left( \widetilde{Q}^{rv} + \tilde{q}^{rv} + \mathcal{B}_{(0)} \right) \, ,  ~~~~~~~~
		\mathcal{J}^i = -\dfrac{1}{2} \left( \widetilde{Q}^{ri} + \tilde{q}^{ri} - J^i_{(1)} - j^i_{(1)} \right) \, .
	\end{split}
\end{equation}
We will now verify that these general expressions match the brute force computations of $E_{vv}$ for particular theories.

\subsection*{(2+1)D pure gauge Chern-Simons term:}
The Chern-Simons Lagrangian is 
\begin{equation}
	L = \tensor{\epsilon}{^\mu^\nu^\lambda} \tensor{F}{_\mu_\nu} \tensor{A}{_\lambda} \, .
\end{equation}
The various structures are \footnote{One can use the generic formulae in eq.\eqref{eq:cseom}, eq.\eqref{eq:defS}, eq.\eqref{eq:defQ} and eq.\eqref{eq:defxi} to cross-check these answers.}
\begin{equation}
	\begin{split}
		\tensor{G}{^\mu} = 2 \tensor{\epsilon}{^\mu^\nu^\lambda} \tensor{F}{_\nu_\lambda} \, , ~~
		\tensor{\Theta}{^\mu} = 2 \tensor{\epsilon}{^\mu^\nu^\lambda} \tensor{A}{_\lambda} \delta \tensor{A}{_\nu} \, , ~~
		\tensor{\Xi}{^\mu} = \tensor{\epsilon}{^\mu^\nu^\lambda} \tensor{F}{_\nu_\lambda} \Lambda \, , ~~
		 \tensor{Q}{^\mu^\nu} = 2 \tensor{\epsilon}{^\mu^\nu^\lambda} \tensor{A}{_\lambda} \left[\tensor{A}{_\alpha} \tensor{\xi}{^\alpha} + \Lambda \right] \, .
	\end{split}
\end{equation}
If we evaluate these expressions on the horizon $r=0$, we get
\begin{equation}
	\implies D{_\rho} \tensor{Q}{^r^\rho} - \tensor{\Theta}{^r} + \tensor{\Xi}{^r} = \mathcal{O}\left(\epsilon^2 \right) \, .
\end{equation}
This is a particular case of the generic result established in \S\ref{sec:puregaugetermhorizon} that pure $U(1)$ gauge Chern-Simons terms do not contribute to the entropy current. 

\subsection*{(2+1)D pure gravitational Chern-Simons term:}
The Chern-Simons Lagrangian is eq.\eqref{lag2+1}
\begin{equation}
	\mathcal{L} = \tensor{\epsilon}{^\lambda^\mu^\nu} \tensor{\Gamma}{^\rho_\lambda_\sigma} \left(\tensor{\partial}{_\mu} \tensor{\Gamma}{^\sigma_\rho_\nu} + \frac{2}{3} \tensor{\Gamma}{^\sigma_\mu_\tau} \tensor{\Gamma}{^\tau_\rho_\nu} \right) \, .
\end{equation}
The various structures are given by
\begin{equation}\label{eq:2+1gravstruc}
	\begin{split}
		\tensor{\Theta}{^\mu} &= \tensor{\epsilon}{^\alpha^\rho^\sigma} \tensor{R}{^\mu^\beta_\rho_\sigma} \delta \tensor{g}{_\alpha_\beta} + \tensor{\epsilon}{^\lambda^\mu^\nu} \tensor{\Gamma}{^\beta_\lambda_\alpha} \delta \tensor{\Gamma}{^\alpha_\beta_\nu} \, , ~~~~
		\tensor{\Xi}{^\mu} = \tensor{\epsilon}{^\mu^\nu^\lambda} \left(\tensor{\partial}{_\nu} \tensor{\Gamma}{^\sigma_\rho_\lambda} \right) \left(\tensor{\partial}{_\sigma} \tensor{\xi}{^\rho} \right) \, , \\
		\tensor{Q}{^\mu^\nu} &= \tensor{\epsilon}{^\nu^\rho^\sigma} \tensor{R}{^\mu^\lambda_\rho_\sigma} \tensor{\xi}{_\lambda} - \tensor{\epsilon}{^\mu^\rho^\sigma} \tensor{R}{^\nu_\lambda_\rho_\sigma} \tensor{\xi}{^\lambda} + \tensor{\epsilon}{^\lambda^\rho^\sigma} \tensor{R}{^\mu^\nu_\rho_\sigma} \tensor{\xi}{_\lambda} + \tensor{\epsilon}{^\mu^\nu^\lambda} \tensor{\Gamma}{^\rho_\lambda_\sigma} D{_\rho} \tensor{\xi}{^\sigma} \, .
	\end{split}
\end{equation}
It should be noted that if one uses the general formulae eq.\eqref{eq:cseom}, eq.\eqref{eq:defxi} and eq.\eqref{eq:defQ}, we get the above expression for $\Xi^{\mu}$ in eq.\eqref{eq:2+1gravstruc} up to an ambiguous total derivative $D_{\nu} L^{\mu\nu}$ which cancels out with $-D_{\nu}L^{\mu\nu}$ because of the additional term $-L^{\mu\nu}$ in $Q^{\mu\nu}$ of eq.\eqref{eq:2+1gravstruc}:
\begin{equation}\label{eq:2+1gengravstruc}
    \begin{split}
        \Xi^{\mu} &= \tensor{\epsilon}{^\mu^\nu^\lambda} \left(\tensor{\partial}{_\nu} \tensor{\Gamma}{^\sigma_\rho_\lambda} \right) \left(\tensor{\partial}{_\sigma} \tensor{\xi}{^\rho} \right) + D_{\nu} L^{\mu\nu} \rightarrow \Xi^{\mu} + D_{\nu} L^{\mu\nu} \, , \\
        Q^{\mu\nu} &= \tensor{\epsilon}{^\nu^\rho^\sigma} \tensor{R}{^\mu^\lambda_\rho_\sigma} \tensor{\xi}{_\lambda} - \tensor{\epsilon}{^\mu^\rho^\sigma} \tensor{R}{^\nu_\lambda_\rho_\sigma} \tensor{\xi}{^\lambda} + \tensor{\epsilon}{^\lambda^\rho^\sigma} \tensor{R}{^\mu^\nu_\rho_\sigma} \tensor{\xi}{_\lambda} + \tensor{\epsilon}{^\mu^\nu^\lambda} \tensor{\Gamma}{^\rho_\lambda_\sigma} D{_\rho} \tensor{\xi}{^\sigma} - L^{\mu\nu}  \rightarrow Q^{\mu\nu} - L^{\mu\nu} \, , \\
        &\text{where} ~~ L^{\mu\nu} =  \frac{1}{2} \tensor{\xi}{^\lambda} \left(\tensor{\epsilon}{^\nu^\rho^\sigma} \tensor{\partial}{_\rho} \tensor{\Gamma}{^\mu_\sigma_\lambda} - \tensor{\epsilon}{^\mu^\rho^\sigma} \tensor{\partial}{_\rho} \tensor{\Gamma}{^\nu_\sigma_\lambda} \right) \, .
    \end{split}
\end{equation}
The total derivative in $\Xi^{\mu}$ cancels the additional term in $Q^{\mu\nu}$ in the combination eq.\eqref{eq:csnoethercharge}\footnote{\label{ambiguityfoot} These total derivative terms manifest as ambiguities because of our final result for $\Xi^{\mu}$ in eq.\eqref{eq:defxi} in component notation. Here the derivative with respect to the Christoffel symbol should be symmetrized in the lower indices of the symbol as explained in footnote \ref{foot:xiamb}.}. Thus, for our purposes, we can work with the structures of eq.\eqref{eq:2+1gravstruc}. The various structures in eq.\eqref{eq:2+1gravstruc} evaluated in our gauge on the horizon $r=0$ become
\begin{equation}\label{eq:thetar2+1}
	\begin{split}
	    \tensor{\Theta}{^r}|_{r=0} = \left(1 + v \tensor{\partial}{_v} \right) \left[\tensor{\epsilon}{^r^v^x} \frac{\tensor{\partial}{_v} h}{2h} \left(\omega + \frac{\tensor{\partial}{_x} h}{2h} \right) \right] + \mathcal{O}\left(\epsilon^2 \right) \, , ~~ \implies ~~ \mathcal{B}_{(0)} = 0 \, .
	\end{split}
\end{equation}

\begin{equation}\label{eq:xir2+1}
	\begin{split}
	    &\Xi^r|_{r=0} = - \dfrac{1}{\sqrt{h}} \partial_v \left( \sqrt{h} \epsilon^{rvx} \omega \right) \, , ~~
     \implies ~~ \tilde{q}^{rv} = - \epsilon^{rvx} \omega \, , ~~~~~ \tilde{q}^{rx} = 0 \, , ~~~~~ j^x_{(1)} = 0 \, .
	\end{split}
\end{equation}

\begin{equation}\label{eq:Q2+1}
	\begin{split}
		\widetilde{Q}^{rv} &= - \epsilon^{rvx} \omega \, , ~~ \widetilde{Q}^{rx} = 0 \, , ~~ W^{rx}_v = \partial_v \left[ - \dfrac{2 \epsilon^{rvx}\partial_v h}{h} \right] + \mathcal{O}(\epsilon^2) \,
		\implies \, J^x_{(1)} = -\dfrac{2}{h}\epsilon^{rvx} \partial_v h \, , \\
		W^{rv}_v &= \epsilon^{rvx} \left( 2 \partial_v \omega + \dfrac{\omega}{2h}\partial_v h + \dfrac{1}{4h^2}(\partial_v h)(\partial_x h) \right) \, .
	\end{split}
\end{equation}
Thus, the components of the current are given by
\begin{equation}
	\begin{split}
		\mathcal{J}^v &= -\dfrac{1}{2} \left( \widetilde{Q}^{rv} + \tilde{q}^{rv} + \mathcal{B}_{(0)} \right) = \epsilon^{rvx} \omega \, , \\
		\mathcal{J}^x &= -\dfrac{1}{2} \left( \widetilde{Q}^{rx} + \tilde{q}^{rx} - J^x_{(1)} - j^x_{(1)} \right) = -\epsilon^{rvx} \dfrac{\partial_v h}{h} \, , 
	\end{split}
\end{equation}
which matches with the brute force computation of $E_{vv}$ in eq.\eqref{Evv2+1lin} and eq.\eqref{ec2+1lin}. 

If we evaluate $s_{IWT}$ of eq.\eqref{eq:siwtfinal}, we get
\begin{equation}
    s_{IWT} = -2 \left( \pdv{L}{R^{v}_{~v r v}} - \pdv{L}{R^{r}_{~r r v}} \right) = \epsilon^{rvx} \omega = \mathcal{J}^v \, .
\end{equation}
Thus, $\mathcal{J}^v$ is exactly the Iyer-Wald-Tachikawa entropy eq.\eqref{eq:tachikawaform} that satisfies the first law. From eq.\eqref{eq:finalcurrents}, we see that both $Q^{\mu\nu}$ (through $\widetilde{Q}^{rv}$) and $\Xi^{\mu}$ (through $\widetilde{q}^{rv}$) contribute to $\mathcal{J}^v$ and thus to $s_{IWT}$. 

Note that since we work with the equation of motion with covariant indices (two upper indices), the match between brute force computations and the expressions of the abstract proof is exact. In the \cite{Bhattacharyya:2021jhr, Biswas:2022grc}, the authors worked with equations of motion with two lower indices, so the match was up to an overall sign. This is just a convention.

\subsection*{(4+1)D Mixed Chern-Simons term:}
We first consider the mixed Chern-Simons term with an anomaly in the $U(1)$ sector eq.\eqref{eq:4+1mixGanomaly}:
\begin{equation}
	L = \epsilon^{\mu\nu\rho\sigma\delta} A_{\mu} R^{\alpha}_{~\beta\nu\rho} R^{\beta}_{~\alpha\sigma\delta} \, .
\end{equation}
Our proof in \S\ref{sec:absprf} relies on transforming this Lagrangian to eq.\eqref{eq:cslagform1} such that the anomaly is in the gravitational sector eq.\eqref{eq:4+1example}:
\begin{equation}
	L = \mathcal{L} + D_{\mu} (U^{\mu\nu}A_{\nu}) \, ,
\end{equation}
where, from eq.\eqref{eq:4+1mixGRanomaly},
\begin{equation}
	\mathcal{L} = 2 \tensor{\epsilon}{^\mu^\nu^\lambda^\rho^\sigma} \tensor{F}{_\mu_\nu} \tensor{\Gamma}{^\alpha_\lambda_\beta} \left(\frac{1}{2} \tensor{R}{^\beta_\alpha_\rho_\sigma} - \frac{1}{3} \tensor{\Gamma}{^\beta_\rho_\tau} \tensor{\Gamma}{^\tau_\alpha_\sigma} \right) \, ,
\end{equation}
is the $U(1)$ gauge invariant Lagrangian and $U^{\mu\nu}$ is given by eq.\eqref{eq:Bterm} as
\begin{equation}
	U^{\mu\nu} = -4 \tensor{\epsilon}{^\mu^\nu^\lambda^\rho^\sigma} \tensor{\Gamma}{^\alpha_\lambda_\beta} \left(\frac{1}{2} \tensor{R}{^\beta_\alpha_\rho_\sigma} - \frac{1}{3} \tensor{\Gamma}{^\beta_\rho_\tau} \tensor{\Gamma}{^\tau_\alpha_\sigma} \right) \, .
\end{equation}
By brute force computation of $\Theta^{\mu}$, $Q^{\mu\nu}$ and $\Xi^{\mu}$, one can show that  $D_{\mu}(U^{\mu\nu}A_{\nu})$ term doesn't contribute to the entropy current. This aligns with the analysis done in \S\ref{sec:totaltermhorizon}. The various structures derived from $\mathcal{L}$ are
\begin{equation}\label{eq:4+1struc}
	\begin{split}
		\tensor{\Theta}{^\mu} &= -\tensor{U}{^\mu^\nu} \delta \tensor{A}{_\nu} + 2 \tensor{\epsilon}{^\mu^\nu^\tau^\rho^\sigma} \tensor{F}{_\rho_\sigma} \tensor{\Gamma}{^\beta_\tau_\alpha} \delta \tensor{\Gamma}{^\alpha_\beta_\nu} + 2 \tensor{\epsilon}{^\rho^\alpha^\beta^\nu^\lambda} \tensor{R}{^\mu^\sigma_\alpha_\beta} \tensor{F}{_\nu_\lambda} \delta \tensor{g}{_\rho_\sigma} \, ,
		\\
		\tensor{\Xi}{^\mu} &= 2 \tensor{\epsilon}{^\mu^\nu^\lambda^\rho^\sigma} \tensor{F}{_\rho_\sigma} \left(\tensor{\partial}{_\nu} \tensor{\Gamma}{^\alpha_\beta_\lambda} \right) \left(\tensor{\partial}{_\alpha} \tensor{\xi}{^\beta} \right) \, ,
		\\
		\tensor{Q}{^\mu^\nu} &= -\tensor{U}{^\mu^\nu} \left(\tensor{A}{_\eta} \tensor{\xi}{^\eta} + \Lambda \right) + 2 \tensor{\epsilon}{^\mu^\nu^\lambda^\rho^\sigma} \tensor{F}{_\rho_\sigma} \tensor{\Gamma}{^\alpha_\lambda_\beta} D{_\alpha} \tensor{\xi}{^\beta} \\
		&\quad + 2 \tensor{\xi}{_\lambda} \tensor{F}{_\rho_\sigma} \left(\tensor{\epsilon}{^\nu^\alpha^\beta^\rho^\sigma} \tensor{R}{^\mu^\lambda_\alpha_\beta} - \tensor{\epsilon}{^\mu^\alpha^\beta^\rho^\sigma} \tensor{R}{_\alpha_\beta^\nu^\lambda} + \tensor{\epsilon}{^\lambda^\alpha^\beta^\rho^\sigma} \tensor{R}{^\mu^\nu_\alpha_\beta} \right) \, .
	\end{split}
\end{equation}
Similar to the case of the $(2+1)$-dimensional pure gravity CS theory, if we use the general formulae of eq.\eqref{eq:cseom}, eq.\eqref{eq:defxi} and eq.\eqref{eq:defQ}, we have ambiguous terms that cancel out in the combination of eq.\eqref{eq:csnoethercharge}:
\begin{equation}\label{eq:4+1genstruc}
    \begin{split}
        \Theta^{\mu} &\to \Theta^{\mu} + D_{\nu} M^{\mu\nu}_1[\delta g_{\alpha\beta}] \, ,\\
        \Xi^{\mu} &\to \Xi^{\mu} + D_{\nu} M^{\mu\nu}_2[\xi] \, , \\
        Q^{\mu\nu} &\to Q^{\mu\nu} +M^{\mu\nu}_1[\mathcal{L}_{\xi}g_{\alpha\beta}] - M^{\mu\nu}_2[\xi] \, , \\
        \text{where} ~~ M^{\mu\nu}_1[\delta g_{\alpha\beta}] &= 4 \tensor{\epsilon}{^\mu^\nu^\tau^\rho^\sigma} \tensor{A}{_\rho} \tensor{\Gamma}{^\beta_\tau_\alpha} \delta \tensor{\Gamma}{^\alpha_\beta_\sigma} \, , \\
        M^{\mu\nu}_2[\xi] &= \tensor{\xi}{^\lambda} \tensor{F}{_\rho_\sigma} \left[\tensor{\epsilon}{^\nu^\alpha^\beta^\rho^\sigma} \tensor{\partial}{_\alpha} \tensor{\Gamma}{^\mu_\beta_\lambda} - \tensor{\epsilon}{^\mu^\alpha^\beta^\rho^\sigma} \tensor{\partial}{_\alpha} \tensor{\Gamma}{^\nu_\beta_\lambda} \right] \, .
    \end{split}
\end{equation}
All the additional ambiguities cancel out in the combination eq.\eqref{eq:csnoethercharge}. Thus, we can use the structures of eq.\eqref{eq:4+1struc} in our gauge on the horizon $r=0$ to get
\begin{equation}\label{eq:thetar4+1}
	\begin{split}
	    \tensor{\Theta}{^r}|_{r=0} = -\tensor{U}{^r^v} \tensor{\partial}{_v} \left(v \tensor{A}{_v} + \Lambda \right) + &\left(1 + v \tensor{\partial}{_v} \right) \left[\tensor{\epsilon}{^r^v^i^j^k} \tensor{F}{_j_k} \left(\tensor{\omega}{^l} \tensor{\partial}{_v} \tensor{h}{_i_l} + 2 \tensor{\Gamma}{^m_i_n} \tensor{\Gamma}{^n_m_v} \right) \right] + \mathcal{O}\left(\epsilon^2 \right) \, , \\
     &\implies \mathcal{B}_{(0)} = 0 \, .
	\end{split}
\end{equation}

\begin{equation}\label{eq:xir4+1}
      \begin{split}
          \Xi^r|_{r=0} = &\dfrac{1}{\sqrt{h}} \partial_v \left( \sqrt{h} [ -2 \epsilon^{rvijk} F_{jk} \omega_i ] \right) + \dfrac{1}{\sqrt{h}} \partial_i \left( \sqrt{h} [ -4 \epsilon^{rvijk} F_{vj} \omega_k  ] \right) \, , \\
          & \implies \tilde{q}^{rv} = -2 \epsilon^{rvijk} F_{jk} \omega_i \, , ~~~~~ \tilde{q}^{ri} = -4 \epsilon^{rvijk} F_{vj} \omega_k \, , ~~~~~ j^i_{(1)} = 0  \, .
      \end{split}
\end{equation}

\begin{equation}\label{eq:Qrnu4+1}
	\tensor{Q}{^r^\nu} = \tensor{\widetilde{Q}}{^r^\nu} + v \tensor{W}{_v^r^\nu} - \tensor{U}{^r^\nu} \left(v \tensor{A}{_v} + \Lambda \right) \, .
\end{equation}
\begin{equation}\label{eq:Q4+1}
	\begin{split}
		\widetilde{Q}^{rv} &= - 2 \epsilon^{rvijk} F_{jk} \omega_i \, , ~~~~~ \widetilde{Q}^{ri} = -4 \epsilon^{rvijk} F_{vk} \omega_j \, ,  \\
		W^{ri}_v &= - 4 \partial_v \left( h^{im}\epsilon^{rvljk} F_{jk} \partial_v h_{ml} \right) + \mathcal{O}(\epsilon^2) \,
		\implies \, J^i_{(1)} = -4 h^{im}\epsilon^{rvljk} F_{jk} \partial_v h_{ml} \, , \\
  W^{rv}_v &= 8 \epsilon^{rvijk}( F_{jk} R_{vrvi} + F_{vi} R_{vrjk} ) + 2 \epsilon^{rvijk} F_{jk} \Gamma^{\alpha}_{i\beta} \Gamma^{\beta}_{\alpha v} \, .
	\end{split}
\end{equation}
Thus, the components of the entropy current are given by
\begin{equation}
	\begin{split}
		\mathcal{J}^v &= -\dfrac{1}{2} \left( \widetilde{Q}^{rv} + \tilde{q}^{rv} + \mathcal{B}_{(0)} \right) = 2 \epsilon^{rvijk} F_{jk} \omega_i \, , \\
		\mathcal{J}^i &= -\dfrac{1}{2} \left( \widetilde{Q}^{ri} + \tilde{q}^{ri} - J^i_{(1)} - j^i_{(1)} \right) = 4 \epsilon^{rvijk} F_{vk} \omega_j - 2 h^{im} \epsilon^{rvljk} F_{jk} \partial_v h_{ml} \, .
	\end{split}
\end{equation}
This matches with the brute force computation of $E_{vv}$ given in eq.\eqref{Evv4+1d} and eq.\eqref{JvJi4+1} \footnote{Once again, we see that this exactly matches because we work with equations of motion with upper indices.}

If we evaluate $s_{IWT}$ of eq.\eqref{eq:siwtfinal}, we get
\begin{equation}
    s_{IWT} = -2 \left( \pdv{L}{R^{v}_{~v r v}} - \pdv{L}{R^{r}_{~r r v}} \right) = 2\tensor{\epsilon}{^r^v^i^j^k} \tensor{F}{_j_k} \tensor{\omega}{_i} = \mathcal{J}^v \, .
\end{equation}
Thus, $\mathcal{J}^v$ is exactly the Iyer-Wald-Tachikawa entropy eq.\eqref{eq:tachikawaform} that satisfies the first law. Again, from eq.\eqref{eq:finalcurrents}, we see that both $Q^{\mu\nu}$ (through $\widetilde{Q}^{rv}$) and $\Xi^{\mu}$ (through $\widetilde{q}^{rv}$) contribute to $\mathcal{J}^v$ and thus to $s_{IWT}$.

\section{Conclusions} \label{sec:concl}

In this section, we conclude by summarizing our results and highlighting the crucial lessons from our analysis. Our primary aim in this paper has been to show that the classical second law of thermodynamics is satisfied for dynamical black hole solutions in CS theories of gravity (and mixed gauge gravity CS theories). We have used the setup developed in \cite{Bhattacharyya:2021jhr, Biswas:2022grc} where the dynamics are assumed to be small perturbations around stationary black holes. To argue the second law, we have looked at the off-shell structure of the EoM ($E_{vv}$) in an adapted coordinate system around the Killing horizon, eq.\eqref{nhmetric}. This enables us to obtain the entropy current defined on the horizon, whose components give us the entropy density and a local in/out flow of it on a horizon slice. We constructed a local version of the second law with an extra assumption of the matter sector satisfying the null energy condition. 

Firstly, we considered specific examples of Chern-Simons theories and obtained the components of entropy current through a brute-force computation of the off-shell structure of $E_{vv}$. For $(2+1)$-dimensional pure gravity CS theory, the result is given in eq.\eqref{Evv2+1lin}, and eq.\eqref{ec2+1lin}, which is limited to linearized perturbations. Furthermore, considering the $(2+1)$-dimensional pure gravity CS theory as a low energy EFT, the quadratic perturbations have been incorporated in the off-shell structure of $E_{vv}$ in eq.\eqref{EvvEFT}, and the entropy current is written in eq.\eqref{JvJiEFT}. In this particular case, we find an example where the second law is established beyond linear order and thus signifies actual entropy production. Next, in $(4+1)$-dimensions for mixed gauge gravity CS theories, working to linearized order in the perturbations, the off-shell $E_{vv}$ and the entropy current has been obtained as written in eq.\eqref{Evv4+1d} and eq.\eqref{JvJi4+1} respectively. 

Once we have obtained these explicit expressions of entropy current for each of the examples mentioned above, we have also explicitly verified that they transform covariantly under a reparametrization of the coordinates maintaining the gauge, eq.\eqref{nhmetric}. We have explicitly obtained how the entropy currents change under this reparametrization (see eq.\eqref{JvJi2+1c} for $(2+1)$-dimensional CS theory as EFT, and eq.\eqref{eq:newJvJi4+1} for $(4+1)$-dimensional CS theory). Most importantly, in both cases, we have seen how a non-trivial cancellation was required to assure the covariance as mentioned in eq.\eqref{Evvtrans}. This also justifies the need to have the spatial components of the entropy current ($\mathcal{J}^i$) on the horizon. Interestingly, we have seen for both the $(2+1)$-dimensional and $(4+1)$-dimensional CS theories that the reparametrization may lead to ambiguities in the entropy current components, see eq.\eqref{eq:4+1repamb}. These ambiguities may be present in the local description of $\mathcal{J}^v$ and $\mathcal{J}^i$; however, they do not contribute to off-shell $E_{vv}$ or to the total entropy when integrated over a compact horizon slice. 

Next, we have developed an algorithm to construct an entropy current consistent with the linearized second law for a generic CS theory. It is like developing a formalism rather than working brute force with a particular example. In \S\ref{sec:gaugedepcslagrangian}, we have shown that the Lagrangian density of any generic CS theory can be written as eq.\eqref{eq:finalcslagrangian}. Then, we have extended the formalism developed in \cite{Bhattacharyya:2021jhr, Biswas:2022grc} to make it applicable to CS theories. We have also used the covariant phase space formalism studied earlier in \cite{Tachikawa:2006sz, Bonora:2011gz, Azeyanagi:2014sna} to relate off-shell $E_{vv}$ with pre-symplectic potential, Noether charge eq.\eqref{eq:main}. We tracked the essential difference between CS theories and diffeomorphism invariant theories: the contribution of the non-covariant $\Xi^\mu$ (which is defined in eq.\eqref{eq:diffvarl}) to eq.\eqref{eq:main}. By analyzing their expressions in our metric gauge, we finally obtained the entropy current in terms of the elements of covariant phase space formalism; see the final result in eq.\eqref{eq:evvfinal} and eq.\eqref{eq:finalcurrents}. In order to justify that the technical arguments indeed produce a consistent algorithm to construct the entropy current, in \S\ref{sec:proofverf}, we have computed the $\mathcal{J}^v$ and $\mathcal{J}^i$ for several examples of CS theories and checked that they exactly match the expressions obtained via brute force analysis. 

Some comments regarding further implications of our results and possible future directions are as follows. First, we must note that the setup to establish linearized second law can be used to argue for the physical process version of the first law. Previously, for model-specific theories, this has been worked out in \cite{Gao:2001ut, Amsel:2007mh, Chatterjee:2011wj, Kolekar:2012tq, Bhattacharjee:2014eea, Mishra:2017sqs}. For arbitrary diffeomorphism invariant theories, this has been established in \cite{Bhattacharyya:2021jhr}, see \S5 therein, and in \cite{Biswas:2022grc} to include the cases of non-minimally coupled matter fields. Mainly, the off-shell $E_{vv}$ as mentioned in eq.\eqref{eom1} is sufficient to extend the proof of the physical process version of the first law in CS theories. 

It is also straightforward to include a cosmological constant in our analysis. More precisely, the result involving the off-shell structure of $E_{vv}$, e.g., eq.\eqref{eom1}, remains unaffected due to the inclusion of a cosmological constant. This is easy to understand since the EoM should change as follows $E_{vv} \sim \Lambda \, g_{vv}$, and in our horizon adapted coordinates, $g_{vv}$ vanishes on the horizon. Thus, all our results can be directly applied to black holes in AdS space-time. Furthermore, our analysis can be directly applied to any Killing horizons, including cosmological horizons in de-Sitter space-times. 

For the examples of CS theories we considered, we see that the JKM ambiguity vanishes, which can be verified from the expressions of $\mathcal{J}^v$. For the $(2+1)$-dimensional CS theory, it is zero till the quadratic order, see in eq.\eqref{JvJiEFT}, whereas for the $(4+1)$-dimensional CS theory, it is zero at the linear order, see in eq.\eqref{JvJi4+1}. In the $2+1$-dimensional case, the vanishing of the JKM ambiguity term can be traced back to the counting the number of derivatives in the off-shell structure of the $vv$-component of equations of motion (EoM), i.e. $E_{vv}$. In turn, this is dictated by the number of derivatives present in the Lagrangian of the theory under consideration, which is three in $2+1$-dimensions. Therefore, in $E_{vv}$ we should have three derivatives, and hence, in $\mathcal{J}^v$ there can be only one derivative. From here, it is easy to justify that the JKM ambiguity in $\mathcal{J}^v$ has to vanish identically in this case. However, this argument based on counting number of derivatives and boost weight will not go through in higher dimensions.  On the other hand, in our abstract proof for the generic case, we have not seen any such obstruction, as there is, at least in principle, a possible JKM piece $\mathcal{B}_{(0)}$ which contributes to $\mathcal{J}^v$ of eq.\eqref{eq:finalcurrents} through eq.\eqref{eq:thetarfinal}. It would be interesting to see if the vanishing of the JKM term in our explicit examples is a coincidence or if some universal statement regarding its existence of it can be made more abstractly. 

As mentioned in footnote \ref{foot:nonabelian}, our algorithm to construct an entropy current is valid only for $U(1)$ gauge fields. Since the analysis in \S\ref{sec:gaugedepcslagrangian} crucially depends on having an abelian gauge field, it would be interesting to see if our proof can be generalized to include non-abelian gauge fields. 

For $(2+1)$-dimensional CS theory, when considered as an EFT, we have seen that the second law can be extended to quadratic order in the perturbations around the stationary black holes. However, recently, in \cite{Hollands:2022fkn}, a formalism has been developed to study the second law to quadratic orders for generic diffeomorphism invariant theories. The result of the $(2+1)$-dimensional CS theory is consistent with Lemma 3.1 and Lemma 4.1 of \cite{Hollands:2022fkn}. Furthermore, the reparameterization of $\mathcal{J}^v$ for both the theories (see eq.\eqref{eq:Jv2+1extra} and eq.\eqref{eq:4+1repamb}) is consistent with Proposition 1 of \cite{Hollands:2022fkn} which previously did not cover the case of CS Lagrangians. It would be interesting to see if Proposition 1 and Lemma 4.1 can be extended to include CS theories. Our analysis in \S\ref{sec:absprf} suggests that at least the linear order analysis (a generalization of Lemma 3.1 of \cite{Hollands:2022fkn}) can be organized similarly to what has previously been done for diffeomorphism invariant theories.

\section*{Acknowledgements} 

We thank Parthajit Biswas for the initial collaboration. We are especially grateful to Sayantani Bhattacharyya for various enlightening discussions and collaboration on related projects. We would like to thank Jyotirmoy Bhattacharya and Harvey Reall for their useful comments on our draft. We would also like to thank Diptarka Das, Anirban Dinda, S. Shankaranarayanan, and Yogesh Kumar Srivastava for useful discussions. PD would like to thank NISER Bhubaneshwar for their warm hospitality during a visit where partial progress of this work was presented in a talk. PD also thanks the organizers of FTAG 2023 for giving the opportunity to present the results of this work in a poster. PD duly acknowledges the Council of Scientific and Industrial Research (CSIR), New Delhi, for financial assistance through the Senior Research Fellowship (SRF) scheme. The work of NK is supported by a MATRICS grant (MTR/2022/000794) 
from the Science and Engineering Research Board (SERB), India. We acknowledge our debt to the people of India for their steady support of research in basic sciences.

\appendix

\section{Intricate details of the calculations}

\subsection{Review of constructing entropy current for diffeomorphism invariant theories of gravity}
\label{sec:reviewcurrent}

In this Appendix, we will quickly summarize the major steps in the construction of entropy current for diffeomorphism invariant theories of gravity \cite{Bhattacharyya:2021jhr, Biswas:2022grc} of the form
\begin{equation}\label{eq:difflag}
    L = L(g_{\mu\nu},R_{\mu\nu\rho\sigma},D_{\alpha_1} R_{\mu\nu\rho\sigma},D_{(\alpha_1}D_{\alpha_2)} R_{\mu\nu\rho\sigma},\dots,\phi,D_{\alpha_1}\phi,D_{(\alpha_1}D_{\alpha_2)}\phi,\dots,F_{\mu\nu},D_{\alpha_1} F_{\mu\nu},\dots) \, .
\end{equation}
We argued on general boost weight arguments that structures of $\Theta^r$, $Q^{rv}$ and $Q^{ri}$ take the form of eq.\eqref{eq:thetaQinitial}. To further constrain the quantities in the RHS of eq.\eqref{eq:thetaQinitial}, we must understand the structure of a generic covariant tensor with a positive boost weight. A typical covariant object with a boost weight $w= a+1 >0$ on the horizon $r=0$ takes the form (see eq.(3.13) of \cite{Bhattacharyya:2021jhr})
\begin{equation}\label{eq:a+1boostwt}
    t^{(k)}_{(a+1)}|_{r=0} = \widetilde{T}_{(-k)} \partial^{k+a+1}_v T_{(0)}|_{r=0} + \mathcal{O}(\epsilon^2) \, .
\end{equation}
One can show that (see Appendix E of \cite{Bhattacharyya:2021jhr}) we can rearrange the $\partial_v$ derivatives to recast it in the form given by Result:1 (see eq.(3.14) of \cite{Bhattacharyya:2021jhr})
\begin{equation}\label{eq:genboostwt}
    \begin{split}
        t^{(k)}_{(a+1)} = \partial^{a+1}_v \left[ \sum_{m=0}^{k-1} (-1)^m \, [^{m+a}C_m ] \, \widetilde{T}_{(-k+m)} \partial^{(k-m)}_v T_{(0)} \right] + (-1)^k \, [^{k+a}C_a] \, \widetilde{T}_{(0)} \partial^{a+1}_v T_{(0)} + \mathcal{O}(\epsilon^2) \, .
    \end{split}
\end{equation}
If one includes $U(1)$ gauge fields in the analysis, then $F_{vi}$ and $F_{ri}$ are additional quantities with non trivial boost weights. Thus, one has to include these in the above analysis of eq.\eqref{eq:genboostwt} carefully \cite{Biswas:2022grc} (see eq.(6.17) and eq.(C.8) of \cite{Biswas:2022grc}). Though we mention that eq.\eqref{eq:a+1boostwt} is valid for covariant tensors \cite{Bhattacharyya:2021jhr}, one can check that it is valid for Christoffel symbols as well. For the boost weight argument of eq.\eqref{eq:a+1boostwt}, we can treat the Christoffel symbol as a tensor with one upper index and two lower indices. This is because under eq.\eqref{boosttrns}, the Christoffel symbol transforms like a tensor. One can also verify this statement from the expressions of Appendix \ref{ap:christoffel}. This will prove to be very useful for the CS theories that we are interested in.

The proof follows by using the general structures of $\Theta^{\mu}$ and $Q^{\mu\rho}$ derived for eq.\eqref{eq:difflag} in \cite{Iyer:1994ys} and then implementing Result:1 of eq.\eqref{eq:genboostwt} in the terms of those expressions. Since $\Theta^{\mu}$ is obtained from the variation of the Lagrangian in eq.\eqref{eq:varL}, it generically contains $\delta \mathcal{S}_{\alpha_1 \alpha_2 \dots \alpha_k}$ where $\mathcal{S}$ is a covariant tensor. In order to obtain eq.\eqref{eq:main}, we had to use the fact that the variation is given by a diffeomorphism and thus $\delta = \mathcal{L}_{\xi}$. Hence, in order to further process the terms in $\Theta^r$, we have to use the input of diffeomorphism invariance to write (eq.(3.23) of \cite{Bhattacharyya:2021jhr})
\begin{equation}
    \delta \mathcal{S}_{\alpha_1 \alpha_2 \dots \alpha_k}[\delta g_{\alpha\beta} \to \mathcal{L}_{\xi} g_{\alpha\beta} ] = \mathcal{L}_{\xi} \mathcal{S}_{\alpha_1 \alpha_2 \dots \alpha_k} \, .
\end{equation}
This implies that the general structure of $\Theta^{\mu}$ to be evaluated on horizon adapted coordinates is given by
\begin{equation}\label{eq:thetamugen}
    \Theta^{\mu} = 2 E^{\mu\nu\alpha\beta}_R D_{\beta} \left( \mathcal{L}_{\xi} g_{\nu\alpha} \right) + \sum_k \mathcal{T}^{\mu\alpha_1 \alpha_2 \dots \alpha_k} \mathcal{L}_{\xi} \mathcal{S}_{\alpha_1 \dots \alpha_k} \, .
\end{equation}
Here $\mathcal{T}$ is a covariant tensor and $E^{\mu\nu\alpha\beta}_R$ is given by
\begin{equation}\label{eq:ERdiff}
    E^{\mu\nu\alpha\beta}_R = \pdv{L}{R_{\mu\nu\alpha\beta}} - D_{\rho_1} \pdv{L}{D_{\rho_1} R_{\mu\nu\alpha\beta} } + \dots + (-1)^m D_{(\rho_1} \dots D_{\rho_m)} \pdv{L}{D_{(\rho_1} \dots D_{\rho_m)} R_{\mu\nu\alpha\beta}} \, ,
\end{equation}
where $L$ is the Lagrangian of the theory eq.\eqref{eq:difflag}. The appearance of $\mathcal{L}_{\xi}$ in eq.\eqref{eq:thetamugen} makes it clear that $\Theta^r$ is linear in $v$ as in eq.\eqref{eq:thetaQinitial}. One can now implement eq.\eqref{eq:genboostwt} in eq.\eqref{eq:thetamugen} to show that on the horizon $r=0$,
\begin{equation}\label{eq:thetardiffgenap}
    \Theta^r |_{r=0} = (1+ v \partial_v) \mathcal{A}_{(1)} + v \partial^2_v \mathcal{B}_{(0)} \, .
\end{equation}
Here $\mathcal{B}_{(0)}$ denotes the JKM ambiguity and it is $\mathcal{O}(\epsilon)$ even though it has boost weight zero. This is because the $\mathcal{B}_{(0)}$ takes the form of a product of two terms that are individually not boost-invariant:
\begin{equation}
    \mathcal{B}_{(0)} \sim X_{(-k+m)} \partial^{k-m}_v Y_{(0)} \sim \mathcal{O}(\epsilon) \, .
\end{equation}

The general structure of $Q^{\mu\nu}$ for diffeomorphism invariant theories is of the form \cite{Iyer:1994ys}:
\begin{equation}
    Q^{\mu\nu} = W^{\mu\nu\rho}\xi_{\rho} - 2 E^{\mu\nu\alpha\beta}_R D_{[\alpha} \xi_{\beta]} \, .
\end{equation}
Following this, $Q^{r\rho}$ takes the form of eq.\eqref{eq:thetaQinitial} given by
\begin{equation}
    Q^{r\rho} = \widetilde{Q}^{r\rho} + v \, W^{r\rho}_v \, .
\end{equation}
Thus, the divergence of $Q^{r\rho}$ on the horizon $r=0$ is given by
\begin{equation}\label{eq:divQdiffgen}
    D_{\mu} Q^{r\mu} = \dfrac{1}{\sqrt{h}} \partial_v \left( \sqrt{h} \widetilde{Q}^{rv}\right) + \nabla_i \widetilde{Q}^{ri} + v \partial_v \left( \nabla_i J^i_{(1)}  \right) + (1 + v \partial_v) W^{rv}_v + \mathcal{O}(\epsilon^2) \, ,
\end{equation}
where $J^i_{(1)}$ is defined through
\begin{equation}
    W^{ri}_v = \partial_v J^i_{(1)} + \mathcal{O}(\epsilon^2) \, .
\end{equation}
We can now substitute eq.\eqref{eq:divQdiffgen} and eq.\eqref{eq:thetardiffgenap} in eq.\eqref{eq:main} with $\Xi^r = 0$ to get
\begin{equation}\label{eq:evvdiffgen}
    \begin{split}
        2 \, v \, E_{vv} &= \left( - \Theta^r + D_{\mu} Q^{r\mu} \right) |_{r=0} \\
        &= - \mathcal{A}_{(1)} + \dfrac{1}{\sqrt{h}} \partial_v \left( \sqrt{h} \, \widetilde{Q}^{rv} \right) + \nabla_i \widetilde{Q}^{ri} + W^{rv}_v \\
        & \quad \quad + v \partial_v \left[ - \mathcal{A}_{(1)} + W^{rv}_v + \nabla_i J^i_{(1)} - \partial_v \mathcal{B}_{(0)} \right] + \mathcal{O}(\epsilon^2) \, .
    \end{split}
\end{equation}
Since we have explicitly accounted for all the factors of $v$, we can compare the coefficients of $v^0$ and $v$ of eq.\eqref{eq:evvdiffgen} to write
\begin{equation}\label{eq:matchvidentity}
    \begin{split}
        \mathcal{A}_{(1)} &= \dfrac{1}{\sqrt{h}} \partial_v \left( \sqrt{h} \, \widetilde{Q}^{rv} \right) + \nabla_i \widetilde{Q}^{ri} + W^{rv}_v \, , \\
        2 E_{vv} &= \partial_v \left[ - \mathcal{A}_{(1)} + W^{rv}_v + \nabla_i J^i_{(1)} - \partial_v \mathcal{B}_{(0)} \right] \, .
    \end{split}
\end{equation}
We emphasize that eq.\eqref{eq:matchvidentity} should be thought of as an identity rather than as an algebraic equation equating different powers of $v$. Once the factors of $v$ are made explicit, eq.\eqref{eq:matchvidentity} holds identically. However, there is one potential source of confusion that we would like to highlight. Apparently, apart from explicit factors of $v$ present on the RHS of eq.\eqref{eq:evvdiffgen}, one may consider various objects appearing on the RHS of eq.\eqref{eq:evvdiffgen} as functions $v$. This may lead to doubting the derivation of eq.\eqref{eq:matchvidentity}. In \S\ref{ap:subsubidentity} we will provide some justifications and checks in some particular model Lagrangians supporting the manipulations leading to eq.\eqref{eq:matchvidentity}.

Combining both the equations in eq.\eqref{eq:matchvidentity} we finally get
\begin{equation}\label{eq:evvfinaldiffgenap}
    2 E_{vv}|_{r=0} = - \partial_v \left( \dfrac{1}{\sqrt{h}} \partial_v \left[ \sqrt{h} \left( \widetilde{Q}^{rv} + \mathcal{B}_{(0)} \right) \right] + \nabla_i \left[ \widetilde{Q}^{ri} - J^i_{(1)} \right] \right) + \mathcal{O}(\epsilon^2) \, .
\end{equation}
Here we have used the fact that crucially $\mathcal{B}_{(0)} \sim \mathcal{O}(\epsilon)$ which is 
\begin{equation}
    \partial^2_v \mathcal{B}_{(0)} = \partial_v \left(\dfrac{1}{\sqrt{h}} \partial_v \left[ \sqrt{h} \, \mathcal{B}_{(0)} \right] \right) + \mathcal{O}(\epsilon^2) \, .
\end{equation}
From eq.\eqref{eq:evvfinaldiffgenap}, we get the components of the entropy current to be
\begin{equation}
    \mathcal{J}^v = - \dfrac{1}{2} \left( \widetilde{Q}^{rv} + \mathcal{B}_{(0)} \right) \, , ~~~ \text{and} ~~ \mathcal{J}^i = - \dfrac{1}{2} \left( \widetilde{Q}^{ri} - J^i_{(1)} \right) \, .
\end{equation}
The components of the current are $U(1)$ gauge invariant if we consider $U(1)$ gauge invariant Lagrangians \cite{Biswas:2022grc}. This is because the structures of eq.\eqref{eq:matchvidentity} are $U(1)$ gauge invariant. This completes our review of the proof of the existence of entropy current for diffeomorphism invariant theories of gravity. 

\subsubsection{More justifications and explicit checks of eq.\eqref{eq:matchvidentity}}
\label{ap:subsubidentity}

Let us write eq.\eqref{eq:evvdiffgen} schematically as follows \begin{equation}\label{eq:maineqapp}
        2\, v \, E_{vv} |_{r=0} 
        =(-\Theta^r + D_{\mu}Q^{r\mu})|_{r=0} + \mathcal{O}(\epsilon^2) 
        = \mathcal{U}_{(1)} + v \, \mathcal{V}_{(2)} + \mathcal{O}(\epsilon^2) \, , 
\end{equation}
where detail forms of $\mathcal{U}_{(1)}$ and $\mathcal{V}_{(2)}$ are omitted. Note that the last equality in eq.\eqref{eq:maineqapp} is not a Taylor series expansion in $v$. Since $\Theta^\mu$ and $Q^{\mu\nu}$ are linear in $\xi$, it is evident that there are no terms involving quadratic or higher powers of $v$ (i.e., $\mathcal{U}_{(1)}$ and $\mathcal{V}_{(2)}$ have no explicit $v$ dependence), and we can get at most a linear piece in $v$. Then, eq.\eqref{eq:matchvidentity} becomes 
 \begin{equation}\label{eq:relations}
        \mathcal{U}_{(1)} = \mathcal{O}(\epsilon^2), \quad 2 E_{vv}|_{r=0} = \mathcal{V}_{(2)} + \mathcal{O}(\epsilon^2)  \, .
 \end{equation}

Now, $\mathcal{U}_{(1)}$ and $\mathcal{V}_{(2)}$ should be thought of as structures constructed out of the metric components $X, \omega_i, h_{ij}$ and various differential operators $\partial_v,\partial_r,\nabla_i$ acting on them. While comparing coefficients of various powers of $v$ on both sides of eq.\eqref{eq:maineqapp}, we are actually working in the configuration space or phase space. For this reason, this is a structural rearrangement of $E_{vv}$, and we never needed to worry about the explicit form of the metric coefficients $X,\omega_i,h_{ij}$ as functions of the coordinates $v,r,x^i$ for this comparison of coefficients of $v$. Hence for us, $\mathcal{U}_{(1)}$ and $\mathcal{V}_{(2)}$ are just structures involving $X,\omega_i,h_{ij}$ and their derivatives, and not explicit functions of $v,r,x^i$. In other words, we are trying to argue that the relation \eqref{eq:maineqapp} is an identity but not an extra restriction in the the form of a differential equation for $X,\omega_i,h_{ij}$.

Recently this issue has been dealt in \cite{Hollands:2022fkn} for diffeomorphism invariant theories by using a one parameter family of interpolating metrics  $g_{\mu\nu}(\lambda,x^{\alpha})$ where $\lambda \in [0,1]$. $\lambda = 0$ corresponds to the stationary metric $\Bar{g}_{\mu\nu}(rv,x^i)$. The choice of parametrization of \cite{Hollands:2022fkn} is
\begin{equation}\label{eq:hkrpara}
    \psi(\lambda) := \Bar{\psi} + \lambda \delta \psi \, ,
\end{equation}
where $\psi$ denotes the components of metric fields. There the following equation was obtained (see eq.(101) of \cite{Hollands:2022fkn})   
\begin{equation}\label{eq:hkrinter}
    \partial_v \left( \dfrac{1}{\sqrt{\mu}} \dfrac{\partial}{\partial \lambda} (\ast Q^{vr}_K \sqrt{\mu}) - v (\ast \theta^r) \left[ g; \dfrac{\partial}{\partial \lambda} g \right] \right)_{\lambda = 0} + D_A \dfrac{\partial}{\partial \lambda} (\ast Q^{A r}_K) |_{\lambda = 0} = 2 v \dfrac{\partial}{\partial \lambda} E_{vv} |_{\lambda = 0} \, .
\end{equation}
Here $\ast$ denotes the hodge dual (since $\mathbf{Q}_K$ is a $d-2$ form) and $\mu = h$ (the determinant of $h_{ij}$) in our notation for the metric. In \cite{Hollands:2022fkn} the background fields have also been taken as given in eq.(98) by construction. We may think eq.\eqref{eq:hkrinter} to be an alternate interpretation of eq.\eqref{eq:maineqapp} as an off-shell statement in the variational phase space of solutions. In the parametrization eq.\eqref{eq:hkrpara}, the interpretation of eq.\eqref{eq:hkrinter} implies that the positive boost weight quantities quantify the tangential derivatives (denoted by $\lambda$ derivative in eq.\eqref{eq:hkrinter}) in the phase space of solutions. Once we have accounted for the explicit factors of $v$, we are essentially comparing the variation of quantities in the phase space of solutions on both the sides of eq.\eqref{eq:hkrinter}. We mentioned $\mathcal{U}_{(1)}$ and $\mathcal{V}_{(2)}$ should be thought of as structures constructed out of the metric components $X,\omega_i,h_{ij}$ and various differential operators $\partial_v,\partial_r,\nabla_i$ acting on them. This, in the language of \cite{Hollands:2022fkn}, corresponds to considering the primitive monomials constructed out of the primitive factors. They ($\mathcal{U}_{(1)}$ and $\mathcal{V}_{(2)}$) are objects in the configuration space. 

It should be possible to formally come up with an extension of this rigorous and formal justification of what we call as comparing structures (not functions of the coordinate $v$) on both sides of eq.\eqref{eq:maineqapp} to Chern-Simons theories. This is because the presence of Christoffel symbols will not affect the boost weight analysis. Since the boost transformation is a linear transformation, the Christoffel symbol $\Gamma^{\mu}_{\nu\rho}$ transforms like a covariant tensor with one upper index and two lower indices. The gauge field terms from the matter sector should contribute something new as the origin of boost weight $+1$ quantity in our basic building blocks (primitive factors in the language of \cite{Hollands:2022fkn}). This has been the case in \cite{Biswas:2022grc}, dealing with the non-minimal coupling between gauge fields and gravity.

Finally, we would like to provide explicit checks of the identities mentioned in eq.\eqref{eq:relations}. First we demonstrate eq.\eqref{eq:relations} in an example of diffeomorphism invariant situation, namely for Riemann squared theory. Next we work out examples of both the specific CS theories that we worked out in this paper. We will show that $\mathcal{U}_{(1)} = 0$ and $\mathcal{V}_{(2)}$ has the entropy current structure for the following examples. 

\paragraph{Diffeomorphism invariant Riemann squared theory: }
We start with the Lagrangian $L= R^{\mu\nu\rho\sigma}R_{\mu\nu\rho\sigma}$. The following objects can be obtained (we refer the readers to \cite{Bhattacharyya:2021jhr} for derivations of them)
        \begin{equation}
            \begin{split}
                \Theta^r &= (1+v\partial_v) ( 8 R_{vijr} K^{ij} - 8 \Bar{K}^{ij}\partial_v K_{ij} ) + v \partial^2_v (8 \Bar{K}^{ij} K_{ij} ) + \mathcal{O}(\epsilon^2) \\
                Q^{rv} &= 8 R_{rvrv} - 8 v \dfrac{1}{\sqrt{h}} \partial_v (\sqrt{h} R_{rvrv}) - 4 v \nabla_i \left( h^{ij}\partial_v \omega_j + K^i_{~k} \omega^k \right) \\
                &~+ 8 v K^{ij} R_{vijr} - 8 v \Bar{K}^{ij} \partial_v K_{ij} + \mathcal{O}(\epsilon^2) \\
                Q^{ri} &= 4 h^{ij} \partial_v \omega_j + 4 K^i_{~k}\omega^k + 4 v \partial_v ( K^i_{~k}\omega^k ) \\
                &~+ v \partial_v ( - 4h^{ij} \partial_v \omega_j + 8 \nabla_m K^{im}) + \mathcal{O}(\epsilon^2)
            \end{split}
        \end{equation}
        \begin{equation}
            \begin{split}
                D_{\mu}Q^{r\mu} &= \dfrac{1}{\sqrt{h}}\partial_v ( \sqrt{h} \, 8 R_{rvrv} ) + \nabla_i ( 4h^{ij}\partial_v \omega_j + 4 K^i_{~k} \omega^k ) \\
                &~~ + v \partial_v \left[ \nabla_i ( 4 K^i_{~k} \omega^k - 4 h^{ij} \partial_v \omega_j + 8 \nabla_m K^{im} ) \right] \\
                &~~ + (1+ v\partial_v) \left[ -8 \dfrac{1}{\sqrt{h}}\partial_v (\sqrt{h} \, R_{rvrv}) - 4 \nabla_i (h^{ij} \partial_v \omega_j + K^i_{~k}\omega^k ) \right. \\
                &~~ \left. + 8 K^{ij} R_{vijr} - 8 \Bar{K}^{ij} \partial_v K_{ij} \right] + \mathcal{O}(\epsilon^2)
            \end{split}
        \end{equation}
        In this case, we have,
        \begin{equation}
           \begin{split}
            \tilde{Q}^{rv} &= 8 R_{rvrv} ~~~~ , ~~~~ \tilde{Q}^{ri} = 4 h^{ij} \partial_v \omega_j + 4 K^i_{~k} \omega^k \, , \\
            J^i_{(1)} &= 4 K^i_{~k} \omega^k - 4 h^{ij} \partial_v \omega_j + 8 \nabla_m K^{im} \, , \\
            W^{rv}_v &= -8 \dfrac{1}{\sqrt{h}}\partial_v (\sqrt{h} \, R_{rvrv}) - 4 \nabla_i (h^{ij} \partial_v \omega_j + K^i_{~k}\omega^k ) \\
                &~~ + 8 K^{ij} R_{vijr} - 8 \Bar{K}^{ij} \partial_v K_{ij} \, , \\
                \mathcal{A}_{(1)} &= 8 R_{vijr} K^{ij} - 8 \Bar{K}^{ij} \partial_v K_{ij} ~~~~, ~~~~  \mathcal{B}_{(0)} = 8 \Bar{K}^{ij} K_{ij}
             \end{split}
         \end{equation}
         Thus,
         \begin{equation}
             \begin{split}
                 \mathcal{U}_{(1)} &= - 8 R_{vijr} K^{ij} + 8 \Bar{K}^{ij} \partial_v K_{ij} + \dfrac{1}{\sqrt{h}}\partial_v ( \sqrt{h} \, 8 R_{rvrv} ) + \nabla_i ( 4h^{ij}\partial_v \omega_j + 4 K^i_{~k} \omega^k ) \\
                 & ~~ -8 \dfrac{1}{\sqrt{h}}\partial_v (\sqrt{h} \, R_{rvrv}) - 4 \nabla_i (h^{ij} \partial_v \omega_j + K^i_{~k}\omega^k )
                 + 8 K^{ij} R_{vijr} - 8 \Bar{K}^{ij} \partial_v K_{ij} \\
                 &= 0 \, , \\
                 \mathcal{V}_{(2)} &= \partial_v \left[ \dfrac{1}{\sqrt{h}} \partial_v [ \sqrt{h} ( - 8 R_{rvrv} - 8 \Bar{K}^{ij}K_{ij} ) ] + \nabla_i ( -8h^{ij}\partial_v\omega_j + 8 \nabla_m K^{im} ) \right] = 2 E_{vv}
             \end{split}
         \end{equation}

\paragraph{Gravitational Chern-Simons theories in $(2+1)$-dimensions: }
We have obtained
\begin{equation}
    2 v \tensor{E}{_v_v} |_{r=0} = \eval{\left(\frac{1}{\sqrt{h}} \tensor{\partial}{_\nu} \left(\sqrt{h} \tensor{Q}{^r^\nu} \right) - \tensor{\Theta}{^r} + \tensor{\Xi}{^r} \right)}_{r = 0} = \mathcal{U}_{(1)} + v \mathcal{V}_{(2)} \, .
\end{equation}
For this example we have the Lagrangian
\begin{equation}
    \mathcal{L} = \tensor{\epsilon}{^\lambda^\mu^\nu} \tensor{\Gamma}{^\rho_\lambda_\sigma} \left(\tensor{\partial}{_\mu} \tensor{\Gamma}{^\sigma_\rho_\nu} + \frac{2}{3} \tensor{\Gamma}{^\sigma_\mu_\tau} \tensor{\Gamma}{^\tau_\rho_\nu} \right) \, .
\end{equation}

As mentioned in equations eq.\eqref{eq:thetar2+1}, eq.\eqref{eq:xir2+1} and eq.\eqref{eq:Q2+1}, one can get
\begin{align}
    \tensor{\Theta}{^r}|_{r=0} &= \left(1 + v \tensor{\partial}{_v} \right) \left[\tensor{\epsilon}{^r^v^x} \frac{\tensor{\partial}{_v} h}{2h} \left(\omega + \frac{\tensor{\partial}{_x} h}{2h} \right) \right] + \mathcal{O}\left(\epsilon^2 \right) \, , 
    \\
    \tensor{\Xi}{^r}|_{r=0} &= - \dfrac{1}{\sqrt{h}} \tensor{\partial}{_v} \left( \sqrt{h} \tensor{\epsilon}{^r^v^x} \omega \right) \, , 
    \\
    \tensor{Q}{^r^v} &= - \tensor{\epsilon}{^r^v^x} \omega + v \tensor{\epsilon}{^r^v^x} \left( 2 \tensor{\partial}{_v} \omega + \dfrac{\omega}{2h}\tensor{\partial}{_v} h + \dfrac{1}{4h^2}(\tensor{\partial}{_v} h)(\tensor{\partial}{_x} h) \right) \, ,
    \\
    \tensor{Q}{^r^x} &= v \tensor{\partial}{_v} \left[ - \dfrac{2}{h} \tensor{\epsilon}{^r^v^x}\tensor{\partial}{_v} h \right] + \mathcal{O}(\epsilon^2) \, .
\end{align}

Using these values of $\tensor{Q}{^r^v}$ and $\tensor{Q}{^r^i}$, we can also get
\begin{equation}
    \begin{split}
        \frac{1}{\sqrt{h}} \tensor{\partial}{_\nu} \left(\sqrt{h} \tensor{Q}{^r^\nu} \right) &= \left(1 + v \tensor{\partial}{_v} \right) \left(\tensor{\epsilon}{^r^v^x} \left( 2 \tensor{\partial}{_v} \omega + \dfrac{\omega}{2h}\tensor{\partial}{_v} h + \dfrac{1}{4h^2}(\tensor{\partial}{_v} h)(\tensor{\partial}{_x} h) \right) \right) \\
        &\quad + \dfrac{1}{\sqrt{h}} \tensor{\partial}{_v} \left( \sqrt{h} [- \tensor{\epsilon}{^r^v^x} \omega ] \right) + v \tensor{\partial}{_v} \left(\frac{1}{\sqrt{h}} \tensor{\partial}{_x} \left(\sqrt{h} \left[- \dfrac{2}{h} \tensor{\epsilon}{^r^v^x}\tensor{\partial}{_v} h \right] \right)  \right) + \mathcal{O}(\epsilon^2) \, .
    \end{split}
\end{equation}

From this, we can see that
\begin{equation}
    \begin{split}
        \mathcal{A}_{(1)} &= \tensor{\epsilon}{^r^v^x} \frac{\tensor{\partial}{_v} h}{2h} \left(\omega + \frac{\tensor{\partial}{_x} h}{2h} \right) \qquad,\qquad \mathcal{B}_{(0)} = 0 \, ,
        \\
        \tensor{\tilde{q}}{^r^v} &= - \tensor{\epsilon}{^r^v^x} \omega \, , ~~~~~ \tensor{\tilde{q}}{^r^x} = 0 \, , ~~~~~ j^x_{(1)} = 0 \, , ~~~~~ \tensor{w}{_v^r^v} = 0 \, ,
        \\
        \tensor{\widetilde{Q}}{^r^v} &= - \tensor{\epsilon}{^r^v^x} \omega \, , ~~~~~ \tensor{\widetilde{Q}}{^r^x} = 0 \, , ~~~~~ W^{rx}_v = \tensor{\partial}{_v} \left[ - \dfrac{2}{h} \tensor{\epsilon}{^r^v^x}\tensor{\partial}{_v} h \right] + \mathcal{O}(\epsilon^2) \, \implies \, J^x_{(1)} = - \dfrac{2}{h} \tensor{\epsilon}{^r^v^x}\tensor{\partial}{_v} h \, , \\
        W^{rv}_v &= \tensor{\epsilon}{^r^v^x} \left( 2 \tensor{\partial}{_v} \omega + \dfrac{\omega}{2h}\tensor{\partial}{_v} h + \dfrac{1}{4h^2}(\tensor{\partial}{_v} h)(\tensor{\partial}{_x} h) \right) \, .
    \end{split}
\end{equation}

Thus, we get
\begin{align}
    \begin{split}
        \mathcal{U}_{(1)} &= \dfrac{1}{\sqrt{h}} \tensor{\partial}{_v} \left( \sqrt{h} [- \tensor{\epsilon}{^r^v^x} \omega ] \right) + \tensor{\epsilon}{^r^v^x} \left( 2 \tensor{\partial}{_v} \omega + \dfrac{\omega}{2h}\tensor{\partial}{_v} h + \dfrac{1}{4h^2}(\tensor{\partial}{_v} h)(\tensor{\partial}{_x} h) \right) \\
        &\quad - \tensor{\epsilon}{^r^v^x} \frac{\tensor{\partial}{_v} h}{2h} \left(\omega + \frac{\tensor{\partial}{_x} h}{2h} \right) - \dfrac{1}{\sqrt{h}} \tensor{\partial}{_v} \left( \sqrt{h} \tensor{\epsilon}{^r^v^x} \omega \right) \\
        &= 0 \, ,
    \end{split}
    \\
    \begin{split}
        \mathcal{V}_{(2)} &= \tensor{\partial}{_v} \left(\tensor{\epsilon}{^r^v^x} \left( 2 \tensor{\partial}{_v} \omega + \dfrac{\omega}{2h}\tensor{\partial}{_v} h + \dfrac{1}{4h^2}(\tensor{\partial}{_v} h)(\tensor{\partial}{_x} h) \right) + \frac{1}{\sqrt{h}} \tensor{\partial}{_x} \left(\sqrt{h} \left[- \dfrac{2}{h} \tensor{\epsilon}{^r^v^x} \tensor{\partial}{_v} h \right] \right) \right) \\
        &\quad - \tensor{\partial}{_v} \left(\tensor{\epsilon}{^r^v^x} \frac{\tensor{\partial}{_v} h}{2h} \left(\omega + \frac{\tensor{\partial}{_x} h}{2h} \right) \right) 
        \\
        &= \tensor{\partial}{_v} \left(\frac{1}{\sqrt{h}} \tensor{\partial}{_x} \left(\sqrt{h} \left[- \dfrac{2}{h} \tensor{\epsilon}{^r^v^x}\tensor{\partial}{_v} h \right] \right) + \frac{1}{\sqrt{h}} \tensor{\partial}{_v} \left(\sqrt{h} \left[2 \tensor{\epsilon}{^r^v^x} \omega \right] \right) \right)
        \\
        &= 2 \tensor{E}{_v_v} \, .
    \end{split}
\end{align}

\paragraph{Mixed gauge gravity Chern-Simons theories in $(4+1)$  dimensions:} 
This has the Lagrangian
\begin{equation}
    \mathcal{L} = 2 \tensor{\epsilon}{^\mu^\nu^\lambda^\rho^\sigma} \tensor{F}{_\mu_\nu} \tensor{\Gamma}{^\alpha_\lambda_\beta} \left(\frac{1}{2} \tensor{R}{^\beta_\alpha_\rho_\sigma} - \frac{1}{3} \tensor{\Gamma}{^\beta_\rho_\tau} \tensor{\Gamma}{^\tau_\alpha_\sigma} \right) \, .
\end{equation}

As mentioned in equations eq.\eqref{eq:thetar4+1}, eq.\eqref{eq:xir4+1}, eq.\eqref{eq:Qrnu4+1} and eq.\eqref{eq:Q4+1}, one can get
\begin{align}
    \tensor{\Theta}{^r}|_{r=0} &= - \tensor{U}{^r^v} \tensor{\partial}{_v} \left(v \tensor{A}{_v} + \Lambda \right) + \left(1 + v \tensor{\partial}{_v} \right) \left[\tensor{\epsilon}{^r^v^i^j^k} \tensor{F}{_j_k} \left(\tensor{\omega}{^l} \tensor{\partial}{_v} \tensor{h}{_i_l} + 2 \tensor{\Gamma}{^m_i_n} \tensor{\Gamma}{^n_m_v} \right) \right] + \mathcal{O}\left(\epsilon^2 \right) \, , 
    \\
    \tensor{\Xi}{^r}|_{r=0} &= \dfrac{1}{\sqrt{h}} \tensor{\partial}{_v} \left( \sqrt{h} [ -2 \tensor{\epsilon}{^r^v^i^j^k} F_{jk} \omega_i ] \right) + \dfrac{1}{\sqrt{h}} \tensor{\partial}{_i} \left( \sqrt{h} [ -4 \tensor{\epsilon}{^r^v^i^j^k} F_{vj} \omega_k  ] \right) \, , 
    \\
    \tensor{Q}{^r^v} &= - 2 \tensor{\epsilon}{^r^v^i^j^k} F_{jk} \omega_i + 2 v \left(\tensor{\epsilon}{^r^v^i^j^k}(4 F_{jk} R_{vrvi} + 4 F_{vi} R_{vrjk} + F_{jk} \Gamma^{\alpha}_{i\beta} \Gamma^{\beta}_{\alpha v} ) \right) - \tensor{U}{^r^v} \left(v \tensor{A}{_v} + \Lambda \right) \, ,
    \\
    \tensor{Q}{^r^i} &= - 4 \tensor{\epsilon}{^r^v^i^j^k} F_{vk} \omega_j - 4 v \tensor{\partial}{_v} \left(h^{im} \tensor{\epsilon}{^r^v^l^j^k} F_{jk} \tensor{\partial}{_v} h_{ml} \right) + \mathcal{O}\left(\epsilon^2 \right) \, .
\end{align}

Using these values of $\tensor{Q}{^r^v}$ and $\tensor{Q}{^r^i}$, we can also get
\begin{equation}
    \begin{split}
        \frac{1}{\sqrt{h}} \tensor{\partial}{_\nu} \left(\sqrt{h} \tensor{Q}{^r^\nu} \right) &= - \tensor{U}{^r^v} \tensor{\partial}{_v} \left(v \tensor{A}{_v} + \Lambda \right) + \dfrac{1}{\sqrt{h}} \tensor{\partial}{_v} \left( \sqrt{h} [2 \tensor{\epsilon}{^r^v^i^j^k} F_{jk} \omega_i ] \right) + \dfrac{1}{\sqrt{h}} \tensor{\partial}{_i} \left( \sqrt{h} [ 4 \tensor{\epsilon}{^r^v^i^j^k} F_{vj} \omega_k  ] \right)  \\
        &\quad + \left(1 + v \tensor{\partial}{_v} \right) \left(\tensor{\epsilon}{^r^v^i^j^k} \tensor{F}{_j_k} \left(\tensor{\omega}{^l} \tensor{\partial}{_v} \tensor{h}{_i_l} + 2 \tensor{\Gamma}{^m_i_n} \tensor{\Gamma}{^n_m_v} \right) \right) \\
        &\quad + v \tensor{\partial}{_v} \left(\frac{1}{\sqrt{h}} \tensor{\partial}{_i} \left(\sqrt{h} \left[8 \tensor{\epsilon}{^r^v^i^j^k} \tensor{\omega}{_j} \tensor{F}{_v_k} - 4 \tensor{\epsilon}{^r^v^m^j^k} \tensor{F}{_j_k} \tensor{h}{^i^n} \tensor{\partial}{_v} \tensor{h}{_n_m} \right] \right)  \right) \\
        &\quad + v \tensor{\partial}{_v} \left(\frac{1}{\sqrt{h}} \tensor{\partial}{_v} \left(\sqrt{h} \left[4 \tensor{\epsilon}{^r^v^i^j^k} \tensor{F}{_j_k} \tensor{\omega}{_i} \right] \right) \right) + \mathcal{O}\left(\epsilon^2 \right) \, .
    \end{split}
\end{equation}

From this, we can see that
\begin{equation}
    \begin{split}
        \mathcal{A}_{(1)} &= \tensor{\epsilon}{^r^v^i^j^k} \tensor{F}{_j_k} \left(\tensor{\omega}{^l} \tensor{\partial}{_v} \tensor{h}{_i_l} + 2 \tensor{\Gamma}{^m_i_n} \tensor{\Gamma}{^n_m_v} \right) \qquad,\qquad \mathcal{B}_{(0)} = 0 \, ,
        \\
        \tensor{\tilde{q}}{^r^v} &= -2 \tensor{\epsilon}{^r^v^i^j^k} F_{jk} \omega_i \, , ~~~~~ \tensor{\tilde{q}}{^r^i} = -4 \tensor{\epsilon}{^r^v^i^j^k} F_{vj} \omega_k \, , ~~~~~ j^i_{(1)} = 0 \, , ~~~~~ \tensor{w}{_v^r^v} = 0 \, ,
        \\
        \tensor{\widetilde{Q}}{^r^v} &= - 2 \tensor{\epsilon}{^r^v^i^j^k} F_{jk} \omega_i \, , ~~~~~ \tensor{\widetilde{Q}}{^r^i} = - 4 \tensor{\epsilon}{^r^v^i^j^k} F_{vk} \omega_j \, ,  \\
		W^{ri}_v &= - 4 \tensor{\partial}{_v} \left( h^{im}\tensor{\epsilon}{^r^v^l^j^k} F_{jk} \tensor{\partial}{_v} h_{ml} \right) + \mathcal{O}(\epsilon^2) \, 
		\implies \, J^i_{(1)} = -4 h^{im}\tensor{\epsilon}{^r^v^l^j^k} F_{jk} \tensor{\partial}{_v} h_{ml} \, , \\
        W^{rv}_v &= 8 \tensor{\epsilon}{^r^v^i^j^k}( F_{jk} R_{vrvi} + F_{vi} R_{vrjk} ) + 2 \tensor{\epsilon}{^r^v^i^j^k} F_{jk} \Gamma^{\alpha}_{i\beta} \Gamma^{\beta}_{\alpha v} \, .
    \end{split}
\end{equation}

Thus, we get
\begin{align}
    \begin{split}
        \mathcal{U}_{(1)} &= \dfrac{1}{\sqrt{h}} \tensor{\partial}{_v} \left( \sqrt{h} [2 \tensor{\epsilon}{^r^v^i^j^k} F_{jk} \omega_i ] \right) + \dfrac{1}{\sqrt{h}} \tensor{\partial}{_i} \left( \sqrt{h} [ 4 \tensor{\epsilon}{^r^v^i^j^k} F_{vj} \omega_k  ] \right) \\
        &\quad + \tensor{\epsilon}{^r^v^i^j^k} \tensor{F}{_j_k} \left(\tensor{\omega}{^l} \tensor{\partial}{_v} \tensor{h}{_i_l} + 2 \tensor{\Gamma}{^m_i_n} \tensor{\Gamma}{^n_m_v} \right) - \tensor{\epsilon}{^r^v^i^j^k} \tensor{F}{_j_k} \left(\tensor{\omega}{^l} \tensor{\partial}{_v} \tensor{h}{_i_l} + 2 \tensor{\Gamma}{^m_i_n} \tensor{\Gamma}{^n_m_v} \right) \\
        &\quad + \dfrac{1}{\sqrt{h}} \tensor{\partial}{_v} \left( \sqrt{h} [- 2 \tensor{\epsilon}{^r^v^i^j^k} F_{jk} \omega_i ] \right) + \dfrac{1}{\sqrt{h}} \tensor{\partial}{_i} \left( \sqrt{h} [- 4 \tensor{\epsilon}{^r^v^i^j^k} F_{vj} \omega_k  ] \right) \\
        &= 0 \, ,
    \end{split}
    \\
    \begin{split}
        \mathcal{V}_{(2)} &= \tensor{\partial}{_v} \left(\frac{1}{\sqrt{h}} \tensor{\partial}{_i} \left(\sqrt{h} \left[8 \tensor{\epsilon}{^r^v^i^j^k} \tensor{\omega}{_j} \tensor{F}{_v_k} - 4 \tensor{\epsilon}{^r^v^m^j^k} \tensor{F}{_j_k} \tensor{h}{^i^n} \tensor{\partial}{_v} \tensor{h}{_n_m} \right] \right) + \frac{1}{\sqrt{h}} \tensor{\partial}{_v} \left(\sqrt{h} \left[4 \tensor{\epsilon}{^r^v^i^j^k} \tensor{F}{_j_k} \tensor{\omega}{_i} \right] \right) \right) \\
        &\quad + \tensor{\partial}{_v} \left(\tensor{\epsilon}{^r^v^i^j^k} \tensor{F}{_j_k} \left(\tensor{\omega}{^l} \tensor{\partial}{_v} \tensor{h}{_i_l} + 2 \tensor{\Gamma}{^m_i_n} \tensor{\Gamma}{^n_m_v} \right) \right) - \tensor{\partial}{_v} \left(\tensor{\epsilon}{^r^v^i^j^k} \tensor{F}{_j_k} \left(\tensor{\omega}{^l} \tensor{\partial}{_v} \tensor{h}{_i_l} + 2 \tensor{\Gamma}{^m_i_n} \tensor{\Gamma}{^n_m_v} \right) \right) 
        \\
        &= \tensor{\partial}{_v} \left(\frac{1}{\sqrt{h}} \tensor{\partial}{_i} \left(\sqrt{h} \left[8 \tensor{\epsilon}{^r^v^i^j^k} \tensor{\omega}{_j} \tensor{F}{_v_k} - 4 \tensor{\epsilon}{^r^v^m^j^k} \tensor{F}{_j_k} \tensor{h}{^i^n} \tensor{\partial}{_v} \tensor{h}{_n_m} \right] \right) + \frac{1}{\sqrt{h}} \tensor{\partial}{_v} \left(\sqrt{h} \left[4 \tensor{\epsilon}{^r^v^i^j^k} \tensor{F}{_j_k} \tensor{\omega}{_i} \right] \right) \right)
        \\
        &= 2 \tensor{E}{_v_v} \, .
    \end{split}
\end{align}

\subsection{Details of the calculation of Noether charge for Chern-Simons theories}
\label{ap:noethercs}

In this Appendix, we give the details of the calculation of $\Xi^{\mu}$ from eq.\eqref{eq:varlxider} and the details of the calculation of $Q^{\mu\nu}$ from eq.\eqref{eq:cseom}. 

\subsubsection*{Evaluating $\Xi^{\mu}$:}

We start off with eq.\eqref{eq:varlxider}:
\begin{equation}\label{eq:deltaxiL}
    \delta_{\xi} L = \mathcal{L}_{\xi} L + \pdv{L}{\tensor{A}{_\mu}} D{_\mu} \Lambda + \pdv{L}{\tensor{\Gamma}{^\lambda_\mu_\nu}} \tensor{\partial}{_\mu}\tensor{\partial}{_\nu} \tensor{\xi}{^\lambda} \, .
\end{equation}
The subscript $\xi$ is used to indicate the variation under the combined diffeomorphism and $U(1)$ gauge transformations eq.\eqref{eq:diffeo}. We can rearrange the non-diffeomorphism/$U(1)$ gauge structures of eq.\eqref{eq:deltaxiL}:
\begin{equation}\label{eq:partialLA}
    \pdv{L}{\tensor{A}{_\mu}} D{_\mu} \Lambda = D{_\mu} \left(\pdv{L}{\tensor{A}{_\mu}} \Lambda \right) - \Lambda D{_\mu} \pdv{L}{\tensor{A}{_\mu}} \, .
\end{equation}
A similar integration by parts manipulation for the other term of eq.\eqref{eq:deltaxiL} gives
\begin{equation}\label{eq:partialLG}
    \begin{split}
        \pdv{L}{\tensor{\Gamma}{^\lambda_\mu_\nu}} \partial_{\mu}\partial_{\nu} \tensor{\xi}{^\lambda} &= \tensor{\partial}{_\mu} \left(\pdv{L}{\tensor{\Gamma}{^\lambda_\mu_\nu}} \tensor{\partial}{_\nu} \tensor{\xi}{^\lambda} \right) - \left(\tensor{\partial}{_\nu} \tensor{\xi}{^\lambda} \right) \left(\tensor{\partial}{_\mu} \pdv{L}{\tensor{\Gamma}{^\lambda_\mu_\nu}} \right) \\
    &= D{_\mu} \left(\pdv{L}{\tensor{\Gamma}{^\lambda_\mu_\nu}} \tensor{\partial}{_\nu} \tensor{\xi}{^\lambda} \right) - \tensor{\Gamma}{^\tau_\mu_\tau} \pdv{L}{\tensor{\Gamma}{^\lambda_\mu_\nu}} \tensor{\partial}{_\nu} \tensor{\xi}{^\lambda} - \left(\tensor{\partial}{_\nu} \tensor{\xi}{^\lambda} \right) \left(\tensor{\partial}{_\mu} \pdv{L}{\tensor{\Gamma}{^\lambda_\mu_\nu}} \right) \\
    &= D{_\mu} \left(\pdv{L}{\tensor{\Gamma}{^\lambda_\mu_\nu}} \tensor{\partial}{_\nu} \tensor{\xi}{^\lambda} \right) - \frac{1}{\sqrt{-g}}\left(\tensor{\partial}{_\nu} \tensor{\xi}{^\lambda} \right) \tensor{\partial}{_\mu} \left(\sqrt{-g} \pdv{L}{\tensor{\Gamma}{^\lambda_\mu_\nu}} \right) \\
    &= D{_\mu} \left(\pdv{L}{\tensor{\Gamma}{^\lambda_\mu_\nu}} \tensor{\partial}{_\nu} \tensor{\xi}{^\lambda} \right) - \tensor{\partial}{_\mu} \left[\frac{\tensor{\xi}{^\lambda}}{\sqrt{-g}} \tensor{\partial}{_\nu} \left(\sqrt{-g} \pdv{L}{\tensor{\Gamma}{^\lambda_\mu_\nu}} \right) \right] \\ & \quad \quad + \tensor{\xi}{^\lambda} \tensor{\partial}{_\mu} \left[\frac{1}{\sqrt{-g}} \tensor{\partial}{_\nu} \left(\sqrt{-g} \pdv{L}{\tensor{\Gamma}{^\lambda_\mu_\nu}} \right) \right] \\
    &= D{_\mu} \left(\pdv{L}{\tensor{\Gamma}{^\lambda_\mu_\nu}} \tensor{\partial}{_\nu} \tensor{\xi}{^\lambda} - \frac{1}{\sqrt{-g}} \tensor{\xi}{^\lambda} \tensor{\partial}{_\nu} \left(\sqrt{-g} \pdv{L}{\tensor{\Gamma}{^\lambda_\mu_\nu}} \right) \right) + \frac{\tensor{\xi}{^\lambda}}{\sqrt{-g}} \tensor{\partial}{_\mu} \partial_{\nu} \left(\sqrt{-g} \pdv{L}{\tensor{\Gamma}{^\lambda_\mu_\nu}} \right) \, .
    \end{split}
\end{equation}
Substituting eq.\eqref{eq:partialLA} and eq.\eqref{eq:partialLG} in eq.\eqref{eq:deltaxiL}, we get eq.\eqref{eq:finaldiffvarL}: 
\begin{equation}
    \delta_{\xi}\left(\sqrt{-g} L \right) = \sqrt{-g} D{_\mu} \left(\tensor{\xi}{^\mu} L \right) + \sqrt{-g} D{_\mu} \tensor{\Xi}{^\mu} - \sqrt{-g} \Lambda D{_\mu} \pdv{L}{\tensor{A}{_\mu}} + \tensor{\xi}{^\lambda} \tensor{\partial}{_\mu}\partial_{\nu} \left(\sqrt{-g} \pdv{L}{\tensor{\Gamma}{^\lambda_\mu_\nu}} \right) \, .
\end{equation}

\subsubsection*{Evaluating $Q^{\mu\nu}$:}

We start off with the definition of $Q^{\mu\nu}$ in eq.\eqref{eq:csnoethercharge}:
\begin{equation}\label{eq:Jmu}
    \tensor{J}{^\mu} = 2 \tensor{E}{^\mu^\nu} \tensor{\xi}{_\nu} + \tensor{G}{^\mu} \left(\tensor{A}{_\nu} \tensor{\xi}{^\nu} + \Lambda \right) + \tensor{\Theta}{^\mu} - \tensor{\xi}{^\mu} L - \tensor{\Xi}{^\mu} \, .
\end{equation}
The current $J^{\mu}$ is conserved by definition and it gives $Q^{\mu\nu}$:
\begin{equation}
    \tensor{J}{^\mu} = D{_\nu} \tensor{Q}{^\mu^\nu} \, .
\end{equation}
We now substitute eq.\eqref{eq:diffeo} for $\delta g_{\alpha\beta}$ and $\delta A_{\nu}$ in $\Theta^{\mu}$ of eq.\eqref{eq:cseom} to get
\begin{equation}
    \begin{split}
        \tensor{\Theta}{^\mu} &= \left(\tensor{g}{^\alpha^\lambda} \pdv{L}{\tensor{R}{^\lambda_\beta_\mu_\nu}} + \tensor{g}{^\alpha^\lambda} \pdv{L}{\tensor{R}{^\lambda_\nu_\mu_\beta}} - \tensor{g}{^\nu^\lambda} \pdv{L}{\tensor{R}{^\lambda_\beta_\mu_\alpha}} \right) D{_\nu} \left(D{_\alpha} \tensor{\xi}{_\beta} + D{_\beta} \tensor{\xi}{_\alpha} \right) \\
    &\quad + 2 \pdv{L}{\tensor{F}{_\mu_\nu}} \left[\tensor{\xi}{^\lambda} \tensor{F}{_\lambda_\nu} + D{_\nu} \left(\tensor{A}{_\lambda} \tensor{\xi}{^\lambda} + \Lambda \right) \right] + 2 \tensor{S}{^\mu^\alpha^\beta} D{_\alpha} \tensor{\xi}{_\beta}
    \end{split} \, .
\end{equation}
Making further simplifications, we get
\begin{equation}
    \begin{split}
        \tensor{\Theta}{^\mu} &= D{_\nu} \left(2 \pdv{L}{\tensor{R}{^\alpha_\beta_\mu_\nu}} D{_\beta} \tensor{\xi}{^\alpha} \right) - 2 \left(D{_\beta} \tensor{\xi}{^\alpha} \right) D{_\nu} \pdv{L}{\tensor{R}{^\alpha_\beta_\mu_\nu}} + 2 \pdv{L}{\tensor{R}{^\alpha_\beta_\nu_\mu}} \tensor{R}{^\alpha_\beta_\nu_\eta} \tensor{\xi}{^\eta} \\
    &\quad + 2 \pdv{L}{\tensor{F}{_\mu_\nu}} \left[\tensor{\xi}{^\lambda} \tensor{F}{_\lambda_\nu} + D{_\nu} \left(\tensor{A}{_\lambda} \tensor{\xi}{^\lambda} + \Lambda \right) \right] + 2 \tensor{S}{^\mu^\alpha^\beta} D{_\alpha} \tensor{\xi}{_\beta} \, .
    \end{split}
\end{equation}

Collecting the terms in the expression of $\tensor{J}{^\mu}$ of eq.\eqref{eq:Jmu} dependent on $\Lambda$, we have
\begin{equation}
    \tensor{G}{^\mu} \Lambda - \pdv{L}{\tensor{A}{_\mu}} \Lambda + 2 \pdv{L}{\tensor{F}{_\mu_\nu}} D{_\nu} \Lambda = 2 D{_\nu} \left(\pdv{L}{\tensor{F}{_\mu_\nu}} \Lambda \right) \, .
\end{equation}
This gives rise to the ``gauge" contributions (the terms with the partial derivatives with respect to the gauge fields) of eq.\eqref{eq:Jmu}:
\begin{equation}\label{eq:Jmugauge}
    J^{\mu}_{\text{gauge}} = D{_\nu} \left(2 \pdv{L}{\tensor{F}{_\mu_\nu}} \left(\tensor{A}{_\lambda} \tensor{\xi}{^\lambda} + \Lambda \right) \right) + \tensor{\xi}{^\lambda} \left(\pdv{L}{\tensor{A}{_\mu}} \tensor{A}{_\lambda} + 2 \pdv{L}{\tensor{F}{_\mu_\nu}} \tensor{F}{_\lambda_\nu} \right) \, .
\end{equation}

The ``metric" contribution (the terms with the partial derivatives with respect to the gravitational fields) of $J^{\mu}$ in eq.\eqref{eq:Jmu} is:
\begin{equation}
    \begin{split}
        J^{\mu}_{\text{metric}} &= 2 \tensor{E}{^\mu^\alpha} \tensor{\xi}{_\alpha} - \tensor{\xi}{^\mu} L + 2 \tensor{S}{^\mu^\nu^\alpha} D{_\nu} \tensor{\xi}{_\alpha} - \pdv{L}{\tensor{\Gamma}{^\lambda_\mu_\nu}} \tensor{\partial}{_\nu} \tensor{\xi}{^\lambda} + \frac{1}{\sqrt{-g}} \tensor{\xi}{^\lambda} \tensor{\partial}{_\nu} \left(\sqrt{-g} \pdv{L}{\tensor{\Gamma}{^\lambda_\mu_\nu}} \right) \\
    &\quad + \left(\tensor{g}{^\alpha^\lambda} \pdv{L}{\tensor{R}{^\lambda_\beta_\mu_\nu}} + \tensor{g}{^\alpha^\lambda} \pdv{L}{\tensor{R}{^\lambda_\nu_\mu_\beta}} - \tensor{g}{^\nu^\lambda} \pdv{L}{\tensor{R}{^\lambda_\beta_\mu_\alpha}} \right) D{_\nu} \left(D{_\alpha} \tensor{\xi}{_\beta} + D{_\beta} \tensor{\xi}{_\alpha} \right)
    \\
    &= 2 \tensor{S}{^\mu^\nu^\alpha} D{_\nu} \tensor{\xi}{_\alpha} - 2 \tensor{\xi}{_\alpha} D{_\nu} \tensor{S}{^\nu^\mu^\alpha} + D{_\nu} \left(2 \pdv{L}{\tensor{R}{^\alpha_\beta_\mu_\nu}} D{_\beta} \tensor{\xi}{^\alpha} \right) - 2 \left(D{_\beta} \tensor{\xi}{^\alpha} \right) D{_\nu} \pdv{L}{\tensor{R}{^\alpha_\beta_\mu_\nu}} \\
    &\quad + 2 \pdv{L}{\tensor{R}{^\alpha_\beta_\nu_\mu}} \tensor{R}{^\alpha_\beta_\nu_\eta} \tensor{\xi}{^\eta} - \pdv{L}{\tensor{\Gamma}{^\lambda_\mu_\nu}} \tensor{\partial}{_\nu} \tensor{\xi}{^\lambda} + \frac{1}{\sqrt{-g}} \tensor{\xi}{^\lambda} \tensor{\partial}{_\nu} \left(\sqrt{-g} \pdv{L}{\tensor{\Gamma}{^\lambda_\mu_\nu}} \right) - 2 \tensor{g}{^\mu^\nu} \pdv{L}{\tensor{g}{^\nu^\lambda}} \tensor{\xi}{^\lambda} 
    \\
    &= D{_\nu} \left(2 \pdv{L}{\tensor{R}{^\alpha_\beta_\mu_\nu}} D{_\beta} \tensor{\xi}{^\alpha} + \tensor{\xi}{_\alpha} \left(\tensor{S}{^\mu^\nu^\alpha} - \tensor{S}{^\nu^\mu^\alpha} \right) \right) + \left(\tensor{S}{^\mu^\nu^\alpha} + \tensor{S}{^\nu^\mu^\alpha} \right) D{_\nu} \tensor{\xi}{_\alpha} - \tensor{\xi}{_\alpha} D{_\nu} \left(\tensor{S}{^\mu^\nu^\alpha} + \tensor{S}{^\nu^\mu^\alpha} \right) \\
    &\quad - 2 \left(D{_\beta} \tensor{\xi}{^\alpha} \right) D{_\nu} \pdv{L}{\tensor{R}{^\alpha_\beta_\mu_\nu}} + 2 \pdv{L}{\tensor{R}{^\alpha_\beta_\nu_\mu}} \tensor{R}{^\alpha_\beta_\nu_\eta} \tensor{\xi}{^\eta} - \pdv{L}{\tensor{\Gamma}{^\lambda_\mu_\nu}} \tensor{\partial}{_\nu} \tensor{\xi}{^\lambda} \\
    &\quad + \frac{1}{\sqrt{-g}} \tensor{\xi}{^\lambda} \tensor{\partial}{_\nu} \left(\sqrt{-g} \pdv{L}{\tensor{\Gamma}{^\lambda_\mu_\nu}} \right) - 2 \tensor{g}{^\mu^\nu} \pdv{L}{\tensor{g}{^\nu^\lambda}} \tensor{\xi}{^\lambda} \, . 
    \end{split}
\end{equation}
We use
\begin{equation}
    \tensor{S}{^\mu^\nu^\alpha} + \tensor{S}{^\nu^\mu^\alpha} = \tensor{g}{^\alpha^\lambda} \left(\pdv{L}{\tensor{\Gamma}{^\lambda_\nu_\mu}} + D{_\beta} \left[\pdv{L}{\tensor{R}{^\lambda_\nu_\mu_\beta}} + \pdv{L}{\tensor{R}{^\lambda_\mu_\nu_\beta}} \right] \right) \, ,
\end{equation}
to simplify $J^{\mu}_{\text{metric}}$ as
\begin{equation}\label{eq:Jmumetric}
    \begin{split}
         J^{\mu}_{\text{metric}} &= D{_\nu} \left(2 \pdv{L}{\tensor{R}{^\alpha_\beta_\mu_\nu}} D{_\beta} \tensor{\xi}{^\alpha} + \tensor{\xi}{_\alpha} \left(\tensor{S}{^\mu^\nu^\alpha} - \tensor{S}{^\nu^\mu^\alpha} \right) + \tensor{\xi}{^\lambda} D{_\beta} \left[\pdv{L}{\tensor{R}{^\lambda_\mu_\nu_\beta}} - \pdv{L}{\tensor{R}{^\lambda_\nu_\mu_\beta}} \right] \right) \\
    &\quad + \tensor{\xi}{^\lambda} \left(2 \pdv{L}{\tensor{R}{^\alpha_\beta_\nu_\mu}} \tensor{R}{^\alpha_\beta_\nu_\lambda} - \tensor{R}{^\mu_\eta_\alpha_\beta} \pdv{L}{\tensor{R}{^\lambda_\eta_\alpha_\beta}} - \tensor{R}{_\alpha_\beta_\lambda^\eta} \pdv{L}{\tensor{R}{^\eta_\mu_\alpha_\beta}} \right) \\
    &\quad + \tensor{\xi}{^\lambda} \left(2 \pdv{L}{\tensor{\Gamma}{^\eta_\mu_\nu}} \tensor{\Gamma}{^\eta_\nu_\lambda} - \tensor{\Gamma}{^\mu_\nu_\eta} \pdv{L}{\tensor{\Gamma}{^\lambda_\eta_\nu}} - 2 \tensor{g}{^\mu^\nu} \pdv{L}{\tensor{g}{^\nu^\lambda}} \right) \, .
    \end{split}
\end{equation}

Substituting eq.\eqref{eq:Jmugauge} and eq.\eqref{eq:Jmumetric} in eq.\eqref{eq:Jmu}, we get
\begin{equation}\label{eq:Jmufinal}
    \tensor{J}{^\mu} = D{_\nu} \tensor{Q}{^\mu^\nu} + \tensor{\xi}{^\lambda} \tensor{B}{^\mu_\lambda} \, ,
\end{equation}
where
\begin{equation}
    \begin{split}
        \tensor{Q}{^\mu^\nu} &= 2 \pdv{L}{\tensor{F}{_\mu_\nu}} \left(\tensor{A}{_\lambda} \tensor{\xi}{^\lambda} + \Lambda \right) + 2 \pdv{L}{\tensor{R}{^\alpha_\beta_\mu_\nu}} D{_\beta} \tensor{\xi}{^\alpha} + \tensor{\xi}{_\alpha} \left(\tensor{S}{^\mu^\nu^\alpha} - \tensor{S}{^\nu^\mu^\alpha} \right)  + \tensor{\xi}{^\lambda} D{_\beta} \left[\pdv{L}{\tensor{R}{^\lambda_\mu_\nu_\beta}} - \pdv{L}{\tensor{R}{^\lambda_\nu_\mu_\beta}} \right] \, .
    \end{split}
\end{equation}
This agrees with the result of eq.\eqref{eq:defQ}. 

We also have
\begin{equation}
    \begin{split}
        \tensor{B}{^\mu_\lambda} &= 2 \pdv{L}{\tensor{R}{^\alpha_\beta_\nu_\mu}} \tensor{R}{^\alpha_\beta_\nu_\lambda} - \tensor{R}{^\mu_\eta_\alpha_\beta} \pdv{L}{\tensor{R}{^\lambda_\eta_\alpha_\beta}} - \tensor{R}{_\alpha_\beta_\lambda^\eta} \pdv{L}{\tensor{R}{^\eta_\mu_\alpha_\beta}} + 2 \pdv{L}{\tensor{\Gamma}{^\eta_\mu_\nu}} \tensor{\Gamma}{^\eta_\nu_\lambda} - \tensor{\Gamma}{^\mu_\nu_\eta} \pdv{L}{\tensor{\Gamma}{^\lambda_\eta_\nu}} \\
    &\quad + \pdv{L}{\tensor{A}{_\mu}} \tensor{A}{_\lambda} + 2 \pdv{L}{\tensor{F}{_\mu_\nu}} \tensor{F}{_\lambda_\nu} - 2 \tensor{g}{^\mu^\nu} \pdv{L}{\tensor{g}{^\nu^\lambda}} \, .
    \end{split}
\end{equation}
Taking the divergence of eq.\eqref{eq:Jmufinal} and using the fact that $J^{\mu}$ is divergenceless by construction, we have
\begin{equation}\label{eq:Bmulzero}
    \tensor{B}{^\mu_\lambda} = 0 \, ,
\end{equation}
since $\xi$ is arbitrary. If we had worked with differential forms from the start like \cite{Bonora:2011gz,Azeyanagi:2014sna}, we would have obtained eq.\eqref{eq:Bmulzero} automatically. We see that for explicit theories considered in \S\ref{sec:proofverf}, eq.\eqref{eq:Bmulzero} is satisfied trivially by using certain identities. Since we work with the component form of the Lagrangian in eq.\eqref{eq:cslagrangian}, it is difficult to see those identities which typically involve the $\epsilon^{\alpha_1 \dots \alpha_k}$ tensor in the definition of a CS Lagrangian.

\subsection{Details of the gauge field dependence of the Chern-Simons Lagrangian}
\label{ap:gaugedepcslagrangian}

In this Appendix, we give the details in the steps involved in starting from eq.\eqref{eq:lncs} to reach eq.\eqref{eq:finalcslagrangian}. Computing the derivative of eq.\eqref{eq:lncs} gives
\begin{equation}
\begin{split}
    \pdv{\tensor{L}{_n}}{\tensor{A}{_\mu}} &= \sum_{i=1}^n \tensor{\mathcal{L}}{^{\nu_1}^\cdots^{\nu_{i-1}}^\mu^{\nu_{i+1}}^\cdots^{\nu_n}} \prod_{j=1, j\neq i}^n \tensor{A}{_{\nu_j}} = n \tensor{\mathcal{L}}{^\mu^{\nu_1}^\cdots^{\nu_{n-1}}} \prod_{i=1}^{n-1} \tensor{A}{_{\nu_i}} \, ,
\end{split}
\end{equation}
where the last equality follows because $\tensor{\mathcal{L}}{^{\nu_1}^\cdots^{\nu_n}}$ is totally symmetric in all its indices. This implies that
\begin{equation}
\begin{split}
    D{_\mu} \pdv{\tensor{L}{_n}}{\tensor{A}{_\mu}} &= n \left(\prod_{i=1}^{n-1} \tensor{A}{_{\nu_i}} \right) \left(D{_\mu} \tensor{\mathcal{L}}{^\mu^{\nu_1}^\cdots^{\nu_{n-1}}} \right) + n \tensor{\mathcal{L}}{^\mu^{\nu_1}^\cdots^{\nu_{n-1}}} \sum_{j=1}^{n-1} \left(\left(D{_\mu} \tensor{A}{_{\nu_j}} \right) \left(\prod_{i=1, i\neq j}^{n-1} \tensor{A}{_{\nu_i}} \right) \right) \\
    &= n \left(\prod_{i=1}^{n-1} \tensor{A}{_{\nu_i}} \right) \left(D{_\mu} \tensor{\mathcal{L}}{^\mu^{\nu_1}^\cdots^{\nu_{n-1}}} \right) + n \left(D{_\mu} \tensor{A}{_\nu} \right) \left(\sum_{j=1}^{n-1} \tensor{\mathcal{L}}{^\mu^{\nu_1}^\cdots^{\nu_{j-1}}^\nu^{\nu_j}^\cdots^{\nu_{n-2}}} \left(\prod_{i=1}^{n-2} \tensor{A}{_{\nu_i}} \right) \right) \\
    &= n \left(\prod_{i=1}^{n-1} \tensor{A}{_{\nu_i}} \right) \left(D{_\mu} \tensor{\mathcal{L}}{^\mu^{\nu_1}^\cdots^{\nu_{n-1}}} \right) + n(n-1) \tensor{\mathcal{L}}{^\mu^\nu^{\nu_1}^\cdots^{\nu_{n-2}}} \left(D{_\mu} \tensor{A}{_\nu} \right) \left(\prod_{i=1}^{n-2} \tensor{A}{_{\nu_i}} \right) \, ,
\end{split}
\end{equation}
where the last equality follows because $\tensor{\mathcal{L}}{^{\nu_1}^\cdots^{\nu_n}}$ is totally symmetric in all its indices. The Bianchi identity eq.\eqref{eq:bianchics} implies that 
\begin{equation}\label{eq:csgaugeconds}
    D{_\mu} \pdv{\tensor{L}{_n}}{\tensor{A}{_\mu}} = 0 \implies n D{_\mu} \tensor{\mathcal{L}}{^\mu^{\nu_1}^\cdots^{\nu_{n-1}}} = 0 \quad \text{and} \quad n(n-1) \tensor{\mathcal{L}}{^\mu^\nu^{\nu_1}^\cdots^{\nu_{n-2}}} = 0 \, ,
\end{equation}
as $\tensor{A}{_\lambda}$ and $D{_{(\mu}} A_{\nu)}$ are independent functions, and $\tensor{\mathcal{L}}{^\mu^\nu^{\nu_1}^\cdots^{\nu_{n-2}}}$ is totally symmetric. But, the second condition of eq.\eqref{eq:csgaugeconds} immediately implies that for $n \ge 2$, $\tensor{L}{_n} = 0$ in eq.\eqref{eq:lncs}. Thus, we need only concern ourselves with $\tensor{L}{_0}, \tensor{L}{_1}$ of eq.\eqref{eq:lncs}. 

The first condition of eq.\eqref{eq:csgaugeconds} will allow us to greatly restrict the form of $\tensor{L}{_1}$. Firstly, $\tensor{\mathcal{L}}{^\mu}$ is gauge invariant, and thus it must be of the form
\begin{equation}\label{eq:l1mudef}
    \tensor{\mathcal{L}}{^\mu} = \sum_n \tensor{\mathcal{C}}{^\mu^{\rho_1}^{\sigma_1}^\cdots^{\rho_n}^{\sigma_n}} \prod_{i=1}^n \tensor{F}{_{\rho_i}_{\sigma_i}} \, ,
\end{equation}
where $\tensor{\mathcal{C}}{^\mu^{\rho_1}^{\sigma_1}^\cdots^{\rho_n}^{\sigma_n}}$ is independent of the $U(1)$ gauge field, antisymmetric in $(\rho_i, \sigma_i)$ for all $i$, and symmetric in pairs of $(\rho_i, \sigma_i)$ and $(\rho_j, \sigma_j)$ for any pair $(i, j)$. This implies that
\begin{equation}
\begin{split}
    D{_\mu} \tensor{\mathcal{L}}{^\mu} &= \sum_n \left(D{_\mu} \tensor{\mathcal{C}}{^\mu^{\rho_1}^{\sigma_1}^\cdots^{\rho_n}^{\sigma_n}} \right) \left(\prod_{i=1}^n \tensor{F}{_{\rho_i}_{\sigma_i}} \right) + \sum_n \sum_{j=1}^n \tensor{\mathcal{C}}{^\mu^{\rho_1}^{\sigma_1}^\cdots^{\rho_n}^{\sigma_n}} \left(D{_\mu} \tensor{F}{_{\rho_j}_{\sigma_j}} \right) \left(\prod_{i=1, i\neq j}^n \tensor{F}{_{\rho_i}_{\sigma_i}} \right) \\
    &= \sum_n \left(D{_\mu} \tensor{\mathcal{C}}{^\mu^{\rho_1}^{\sigma_1}^\cdots^{\rho_n}^{\sigma_n}} \right) \left(\prod_{i=1}^n \tensor{F}{_{\rho_i}_{\sigma_i}} \right) + \sum_n n \tensor{\mathcal{C}}{^\mu^\nu^\lambda^{\rho_1}^{\sigma_1}^\cdots^{\rho_{n-1}}^{\sigma_{n-1}}} \left(D{_\mu} \tensor{F}{_\nu_\lambda} \right) \left(\prod_{i=1}^{n-1} \tensor{F}{_{\rho_i}_{\sigma_i}} \right) \, .
\end{split}
\end{equation}
Now, the first condition of eq.\eqref{eq:csgaugeconds} on $\tensor{L}{_1}$ gives $D{_\mu} \tensor{\mathcal{L}}{^\mu} = 0$. As $\tensor{F}{_\rho_\sigma}, D{_\nu} \tensor{F}{_\rho_\sigma}$ are independent functions, this implies that
\begin{equation}\label{eq:lmugaugeconds}
    D{_\mu} \tensor{\mathcal{C}}{^\mu^{\rho_1}^{\sigma_1}^\cdots^{\rho_n}^{\sigma_n}} = 0 \qquad\text{and}\qquad \tensor{\mathcal{C}}{^\mu^\nu^\lambda^{\rho_1}^{\sigma_1}^\cdots^{\rho_{n-1}}^{\sigma_{n-1}}} \text{ is antisymmetric in } (\mu, \nu, \lambda) \, .
\end{equation}
But, by the properties of $\tensor{\mathcal{C}}{^\mu^\nu^\lambda^{\rho_1}^{\sigma_1}^\cdots^{\rho_{n-1}}^{\sigma_{n-1}}}$ in eq.\eqref{eq:l1mudef}, the second condition of eq.\eqref{eq:lmugaugeconds} immediately implies that we must have $\tensor{\mathcal{C}}{^\mu^\nu^\lambda^{\rho_1}^{\sigma_1}^\cdots^{\rho_{n-1}}^{\sigma_{n-1}}}$ totally antisymmetric in all of its indices. Due to this, the first condition of eq.\eqref{eq:lmugaugeconds} implies that $\tensor{\mathcal{C}}{^\mu^{\rho_1}^{\sigma_1}^\cdots^{\rho_n}^{\sigma_n}}$ of eq.\eqref{eq:l1mudef} has two possible choices
\begin{equation}\label{eq:lmufinal}
    \tensor{\mathcal{C}}{^\mu^{\rho_1}^{\sigma_1}^\cdots^{\rho_n}^{\sigma_n}} = 
    \begin{cases}
        D{_\nu} \tensor{\mathcal{B}}{^\nu^\mu^{\rho_1}^{\sigma_1}^\cdots^{\rho_n}^{\sigma_n}} & (i) ~~ 2n+1 < D \\
        a_g \, \tensor{\epsilon}{^\mu^{\rho_1}^{\sigma_1}^\cdots^{\rho_n}^{\sigma_n}} & (ii) ~~ 2n+1 = D
    \end{cases}
\end{equation}
where $D$ is the dimension of spacetime and $\tensor{\mathcal{B}}{^\nu^\mu^{\rho_1}^{\sigma_1}^\cdots^{\rho_n}^{\sigma_n}}$ is independent of the $U(1)$ gauge field, antisymmetric in $(\mu, \nu)$, and thus antisymmetric in all indices, and $a_g$ is a constant as $D{_\mu} a_g = 0$. For Choice $(i)$, eq.\eqref{eq:l1mudef} and eq.\eqref{eq:lmufinal} implies
\begin{equation}\label{eq:lmuamu}
    \begin{split}
        \tensor{\mathcal{L}}{^\mu} \tensor{A}{_\mu} &= \tensor{A}{_\nu} \sum_{n=0}^{N} \left(D{_\mu} \tensor{\mathcal{B}}{^\mu^\nu^{\rho_1}^{\sigma_1}^\cdots^{\rho_n}^{\sigma_n}} \right) \left(\prod_{i=1}^n \tensor{F}{_{\rho_i}_{\sigma_i}} \right) = \tensor{A}{_\nu} D{_\mu} \left[\sum_{n=0}^{N} \tensor{\mathcal{B}}{^\mu^\nu^{\rho_1}^{\sigma_1}^\cdots^{\rho_n}^{\sigma_n}} \left(\prod_{i=1}^n \tensor{F}{_{\rho_i}_{\sigma_i}} \right) \right] \\
    &= D{_\mu} \left[\tensor{A}{_\nu} \sum_{n=0}^{N} \tensor{\mathcal{B}}{^\mu^\nu^{\rho_1}^{\sigma_1}^\cdots^{\rho_n}^{\sigma_n}} \left(\prod_{i=1}^n \tensor{F}{_{\rho_i}_{\sigma_i}} \right) \right] - \frac{1}{2} \left(\tensor{F}{_\mu_\nu} \right) \sum_{n=0}^{N} \tensor{\mathcal{B}}{^\mu^\nu^{\rho_1}^{\sigma_1}^\cdots^{\rho_n}^{\sigma_n}} \left(\prod_{i=1}^n \tensor{F}{_{\rho_i}_{\sigma_i}} \right)
    \end{split}
\end{equation}
and eq.\eqref{eq:lmufinal} implies we have in general an extra term (when the number of indices are equal to the dimension of the spacetime) $a_g \tensor{\epsilon}{^\mu^{\rho_1}^{\sigma_1}^\cdots^{\rho_n}^{\sigma_n}} \tensor{A}{_\mu} \prod_{i=1}^n \tensor{F}{_{\rho_i}_{\sigma_i}}$ and it is clear that this term can only exist in odd dimensions. Choice 
$(ii)$ of eq.\eqref{eq:lmufinal} is non-trivial only when $n \geq 1$. We can now use eq.\eqref{eq:lmuamu} and eq.\eqref{eq:lmufinal} in eq.\eqref{eq:l1mudef} to arrive at eq.\eqref{eq:finalcslagrangian}. Though choice $(i)$ of eq.\eqref{eq:lmufinal} seemingly suggests that this term exists only in even dimensions, it should be noted that there are hidden indices in the gravitational sector that haven't been accounted for. CS Lagrangians are defined in odd dimensions only. We didn't need to consider the gravitational Bianchi identity of eq.\eqref{eq:bianchics} for arguing the entropy current structure.

\subsection{The total derivative term of the Chern-Simons Lagrangian}
\label{sec:totalterm}

In this Appendix, we will analyze the total derivative term of eq.\eqref{eq:finalcslagrangian} which is of the form $D_{\mu} \mathcal{L}^{\mu}$. While this doesn't contribute to the equations of motion $E_{\mu\nu}$ and $G_{\mu}$, it does contribute to $\Theta^{\mu}$ and $Q^{\mu\nu}$. Thus, it may potentially contribute to the entropy current through eq.\eqref{eq:main}. So we carefully analyze this term here. The variation of the action gives
\begin{equation}
    \begin{split}
    \delta \left(\sqrt{-g} D{_\mu} \tensor{\mathcal{L}}{^\mu} \right) = \tensor{\partial}{_\mu} \delta \left(\sqrt{-g} \tensor{\mathcal{L}}{^\mu} \right) = \sqrt{-g} D{_\mu} \left(\delta \tensor{\mathcal{L}}{^\mu} + \frac{1}{2} \tensor{\mathcal{L}}{^\mu} \tensor{g}{^\alpha^\beta} \delta \tensor{g}{_\alpha_\beta} \right) \, .
\end{split}
\end{equation}
Following the analysis of \S\ref{sec:noetherchargecs}, this can be written in terms of the partial derivatives of $\tensor{\mathcal{L}}{^\mu}$ as
\begin{equation}\label{eq:twowaysthetatotal}
    \begin{split}
        \delta \tensor{\mathcal{L}}{^\mu} + \frac{1}{2} \tensor{\mathcal{L}}{^\mu} \tensor{g}{^\alpha^\beta} \delta \tensor{g}{_\alpha_\beta} = \frac{1}{\sqrt{-g}} \left(\sqrt{-g} \delta \tensor{\mathcal{L}}{^\mu} \right) = \tensor{\mathcal{E}}{^\mu^\rho^\sigma} \delta \tensor{g}{_\rho_\sigma} + \tensor{\mathcal{G}}{^\mu^\nu} \delta \tensor{A}{_\nu} + D{_\nu} \tensor{\theta}{^\mu^\nu} \, ,
    \end{split}
\end{equation}
where
\begin{equation}
    \begin{split}
        \tensor{\mathcal{G}}{^\mu^\nu} &= \pdv{\tensor{\mathcal{L}}{^\mu}}{\tensor{A}{_\nu}} + 2 D{_\lambda} \pdv{\tensor{\mathcal{L}}{^\mu}}{\tensor{F}{_\nu_\lambda}} \, , ~~~
    \tensor{\mathcal{E}}{^\mu^\rho^\sigma} = \frac{1}{2} \tensor{\mathcal{L}}{^\mu} \tensor{g}{^\rho^\sigma} - \tensor{g}{^\rho^\alpha} \tensor{g}{^\sigma^\beta} \pdv{\tensor{\mathcal{L}}{^\mu}}{\tensor{g}{^\alpha^\beta}} - D{_\alpha} \tensor{\mathcal{S}}{^\mu^\alpha^\rho^\sigma} \, ,
    \\
    \tensor{\theta}{^\mu^\nu} &= 2 \pdv{\tensor{\mathcal{L}}{^\mu}}{\tensor{F}{_\nu_\lambda}} \delta \tensor{A}{_\lambda} + 2 \pdv{\tensor{\mathcal{L}}{^\mu}}{\tensor{R}{^\alpha_\beta_\nu_\lambda}} \delta \tensor{\Gamma}{^\alpha_\beta_\lambda} + \tensor{\mathcal{S}}{^\mu^\nu^\rho^\sigma} \delta \tensor{g}{_\rho_\sigma} \, ,
    \\
    & \text{with}, \\
    \tensor{\mathcal{S}}{^\mu^\nu^\rho^\sigma} &= \frac{1}{2} D{_\beta} \left[\tensor{g}{^\rho^\alpha} \left(\pdv{\tensor{\mathcal{L}}{^\mu}}{\tensor{R}{^\alpha_\sigma_\nu_\beta}} + \pdv{\tensor{\mathcal{L}}{^\mu}}{\tensor{R}{^\alpha_\nu_\sigma_\beta}} \right) + \tensor{g}{^\sigma^\alpha} \left(\pdv{\tensor{\mathcal{L}}{^\mu}}{\tensor{R}{^\alpha_\rho_\nu_\beta}} + \pdv{\tensor{\mathcal{L}}{^\mu}}{\tensor{R}{^\alpha_\nu_\rho_\beta}} \right) \right. \\ &\left. - \tensor{g}{^\nu^\alpha} \left(\pdv{\tensor{\mathcal{L}}{^\mu}}{\tensor{R}{^\alpha_\sigma_\rho_\beta}} + \pdv{\tensor{\mathcal{L}}{^\mu}}{\tensor{R}{^\alpha_\rho_\sigma_\beta}} \right) \right]
     + \frac{1}{2} \left(\tensor{g}{^\rho^\alpha} \pdv{\tensor{\mathcal{L}}{^\mu}}{\tensor{\Gamma}{^\alpha_\sigma_\nu}} + \tensor{g}{^\sigma^\alpha} \pdv{\tensor{\mathcal{L}}{^\mu}}{\tensor{\Gamma}{^\alpha_\rho_\nu}} - \tensor{g}{^\nu^\alpha} \pdv{\tensor{\mathcal{L}}{^\mu}}{\tensor{\Gamma}{^\alpha_\sigma_\rho}} \right) \, .
    \end{split}
\end{equation}
Thus, the total derivative term has zero equation of motion, but a $\Theta^{\mu}$ equalling
\begin{equation}\label{eq:totaltheta}
    \begin{split}
        \Theta^{\mu}_t = \delta \tensor{\mathcal{L}}{^\mu} + \frac{1}{2} \tensor{\mathcal{L}}{^\mu} \tensor{g}{^\alpha^\beta} \delta \tensor{g}{_\alpha_\beta}
    = \tensor{\mathcal{E}}{^\mu^\rho^\sigma} \delta \tensor{g}{_\rho_\sigma} + \tensor{\mathcal{G}}{^\mu^\nu} \delta \tensor{A}{_\nu} + D{_\nu} \tensor{\theta}{^\mu^\nu} \, .
    \end{split}
\end{equation}

Additionally, as we have
\begin{equation}
    \begin{split}
        \mathcal{L}_{\xi} \tensor{\mathcal{L}}{^\mu} &= \tensor{\xi}{^\nu} D{_\nu} \tensor{\mathcal{L}}{^\mu} - \tensor{\mathcal{L}}{^\nu} D{_\nu} \tensor{\xi}{^\mu} = D{_\nu} \left(\tensor{\mathcal{L}}{^\mu} \tensor{\xi}{^\nu} - \tensor{\mathcal{L}}{^\nu} \tensor{\xi}{^\mu} \right) - \tensor{\mathcal{L}}{^\mu} D{_\nu} \tensor{\xi}{^\nu} + \tensor{\xi}{^\mu} D{_\nu} \tensor{\mathcal{L}}{^\nu} \, ,
    \end{split}
\end{equation}
we get
\begin{equation}
    \delta \left(\sqrt{-g} D{_\mu} \tensor{\mathcal{L}}{^\mu} \right) = \sqrt{-g} D{_\mu} \left(\tensor{\xi}{^\mu} D{_\nu} \tensor{\mathcal{L}}{^\nu} \right) + \sqrt{-g} D{_\mu} \left(\left(\delta - \mathcal{L}_{\xi} \right) \tensor{\mathcal{L}}{^\mu} \right) \, .
\end{equation}
Thus from eq.\eqref{eq:diffvarl}, the $\Xi^{\mu}$ term equals
\begin{equation}\label{eq:totalxi}
    \begin{split}
        \Xi^{\mu}_t &= \delta \tensor{\mathcal{L}}{^\mu} - \mathcal{L}_{\xi} \tensor{\mathcal{L}}{^\mu} = \pdv{\tensor{\mathcal{L}}{^\mu}}{\tensor{A}{_\nu}} D{_\nu} \Lambda + \pdv{\tensor{\mathcal{L}}{^\mu}}{\tensor{\Gamma}{^\tau_\rho_\sigma}} \tensor{\partial}{^2_\rho_\sigma} \tensor{\xi}{^\tau} \, .
    \end{split}
\end{equation}

Finally, from eq.\eqref{eq:csnoethercharge}
\begin{equation}
    \begin{split}
        D_{\nu} Q^{\mu\nu}_{t} &= 2 \tensor{E}{^\mu^\nu} \tensor{\xi}{_\nu} + \tensor{G}{^\mu} \left(\tensor{A}{_\nu} \tensor{\xi}{^\nu} + \Lambda \right) + \tensor{\Theta}{^\mu} - \tensor{\xi}{^\mu} L - \tensor{\Xi}{^\mu} \\
    &= \delta \tensor{\mathcal{L}}{^\mu} + \tensor{\mathcal{L}}{^\mu} D{_\nu} \tensor{\xi}{^\nu} - \left(\delta - \mathcal{L}_{\xi} \right) \tensor{\mathcal{L}}{^\mu} - \tensor{\xi}{^\mu} D{_\nu} \tensor{\mathcal{L}}{^\nu} = D{_\nu} \left(\tensor{\mathcal{L}}{^\mu} \tensor{\xi}{^\nu} - \tensor{\mathcal{L}}{^\nu} \tensor{\xi}{^\mu} \right) \, ,
    \end{split}
\end{equation}
where the last equality follows by the relation on $\mathcal{L}_{\xi} \tensor{\mathcal{L}}{^\mu}$ derived above. This gives a Noether charge
\begin{equation}\label{eq:totalQ}
    Q^{\mu\nu}_t = \tensor{\mathcal{L}}{^\mu} \tensor{\xi}{^\nu} - \tensor{\mathcal{L}}{^\nu} \tensor{\xi}{^\mu} \, .
\end{equation}

Hence, we see that a total derivative term has zero equations of motion, but non-zero $\Theta^{\mu}$ eq.\eqref{eq:totaltheta}, non zero $\Xi^{\mu}$ eq.\eqref{eq:totalxi}, and Noether charge eq.\eqref{eq:totalQ}.

\subsection{Details of the structure of $\Xi^{\mu}$}
\label{ap:xistruc}

In this appendix, we will prove that $q^{\mu\nu}$ defined through $\Xi^{\mu}_{\mathcal{L}}$ of eq.\eqref{eq:xitoq} is linear $v$ and is $U(1)$ gauge invariant. To prove this, we will crucially use the general form of the $\Xi$ term eq.\eqref{eq:defxi}
\begin{equation}\label{eq:xildef}
    \Xi^{\mu}_{\mathcal{L}} = \pdv{\mathcal{L}}{\tensor{\Gamma}{^\lambda_\mu_\nu}} \tensor{\partial}{_\nu} \tensor{\xi}{^\lambda} - \frac{1}{\sqrt{-g}} \tensor{\xi}{^\lambda} \tensor{\partial}{_\nu} \left(\sqrt{-g} \pdv{\mathcal{L}}{\tensor{\Gamma}{^\lambda_\mu_\nu}} \right) \, .
\end{equation}

\subsubsection*{Linearity in $v$:}
We first note that $\Xi$ in eq.\eqref{eq:xildef} is linear in $\xi$: $\Xi(\xi_1 + \xi_2) = \Xi(\xi_1) + \Xi(\xi_2)$. As applying $D{_\nu}$ in eq.\eqref{eq:xitoq} does not cause any explicit factors of $\xi$ to appear, they must come, in our gauge, only from $\tensor{q}{^\mu^\nu}$. Additionally, as $\tensor{\partial}{^2_\mu_\nu} \tensor{\xi}{^\lambda} = 0$ in our gauge, the most general form that $\tensor{q}{^\mu^\nu}$ can take is
\begin{equation}
    \tensor{q}{^\mu^\nu} = \tensor{\mathcal{K}}{^\mu^\nu_\lambda} \tensor{\xi}{^\lambda} + \tensor{\mathcal{L}}{^\mu^\nu^\alpha_\beta} \tensor{\partial}{_\alpha} \tensor{\xi}{^\beta} \, .
\end{equation}
Thus, we must have
\begin{equation}\label{eq:xiqdef}
    \tensor{q}{^\mu^\nu} = \tensor{\tilde{q}}{^\mu^\nu} + v \tensor{w}{_v^\mu^\nu} \, ,
\end{equation}
where $\tilde{q}, w$ have no explicit factors of $v$.

\subsubsection*{Establishing gauge invariance:}
We know that $\Xi^{\mu}$ of eq.\eqref{eq:xildef} is $U(1)$ gauge invariant because $\mathcal{L}$ is $U(1)$ gauge invariant. However, it is not entirely obvious that $q^{\mu\nu}$ defined through eq.\eqref{eq:xitoq} is $U(1)$ gauge invariant. This is because we pull out an overall $D_{\nu}$ in eq.\eqref{eq:xitoq}. Thus, $q^{\mu\nu}$ may not be gauge invariant if we pull out a $D_{\nu}$ from $F_{\nu\alpha}$. We thus have to investigate the gauge invariance of $q^{\mu\nu}$ carefully. We will prove that $q^{\mu\nu}$ of eq.\eqref{eq:xiqdef} is $U(1)$ gauge invariant up to a total derivative. The analysis is similar in spirit to how we analyzed the gauge dependent terms in the CS Lagrangian in \S\ref{sec:gaugedepcslagrangian} and Appendix \ref{ap:gaugedepcslagrangian}. The most general form of $q^{\mu\nu}$ at the horizon consists of sums of the form
\begin{equation}\label{eq:qnmunu}
    \tensor{q}{_n^\mu^\nu} = \tensor{\mathcal{Q}}{^\mu^\nu^{\lambda_1}^\cdots^{\lambda_n}} \prod_{i=1}^n \tensor{A}{_{\lambda_i}} \, ,
\end{equation}
where $\tensor{\mathcal{Q}}{^\mu^\nu^{\lambda_1}^\cdots^{\lambda_n}}$ is totally symmetric in the $\lambda_i$ indices, antisymmetric in the $(\mu, \nu)$ indices, $U(1)$ gauge invariant, and linear in $\xi$ because of eq.\eqref{eq:xiqdef}. Taking the divergence of eq.\eqref{eq:qnmunu},
\begin{equation}\label{eq:divqmunu}
    \begin{split}
        D{_\nu} \tensor{q}{_n^\mu^\nu} &= \left(D{_\nu} \tensor{\mathcal{Q}}{^\mu^\nu^{\lambda_1}^\cdots^{\lambda_n}} \right) \left(\prod_{i=1}^n \tensor{A}{_{\lambda_i}} \right) + \sum_{j=1}^n \tensor{\mathcal{Q}}{^\mu^\nu^{\lambda_1}^\cdots^{\lambda_n}} \left(D{_\nu} \tensor{A}{_{\lambda_j}} \right) \left(\prod_{i=1, i\neq j}^n \tensor{A}{_{\lambda_i}} \right) \\
    &= \left(D{_\nu} \tensor{\mathcal{Q}}{^\mu^\nu^{\lambda_1}^\cdots^{\lambda_n}} \right) \left(\prod_{i=1}^n \tensor{A}{_{\lambda_i}} \right) + n \tensor{\mathcal{Q}}{^\mu^\nu^{\lambda\lambda_1}^\cdots^{\lambda_{n-1}}} \left(D{_\nu} \tensor{A}{_\lambda} \right) \left(\prod_{i=1}^{n-1} \tensor{A}{_{\lambda_i}} \right) \, .
    \end{split}
\end{equation}
Now, the functions $\tensor{A}{_\lambda}$ and $D{_{(\nu}} \tensor{A}{_{\lambda)}}$ are gauge non invariant. Thus, for $D{_\nu} \tensor{q}{^\mu^\nu} = \Xi^{\mu}$ of eq.\eqref{eq:xildef} to be gauge invariant, we must have
\begin{equation}
    \tensor{\mathcal{Q}}{^\mu^{(\nu}^{\lambda)}^{\lambda_1}^\cdots^{\lambda_n}} = 0 \, .
\end{equation}
That is, $\tensor{\mathcal{Q}}{^\mu^\nu^\lambda^{\lambda_1}^\cdots^{\lambda_n}}$ must be antisymmetric in $(\nu, \lambda)$. This implies that
\begin{equation}
    \begin{split}
        \tensor{\mathcal{Q}}{^\mu^\nu^\lambda^{\lambda_1}^\cdots^{\lambda_n}} &= - \tensor{\mathcal{Q}}{^\mu^\lambda^\nu^{\lambda_1}^{\lambda_2}^\cdots^{\lambda_n}} = - \tensor{\mathcal{Q}}{^\mu^{\lambda_1}^\nu^\lambda^{\lambda_2}^\cdots^{\lambda_n}} = \tensor{\mathcal{Q}}{^\mu^{\lambda_1}^\lambda^\nu^{\lambda_2}^\cdots^{\lambda_n}} = \tensor{\mathcal{Q}}{^\mu^\lambda^{\lambda_1}^\nu^{\lambda_2}^\cdots^{\lambda_n}} \, , \\
    \tensor{\mathcal{Q}}{^\mu^\nu^\lambda^{\lambda_1}^\cdots^{\lambda_n}} &= \tensor{\mathcal{Q}}{^\mu^\nu^{\lambda_1}^\lambda^\cdots^{\lambda_n}} = - \tensor{\mathcal{Q}}{^\mu^\lambda^{\lambda_1}^\nu^\cdots^{\lambda_n}} \, .
    \end{split}
\end{equation}
Hence, we get
\begin{equation}
    \tensor{\mathcal{Q}}{^\mu^\nu^{\lambda_1}^\cdots^{\lambda_n}} = 0 ~~~ \text{for} ~~~ n \ge 2 \, .
\end{equation}
This is entirely analogous eq.\eqref{eq:csgaugeconds}. Using this in the first term of the second step of eq.\eqref{eq:divqmunu}, we get
\begin{equation}\label{eq:dq123}
    D{_\nu} \tensor{\mathcal{Q}}{^\mu^\nu^\lambda} = 0 \, .
\end{equation}

eq.\eqref{eq:dq123} can be used to greatly constrain the form of $q^{\mu\nu}$ by an analysis similar to eq.\eqref{eq:lmugaugeconds} and eq.\eqref{eq:lmufinal}. To see this, first note that $\tensor{\mathcal{Q}}{^\mu^\nu^\lambda}$, being gauge invariant, has the general form
\begin{equation}\label{eq:defq123}
    \tensor{\mathcal{Q}}{^\mu^\nu^\lambda} = \sum_n \tensor{\mathcal{M}}{^\mu^\nu^\lambda^{\rho_1}^{\sigma_1}^\cdots^{\rho_n}^{\sigma_n}} \prod_{i=1}^n \tensor{F}{_{\rho_i}_{\sigma_i}} \, ,
\end{equation}
where $\tensor{\mathcal{M}}{^\mu^\nu^\lambda^{\rho_1}^{\sigma_1}^\cdots^{\rho_n}^{\sigma_n}}$ is independent of the $U(1)$ gauge field, antisymmetric in $(\rho_i, \sigma_i)$ for all $i$, symmetric in pairs of $(\rho_i, \sigma_i)$ and $(\rho_j, \sigma_j)$ for any pair $(i, j)$, and linear in $\xi$. Taking the derivative of eq.\eqref{eq:defq123}, we get
\begin{equation}
    \begin{split}
        D{_\nu} \tensor{\mathcal{Q}}{^\mu^\nu^\lambda} &= \sum_n \left(D{_\nu} \tensor{\mathcal{M}}{^\mu^\nu^\lambda^{\rho_1}^{\sigma_1}^\cdots^{\rho_n}^{\sigma_n}} \right) \left(\prod_{i=1}^n \tensor{F}{_{\rho_i}_{\sigma_i}} \right) \\ & \quad \quad + \sum_n n \tensor{\mathcal{M}}{^\mu^\nu^\lambda^\rho^\sigma^{\rho_1}^{\sigma_1}^\cdots^{\rho_{n-1}}^{\sigma_{n-1}}} \left(D{_\nu} \tensor{F}{_\rho_\sigma} \right) \left(\prod_{i=1}^{n-1} \tensor{F}{_{\rho_i}_{\sigma_i}} \right) \, .
    \end{split}
\end{equation}
As $\tensor{F}{_\rho_\sigma}, D{_\nu} \tensor{F}{_\rho_\sigma}$ are independent functions, $D{_\nu} \tensor{\mathcal{Q}}{^\mu^\nu^\lambda} = 0$ implies
\begin{equation}\label{eq:casesforq}
    \tensor{\mathcal{M}}{^\mu^\nu^\lambda^{\rho_1}^{\sigma_1}^\cdots^{\rho_n}^{\sigma_n}} = 
    \begin{cases}
        D{_\tau} \tensor{\mathcal{N}}{^\tau^\mu^\nu^\lambda^{\rho_1}^{\sigma_1}^\cdots^{\rho_n}^{\sigma_n}} & (i) ~~ 2n + 3 < D \\
        \mathcal{S} \, \tensor{\epsilon}{^\mu^\nu^\lambda^{\rho_1}^{\sigma_1}^\cdots^{\rho_n}^{\sigma_n}} & (ii) ~~ 2n+3 = D
    \end{cases}
\end{equation}
where $\tensor{\mathcal{N}}{^\tau^\mu^\nu^\lambda^{\rho_1}^{\sigma_1}^\cdots^{\rho_n}^{\sigma_n}}$ is totally antisymmetric and independent of the $U(1)$ gauge field, and $\mathcal{S}$ is a constant as $D{_\mu} \mathcal{S} = 0$. This equation is entirely analogous to eq.\eqref{eq:lmufinal}. But, as $\mathcal{M}$ is linear in $\xi$ because of eq.\eqref{eq:defq123}, we must have $\mathcal{N}, \mathcal{S}$ be linear in $\xi$. Thus, $\mathcal{S} = 0$ in case $(ii)$ of eq.\eqref{eq:casesforq}, and we get a non-zero answer from case $(i)$ of eq.\eqref{eq:casesforq}:
\begin{equation}
    \begin{split}
        \tensor{A}{_\lambda} \tensor{\mathcal{Q}}{^\mu^\nu^\lambda} &= \sum_n \tensor{A}{_\lambda} \left(D{_\tau} \tensor{\mathcal{N}}{^\tau^\mu^\nu^\lambda^{\rho_1}^{\sigma_1}^\cdots^{\rho_n}^{\sigma_n}} \right) \left(\prod_{i=1}^n \tensor{F}{_{\rho_i}_{\sigma_i}} \right) \\
    &= D{_\tau} \left[\sum_n \tensor{A}{_\lambda} \tensor{\mathcal{N}}{^\tau^\mu^\nu^\lambda^{\rho_1}^{\sigma_1}^\cdots^{\rho_n}^{\sigma_n}} \prod_{i=1}^n \tensor{F}{_{\rho_i}_{\sigma_i}} \right] - \frac{1}{2} \sum_n \tensor{\mathcal{N}}{^\tau^\mu^\nu^\lambda^{\rho_1}^{\sigma_1}^\cdots^{\rho_n}^{\sigma_n}} \tensor{F}{_\tau_\lambda} \prod_{i=1}^n \tensor{F}{_{\rho_i}_{\sigma_i}} \, .
    \end{split}
\end{equation}
Substituting the above equation in eq.\eqref{eq:qnmunu}, we have
\begin{equation}\label{eq:qmunufinal}
    \tensor{q}{^\mu^\nu} = \tensor{\mathcal{Q}}{^\mu^\nu} + D{_\tau} \left(\sum_{n=0}^{N-2} \tensor{A}{_\lambda} \tensor{\mathcal{N}}{^\tau^\mu^\nu^\lambda^{\rho_1}^{\sigma_1}^\cdots^{\rho_n}^{\sigma_n}} \prod_{i=1}^n \tensor{F}{_{\rho_i}_{\sigma_i}} \right) \, ,
\end{equation}
where $\tensor{\mathcal{Q}}{^\mu^\nu}$ is gauge invariant, and $\tensor{\mathcal{N}}{^\tau^\mu^\nu^\lambda^{\rho_1}^{\sigma_1}^\cdots^{\rho_n}^{\sigma_n}}$ is independent of the $U(1)$ gauge field and totally antisymmetric. We see that $q^{\mu\nu}$ defined by eq.\eqref{eq:xitoq} and eq.\eqref{eq:xiqdef} is $U(1)$ gauge invariant up to an incosequential total derivative term that drops out from eq.\eqref{eq:xitoq}. This completes our proof that $q^{\mu\nu}$ defined through eq.\eqref{eq:xitoq} is linear in $v$ and $U(1)$ gauge invariant up to a total derivative.

\subsection{Details of the intermediary calculations}

\subsubsection*{Analysis of the pure gauge term in eq.(\ref{eq:cslagform1}):}

The result of eq.\eqref{eq:lgevvgen} is obtained by
\begin{equation}\label{eq:lgevvgenap}
    \begin{split}
        &\quad\ \eval{\left[D{_\rho} Q^{r\rho}_g - \Theta^{r}_g + \Xi^{r}_g \right]}_{r=0} - G^{r}_g \left(\tensor{A}{_\rho} \tensor{\xi}{^\rho} + \Lambda \right) \\
    &= D{_\rho} \left[2 \pdv{L_g}{\tensor{F}{_r_\rho}} \left[v \tensor{A}{_v} + \Lambda \right] \right] - 2 \pdv{L_g}{\tensor{F}{_r_\rho}} \left[v \tensor{F}{_v_\rho} + D{_\rho} \left(v \tensor{A}{_v} + \Lambda \right) \right] + \pdv{L_g}{\tensor{A}{_r}} \Lambda \\ & \quad \quad -\left[\pdv{L_g}{\tensor{A}{_r}} + 2 D{_\rho} \pdv{L_g}{\tensor{F}{_r_\rho}} \right] \left[v \tensor{A}{_v} + \Lambda \right]
    \\
    &= - v \left(\pdv{L_g}{\tensor{A}{_r}} \tensor{A}{_v} + 2 \pdv{L_g}{\tensor{F}{_r_i}} \tensor{F}{_v_i} \right) \, .
    \end{split}
\end{equation}

The result of eq.\eqref{eq:lgevvfinal} is obtained by 
\begin{equation}\label{eq:lgevvfinalap}
    \begin{split}
        &\quad\ \eval{\left[D{_\rho} Q^{r\rho}_g - \Theta^{r}_g + \Xi^{r}_g \right]}_{r=0} - G^{r}_g \left(\tensor{A}{_\rho} \tensor{\xi}{^\rho} + \Lambda \right) \\
    &= - v \left(a_g \tensor{\epsilon}{^r^{\rho_1}^{\sigma_1}^\cdots^{\rho_N}^{\sigma_N}} \tensor{A}{_v} \prod_{i=1}^N \tensor{F}{_{\rho_i}_{\sigma_i}} + 2 N a_g \tensor{\epsilon}{^\mu^r^i^{\rho_1}^{\sigma_1}^\cdots^{\rho_{N-1}}^{\sigma_{N-1}}} \tensor{A}{_\mu} \tensor{F}{_v_i} \prod_{i=1}^{N-1} \tensor{F}{_{\rho_i}_{\sigma_i}} \right) \\
    &= - v \left(2N a_g \tensor{\epsilon}{^r^v^i^{j_1}^{k_1}^\cdots^{j_{N-1}}^{k_{N-1}}} \tensor{A}{_v} \tensor{F}{_v_i} \prod_{a=1}^{N-1} \tensor{F}{_{j_a}_{k_a}} \right. \\ & \quad \left. + 2 N a_g \tensor{\epsilon}{^v^r^i^{j_1}^{k_1}^\cdots^{j_{N-1}}^{k_{N-1}}} \tensor{A}{_v} \tensor{F}{_v_i} \prod_{a=1}^{N-1} \tensor{F}{_{j_a}_{k_a}} \right) = 0 \, .
    \end{split}
\end{equation}
This result is exact even up to $\mathcal{O}(\epsilon^2)$.

\subsubsection*{Contributions of the ``gauge" terms on the horizon:}

The result of eq.\eqref{eq:gaugeLcont} is obtained by
\begin{equation}\label{eq:gaugeLcontap}
    \begin{split}
        &\quad\ \eval{\left[D{_\rho} Q^{r\rho}_{\mathcal{L}} - \Theta^r_{\mathcal{L}} \right]}_{\text{gauge}} - G^r_{\mathcal{L}} \left(\tensor{A}{_\rho} \tensor{\xi}{^\rho} + \Lambda \right) \\
    &= D{_\rho} \left[2 \pdv{\mathcal{L}}{\tensor{F}{_r_\rho}} \left[v \tensor{A}{_v} + \Lambda \right] \right] - 2 \pdv{\mathcal{L}}{\tensor{F}{_r_\rho}} \left[v \tensor{F}{_v_\rho} + D{_\rho} \left(v \tensor{A}{_\lambda} + \Lambda \right) \right] - \left[\pdv{\mathcal{L}}{\tensor{A}{_r}} + 2 D{_\rho} \pdv{\mathcal{L}}{\tensor{F}{_r_\rho}} \right] \left[v \tensor{A}{_v} + \Lambda \right]
    \\
    &= -  \left( \dfrac{\partial\mathcal{L}}{\partial A_r} ( v \tensor{A}{_v} + \Lambda) + 2 \pdv{\mathcal{L}}{\tensor{F}{_r_i}} v\tensor{F}{_v_i} \right) = \mathcal{O}(\epsilon^2) \, .
    \end{split}
\end{equation}
Since $\mathcal{L}$ is $U(1)$ gauge invariant, the first term of the final step is zero. The second term of the final step becomes $\mathcal{O}(\epsilon^2)$ from boost weight analysis. 

\subsubsection*{$\Theta^r$ of ``gravity" terms on the horizon:}

Here we give the details of the calculation of $\Theta^r$ for the ``gravity" terms of eq.\eqref{eq:thetamuLbr}. We show that it is of the form given by eq.\eqref{eq:thetarfinal}. We first use eq.\eqref{eq:ELcs} in the ``gravity" terms of $\Theta^r$ in eq.\eqref{eq:thetamuLbr} to get
\begin{equation}\label{eq:thetarLinter}
    \begin{split}
        \eval{\Theta^r_{\mathcal{L}}}_{\text{gravity}} &= 2 \pdv{\mathcal{L}}{\tensor{R}{^\alpha_\beta_r_\nu}} \delta \tensor{\Gamma}{^\alpha_\beta_\nu} +S^{r\alpha\beta}_{\mathcal{L}} \delta \tensor{g}{_\alpha_\beta} = E^{r\alpha\beta\nu}_{\mathcal{L}} D{_\nu} \delta \tensor{g}{_\alpha_\beta} + S^{r\alpha\beta}_{\mathcal{L}} \delta \tensor{g}{_\alpha_\beta} \\
    &= D{_\nu} \left(E^{r\alpha\beta\nu}_{\mathcal{L}} \delta \tensor{g}{_\alpha_\beta} \right) + \left(S^{r\alpha\beta}_{\mathcal{L}} - D{_\nu} E^{r\alpha\beta\nu}_{\mathcal{L}} \right) \delta \tensor{g}{_\alpha_\beta} \, .
    \end{split}
\end{equation}
Now, we have for any $\tensor{U}{^r^\alpha^\beta}$ on the horizon $r=0$
\begin{equation}\label{eq:Udeltag}
    \tensor{U}{^r^\alpha^\beta} \delta \tensor{g}{_\alpha_\beta} = v \tensor{U}{^r^m^n} \tensor{\partial}{_v} \tensor{h}{_m_n} = \mathcal{O}\left(\epsilon^2 \right) \, .
\end{equation}
Defining
\begin{equation}\label{eq:defPmunu}
    \tensor{P}{^\mu^\nu} = E^{\mu\alpha\beta\nu}_{\mathcal{L}} \delta \tensor{g}{_\alpha_\beta} \, ,
\end{equation}
and using eq.\eqref{eq:Udeltag} in eq.\eqref{eq:thetarLinter}, we get
\begin{equation}\label{eq:thetarP}
    \eval{\Theta^r_{\mathcal{L}}}_{\text{gravity}} = D{_\rho} \tensor{P}{^r^\rho} + \mathcal{O}\left(\epsilon^2 \right) \, .
\end{equation}
From eq.\eqref{eq:Udeltag} and eq.\eqref{eq:defPmunu}, it is clear that $\tensor{P}{^r^v}$ and  $\tensor{P}{^r^r}$ are $\mathcal{O}\left(\epsilon \right)$ whereas $\tensor{P}{^r^i}$ is $\mathcal{O}\left(\epsilon^2 \right)$ at the horizon. Using this, we can compute
\begin{equation}
    \begin{split}
        D{_\rho} \tensor{P}{^r^\rho} = \tensor{\partial}{_\rho} \tensor{P}{^r^\rho} + \tensor{\Gamma}{^r_\lambda_\rho} \tensor{P}{^\lambda^\rho} + \tensor{\Gamma}{^\rho_\lambda_\rho} \tensor{P}{^r^\lambda} = \tensor{\partial}{_r} \tensor{P}{^r^r} + \tensor{\partial}{_v} \tensor{P}{^r^v} + \mathcal{O}(\epsilon^2) \, .
    \end{split}
\end{equation}
This can finally be further simplified by using $E^{rmnr}_{\mathcal{L}} = \tensor{\partial}{_v} \tensor{J}{_{(1)}^m^n} + \mathcal{O}(\epsilon^2)$ (which follows from generic boost weight argument of eq.\eqref{eq:genboostwt}) as
\begin{equation}\label{eq:dPexpr}
    \begin{split}
        \tensor{\partial}{_v} \tensor{P}{^r^v} &= \left(1 + v \tensor{\partial}{_v} \right) \left(E^{rmnv}_{\mathcal{L}} \tensor{\partial}{_v} \tensor{h}{_m_n} \right) \, , \\
    \tensor{\partial}{_r} \tensor{P}{^r^r} &= E^{rmnr}_{\mathcal{L}} \left(v \tensor{\partial}{^2_v_r} - \tensor{\partial}{_r} \right) \tensor{h}{_m_n} \\
    &= - \left(1 + v \tensor{\partial}{_v} \right) \left(E^{rmnr}_{\mathcal{L}} \tensor{\partial}{_r} \tensor{h}{_m_n} \right) + v \tensor{\partial}{_v} \left(E^{rmnr}_{\mathcal{L}} \tensor{\partial}{_r} \tensor{h}{_m_n} \right) + v E^{rmnr}_{\mathcal{L}} \tensor{\partial}{^2_v_r} \tensor{h}{_m_n} \\
    &= - \left(1 + v \tensor{\partial}{_v} \right) \left(E^{rmnr}_{\mathcal{L}} \tensor{\partial}{_r} \tensor{h}{_m_n} \right) + v \tensor{\partial}{_v} \left(\left(\tensor{\partial}{_v} \tensor{J}{_{(1)}^m^n} \right) \left(\tensor{\partial}{_r} \tensor{h}{_m_n} \right) + \tensor{J}{_{(1)}^m^n} \tensor{\partial}{^2_v_r} \tensor{h}{_m_n} \right) + \mathcal{O}\left(\epsilon^2 \right) \\
    &= - \left(1 + v \tensor{\partial}{_v} \right) \left(E^{rmnr}_{\mathcal{L}} \tensor{\partial}{_r} \tensor{h}{_m_n} \right) + v \tensor{\partial}{^2_v} \left(\tensor{J}{_{(1)}^m^n} \tensor{\partial}{_r} \tensor{h}{_m_n} \right) + \mathcal{O}\left(\epsilon^2 \right) \, .
    \end{split}
\end{equation}

Substituting eq.\eqref{eq:dPexpr} in eq.\eqref{eq:thetarP}, we finally get eq.\eqref{eq:thetarfinal}
\begin{equation}\label{eq:thetarfinalap}
    \eval{\Theta^r_{\mathcal{L}}}_{\text{gravity}} = \left(1 + v \tensor{\partial}{_v} \right) \tensor{\mathcal{A}}{_{(1)}} + v \tensor{\partial}{^2_v} \tensor{\mathcal{B}}{_{(0)}} + \mathcal{O} \left(\epsilon^2 \right) \, ,
\end{equation}
where $\tensor{\mathcal{B}}{_{(0)}}$ is clearly $\mathcal{O}\left(\epsilon \right)$ because it is of the form eq.\eqref{eq:B0diffgen} and $\tensor{\mathcal{A}}{_{(1)}}, \tensor{\mathcal{B}}{_{(0)}}$ are $U(1)$ gauge invariant.

\subsection{Useful quantities in our gauge}
\label{ap:christoffel}

The gauge choice we work with is eq.\eqref{nhmetric}. The horizon is located at $r=0$. The inverse metric components are given by:
\begin{equation}
    \begin{split}
        g^{rr} = r^2 X(r,v,x^i) + r^2 \omega^i (r,v,x^i) \omega_i (r,v,x^i) = r^2 X + r^2 \omega^2 \\
        g^{ri} = - r \omega^i \hspace{0.8cm} g^{rv} = 1 \hspace{0.8cm} g^{ij} = h^{ij} \hspace{0.8cm} g^{vv} = g^{vi} =0
    \end{split}
\end{equation}
We define the following (extrinsic curvature) quantities:
\begin{equation}
    \begin{split}
        K_{ij} = \dfrac{1}{2} \partial_v h_{ij} \hspace{0.8cm} \Bar{K}_{ij} = \dfrac{1}{2} \partial_r h_{ij} \hspace{0.8cm} K^{ij} = -\dfrac{1}{2} \partial_v h^{ij} \hspace{0.8cm} \Bar{K}^{ij} = - \dfrac{1}{2} \partial_r h^{ij} \\
        K = \dfrac{1}{2} h^{ij} \partial_v h_{ij} = \dfrac{1}{\sqrt{h}} \partial_v \sqrt{h} \hspace{0.8cm} \Bar{K} = \dfrac{1}{2}h^{ij} \partial_r h_{ij} = \dfrac{1}{\sqrt{h}} \partial_r \sqrt{h}
    \end{split}
\end{equation}
The Christofell symbols for the metric are derived to be:
\begin{gather*}
        \Gamma^v_{vv} = \dfrac{1}{2}(2r X + r^2 \partial_r X) \hspace{0.8cm} \Gamma^v_{vr} = 0 \hspace{0.8cm} \Gamma^v_{vi} = -\dfrac{1}{2}(\omega_i + r \partial_r \omega_i) \\
        \Gamma^r_{vv} = \dfrac{1}{2}(2r^3 X^2 + 2 r^3 X \omega^2 + r^4 X \partial_r X + r^4 \omega^2 \partial_r X - r^2 \partial_v X - 2 r^2 \omega^i \partial_v \omega_i - r^3 \omega^i \partial_i X) \\
        \Gamma^r_{vr} = - \dfrac{1}{2}(2rX + r^2 \partial_r X + r \omega^2 + r^2 \omega^i \partial_r \omega_i) \\
        \Gamma^r_{vi} = -\dfrac{1}{2}(r^3 X \partial_r \omega_i + r^3 \omega^2 \partial_r \omega_i + r^2 X \omega_i + r^2 \omega^2 \omega_i + r^2 \partial_i X + r^2 \omega^j \partial_i \omega_j - r^2 \omega^j \partial_j \omega_i) - r \omega^j K_{ij} \\
        \Gamma^i_{vv} = \dfrac{1}{2}(-2r^2 \omega^i X - r^3 \omega^i \partial_r X + 2 h^{ij}r \partial_v \omega_j + r^2 h^{ij}\partial_j X) \\
        \Gamma^i_{vj} = \dfrac{1}{2}r\omega^i(\omega_j + r \partial_r \omega_j) + \dfrac{1}{2}h^{ik}(r \partial_j \omega_k - r \partial_k \omega_j) + h^{ik} K_{jk} \\
        \Gamma^i_{vr} = \dfrac{1}{2}h^{ij}(r \partial_r \omega_j + \omega_j) \hspace{0.8cm} \Gamma^v_{rr} =0 \hspace{0.8cm} \Gamma^v_{ri}=0 \hspace{0.8cm} \Gamma^v_{ij} = - \Bar{K}_{ij} \hspace{0.8cm} \Gamma^r_{rr} = 0 \\
        \Gamma^i_{rr} = 0 \hspace{0.8cm} \Gamma^r_{ri} = - r \omega^j \Bar{K}_{ij} + \dfrac{1}{2}(\omega_i + r \partial_r \omega_i) \\
        \Gamma^r_{ij} = \dfrac{1}{2}(r \partial_i \omega_j + r \partial_j \omega_i) - r \omega_k \hat{\Gamma}^k_{ij} - (r^2 X + r^2 \omega^2) \Bar{K}_{ij} - K_{ij} \\
        \Gamma^i_{rj} = h^{ik} \Bar{K}_{jk} \hspace{0.8cm} \Gamma^i_{jk} = r \omega^i \Bar{K}_{jk} + \hat{\Gamma}^i_{jk}
\end{gather*}
where $\hat{\Gamma}^i_{jk}$ is the Christoffel symbol with respect to the induced metric on the horizon $h_{ij}$. The relevant expressions for curvature components at the horizon is taken from \cite{Bhattacharya:2019qal}:
\begin{equation}
    \begin{split}
        R_{rvrv} &= X + \dfrac{1}{4} \omega^2 \\
        R_{vjvr} &= \dfrac{1}{2}(\partial_v \omega_j + \omega^k K_{jk}) \\
        R_{vlvm} &= - \partial_v K_{lm} + K_{ln} K^n_{\hspace{0.2cm}m} \\
        R_{rv} &= -X - \dfrac{1}{2} \omega^2 - \partial_r K - \Bar{K}_{ij} K^{ij} + \dfrac{1}{2} \nabla^i \omega_i \\
        R_{jv} &= - \dfrac{1}{2} \partial_v \omega_j + \nabla_n K^{n}_{\hspace{0.2cm}j} - \nabla_j K - \dfrac{1}{2} \omega_j K \\
        R_{vv} &= - \partial_v K - K_{mn} K^{mn} \\
        R &= \hat{R} - 2 X -\dfrac{3}{2} \omega^2 - 4 \partial_r K + 2 (\nabla^i \omega_i) -2 \Bar{K}_{ij} K^{ij} - 2 \Bar{K} K
    \end{split}
\end{equation}

\bibliographystyle{JHEP}
\bibliography{chernsimons}

\end{document}